\documentclass[superscriptaddress,prx,twocolumn,nofootinbib,citeautoscript,longbibliography,notitlepage]{revtex4-1}
\usepackage{graphicx}
\usepackage{dcolumn}
\usepackage{bm}
\usepackage{color}
\usepackage{amsmath}
\usepackage{tabularx,graphicx}
\usepackage{epstopdf}
\usepackage{latexsym}
\usepackage{amssymb}
\usepackage{amsmath}
\usepackage{color, colortbl}
\usepackage{psfrag}
\usepackage{bbm}
\usepackage{booktabs}
\usepackage{bm}
\usepackage{titlesec}
\usepackage{dsfont}
\usepackage{feynmp}
\usepackage{slashed}
\usepackage{multirow}
\usepackage[tight]{subfigure}
\usepackage{comment}
\usepackage{float}
\usepackage{diagbox} 
\usepackage{makecell}
\usepackage{array}
\usepackage[papersize={8.5in,11in}]{geometry}
\usepackage{color}
\usepackage{fmtcount} 
\definecolor{darkblue}{rgb}{0.,0.,0.4}
\definecolor{darkred}{rgb}{0.5,0.,0.}
\definecolor{BlueViolet}{RGB}{138,43,226}
\definecolor{SkyBlue}{RGB}{30,144,255}
\definecolor{DarkGreen}{RGB}{0,100,0}

\usepackage[pdftex,colorlinks=true,linkcolor=darkblue,citecolor=blue,urlcolor=darkred]{hyperref}

\renewcommand{\epsilon}{\varepsilon}

\allowdisplaybreaks[3]

\usepackage{cases}

\allowdisplaybreaks[4]

\UseRawInputEncoding

\begin{document}

%

\title{Nontrivial topological properties from the cooperation of static perturbations and Floquet
fields in type-II semi-Dirac semimetals}

\author{Wen-Hao Bian}
\altaffiliation{whbian@smail.nju.edu.cn}
\affiliation{Department of Physics, Tianjin University, Tianjian 300072, P.R. China}
\affiliation{School of Physics, Nanjing University, Nanjing, Jiangsu 210093, P.R. China}

\author{Jing Wang}
\altaffiliation{jing$\textunderscore$wang@tju.edu.cn}
\affiliation{Department of Physics, Tianjin University, Tianjian 300072, P.R. China}
\affiliation{Tianjin Key Laboratory of Low Dimensional Materials Physics and Preparing Technology,
Tianjin University, Tianjin 300072, P.R. China}

\date{\today}


\begin{abstract}

Two-dimensional type-II semi-Dirac semimetals (SDSMs) host strongly anisotropic and nonreciprocal band structures that intrinsically break both spatial-inversion and time-reversal symmetries, providing a promising platform for topological and nontrivial transport phenomena. In this work, we present a comprehensive theoretical study for the topological and anomalous transport properties
of type-II SDSMs under three scenarios, i.e., static symmetry-breaking perturbations, periodic Floquet driving by two-frequency orthogonally polarized light (TFOPL), and their cooperative effects in the hybrid scenario. We first show that a static mass $m_z$ opens a bulk gap and drives the system into a Chern insulator with Chern number $C^{\rm per}=\pm 2\,{\rm sgn}(m_z)$. In comparison, TFOPL alone only shifts the positions of Dirac points and cannot open a full gap. Nevertheless, it produces a Berry curvature odd in the $x$-direction, yielding a finite Berry curvature dipole (BCD) along $x$ proportional to the sine of the initial phase, while the $y$-component vanishes. The hybrid scenario, by contrast, exhibits a remarkably rich phenomenology. Specifically, its gap-closure criterion becomes cubic and thus admits up to three real roots when the effective parameter along the $x$-direction is negative, thereby producing multiple light-tunable gap-closing points and a series of topological phase transitions. In particular, the BCD exhibits a double-peak structure with strong enhancement near gap-closing manifolds and synchronized sign reversal of the full vector under external tuning. We further analyze the anomalous Hall conductivity (AHC), anomalous Nernst conductivity (ANC), and anomalous thermal Hall conductivity (ATHC). The AHC and ATHC exhibit quantized plateaus that serve as signatures of Chern-insulator phases, while the ANC, being a Fermi-surface probe, provides complementary local information and can identify light-induced gaps even when global topological invariants remain unchanged. Taken together, these results offer useful insights into the topological properties of 2D type-II SDSMs and related systems.

\end{abstract}


\maketitle

\section{Introduction}

Topological phases of matter have emerged as a central theme in condensed-matter physics,
revealing deep connections among band geometry, symmetry, and transport phenomena~~\cite{Thouless1982PRL,Haldane1988PRL,Hatsugai1993PRB,Qi2006PRB,Qi2011RMP,Chang2023RMP,
Nagaosa2010RMP,Xiao2010RMP,Sodemann2015PRL,Ma2019Nature,Vaidya2024PRL}. In particular, Chern
insulators constitute a paradigmatic family of gapped topological quantum phases and represent
one of the most active research frontiers in contemporary condensed-matter physics~\cite{Thouless1982PRL,Haldane1988PRL,Hatsugai1993PRB,Qi2006PRB,Chang2023RMP,Vaidya2024PRL}.
A nonvanishing Chern number yields quantized Hall conductance without external magnetic fields or Landau levels~\cite{Thouless1982PRL,Haldane1988PRL,Hatsugai1993PRB,Qi2006PRB,Chang2023RMP,Vaidya2024PRL}.
These phases can be constructed by opening a full bulk gap at a band-touching node via an appropriate
symmetry-breaking mass term. Within the low-energy continuum description, the sign and magnitude of
the resulting total Chern number are determined by the symmetry properties and momentum-space structure
of this mass term across all valleys~\cite{Haldane1988PRL,Chang2023RMP,Yu2010SCI,Qi2011RMP}.

SDSMs are particularly promising for realizing mass-driven topological
transitions~\cite{Huang2015PRB,Shao2024PRX,Uchoa2026arXiv,Chen2022PRB,Elsayed2026PRR}.
They are conventionally classified into two types~\cite{Huang2015PRB}. Type-I SDSMs possess a low-energy
Hamiltonian containing only quadratic and linear terms along two orthogonal momentum directions.
Type-II SDSMs, by contrast, feature an additional off-diagonal cross term that couples the two
momentum components, fundamentally reshaping the Fermi-surface geometry. This cross term breaks
time-reversal symmetry and, together with the linear term, breaks spatial inversion, producing
nonreciprocal band features~\cite{Huang2015PRB,Valagiannopoulos2020PRR}. Owing to its unique band structure, the type-II
variant has attracted growing attention as a natural setting for nonlinear transport and Berry-curvature-driven phenomena~\cite{Huang2015PRB,Shao2024PRX,Uchoa2026arXiv,Chen2022PRB,Elsayed2026PRR}.
Recent theoretical studies have explored Coulomb interaction effects and the renormalization of
topologically nontrivial semi-Dirac fermions in candidate materials such as $\mathrm{TiO_2/VO_2}$ heterostructures~\cite{Elsayed2026PRR,Huang2015PRB,Pardo2009PRL,WGZ2017PRB}.
However, since its spectrum remains gapless at $\Gamma$, the free type-II SDSM requires additional
tuning parameters to acquire a well-defined Chern number.

Two distinct strategies have been employed. The first comprises static
symmetry-breaking perturbations. A spin-orbit-generated mass term opens a full bulk gap
and induces a Chern-insulating phase, while in-plane perturbations merely shift or split the nodal point~\cite{Huang2015PRB,Shao2024PRX,Uchoa2026arXiv,Chen2022PRB,Elsayed2026PRR}. The second is
periodic Floquet driving, which enables the engineering of
topological band structures without modifying the equilibrium Hamiltonian~\cite{Xiong2023PRB,Rudner2020NRP,Kang2020PRA,Wang2023PRA,Maryam2026arXiv,
Kukhar2022ZTF,Kanega2025PRB,Lesko2025arXiv,Saha2016PRB}.
In SDSMs, the off-resonant light is able to induce topological phase transitions and reshapes thermoelectric transport~\cite{Vargiamidis2025PRB}, while multi-frequency, orthogonally polarized fields
generate effective gauge- and mass-like dressings of the band structure~\cite{Xiong2023PRB,Rudner2020NRP,Kang2020PRA,Wang2023PRA,Maryam2026arXiv,Kukhar2022ZTF,Kanega2025PRB,
Lesko2025arXiv,Saha2016PRB,Vargiamidis2025PRB,Oka2019ARCMP,Inoue2010PRL}. Nevertheless, the interplay
between these two strategies in SDSMs has yet largely unexplored. In particular, whether the Floquet
fields can cooperate with static perturbations to generate new topological phases remains an open question.

Closely tied to band topology is the concept of Berry curvature (BC), which encodes the local geometry
of Bloch wavefunctions in momentum space~\cite{Nagaosa2010RMP,Saha2016PRB,Xiao2010RMP,Sodemann2015PRL}.
Its integral over the Brillouin zone gives the Chern number~\cite{Thouless1982PRL,Haldane1988PRL,Hasan2010RMP,Qi2011RMP}.
Beyond this integral, the momentum-space distribution of BC governs a wide range of transport phenomena.
A particularly important quantity is the BCD, defined as the Fermi-surface-weighted integral of
the Berry-curvature gradient~\cite{Ma2019Nature,Faisal2026arXiv,Sodemann2015PRL,Battilomo2019PRL,Ortix2021AQT,Liao2025Small}.
Although they have been studied in Weyl semimetals and related systems~\cite{Pang2023arXiv,Roy-Narayan2022JPCM,Du-Xie2018PRL},
the behavior in type-II SDSMs under the two strategies presented above has not been systematically investigated.

Anomalous transport coefficients offer complementary probes of band topology and geometry~\cite{Thouless1982PRL,Haldane1988PRL,Qi2006PRB,Nagaosa2010RMP,Chang2013SCI,Chen2022SA,
Ikhlas2017NP,Yokoi2021SCI,Wan2026arXiv,Xiao2010RMP,Xiao2006PRL,Luttinger1964PR}.
On one hand, the anomalous Hall conductivity (AHC) and anomalous
thermal Hall conductivity (ATHC) are tied to the global topological invariant~\cite{Haldane1988PRL,Nagaosa2010RMP,Qin2011PRL},
manifesting as quantized plateaus in Chern-insulator phases. On the other hand, the anomalous Nernst
conductivity (ANC) is a Fermi-surface property that resolves local Berry-curvature
features and can detect gap openings even when the global topological invariant remains
unchanged~\cite{Xiao2006PRL,Saha2018EPJB,Ferreiros2017PRB,Alisultanov2025arXiv}.
Nevertheless, a unified study of these quantities in type-II SDSMs under static,
Floquet, and hybrid conditions is still lacking.

Motivated by these unresolved issues, we develop a comprehensive theoretical
framework for the topological and anomalous transport properties of two-dimensional
type-II SDSMs in three scenarios, namely (i) static perturbations, including
a symmetry-breaking mass term that opens a bulk gap, (ii) periodic Floquet driving
by TFOPL, and (iii) their cooperative
interplay in the hybrid scenario.

We begin by introducing the free type-II SDSM Hamiltonian
and then examine the static perturbative scenario, in which in-plane shifts $(\Delta_x,\Delta_y)$
and a spin-orbit mass term $m_z\sigma_z$ are added to the free Hamiltonian~(\ref{eq:Hamiltonian_per}).
It shows that $\Delta_{x,y}$ only move or split the nodal points,
whereas the mass $m_z\sigma_z$ opens a full gap and drives a Chern insulator
with $C^{\rm per}=\pm 2\,{\rm sgn}(m_z)$.

Next, we turn to the Floquet scenario, where TFOPL generates a momentum-dependent Zeeman term $\gamma k_x\sigma_z$.
Within first-order Floquet theory, TFOPL alone cannot open a full gap, but it produces a BC with odd parity in $k_x$,
yielding a vanishing Chern number and a finite BCD component $D_x\propto\sin\phi$ while eliminating $D_y$.

Furthermore, we consider the hybrid scenario in Eq.~(\ref{eq:H_tot}), which combines
static perturbations and Floquet driving by replacing the in-plane shift $\Delta_x$
with $\tilde{\Delta}_x\equiv \Delta_x+\delta_x$~(\ref{Eq_h_xyz}). This scenario exhibits
the central cooperative effects. The Floquet term $\gamma k_x \sigma_z$
competes with $m_z\sigma_z$ in-plane perturbations. Specifically, we derive
a gap-closure criterion that defines a surface in the three-dimensional
parameter space $(\tilde{\Delta}_x,\Delta_y,m_z)$, as given
in Eq.~(\ref{eq:condition}). We find that this criterion becomes cubic
in $m_z$ and admits up to three distinct real roots when $\tilde{\Delta}_x<0$,
corresponding to multiple gap-closing surfaces and sequential topological phase
transitions as $m_z$ is varied. Notably, a Chern insulator emerges even at $m_z=0$,
with its sign governed by $\mathrm{sgn}(\phi\Delta_y)$. The BCD exhibits a characteristic
double-peak structure, a strong enhancement near gap-closing manifolds,
and a synchronized sign reversal of the full BCD vector under tuning of
external parameters.

Finally, we analyze the anomalous transport coefficients displayed in
Figs.~\ref{fig:AHC}-\ref{fig:ATHC}. The AHC and ATHC exhibit quantized plateaus
that serve as signatures of Chern-insulator phases. By contrast, the ANC, as a
Fermi-surface probe, provides complementary local information and can identify
light-induced gaps even when the global topological invariants remain unchanged.
We further show that combined measurements of the AHC, ANC, and ATHC can confirm
the cooperative topological mechanism and distinguish it from purely perturbative
or Floquet scenarios.

The remainder of this paper is organized as follows. The free model and its symmetry properties
are introduced in Sec.~\ref{Sec_free_model}. In Sec.~\ref{Sec_effectie_model}, the static perturbative,
Floquet, and perturbative-Floquet hybrid scenarios are presented, their topological responses are formulated,
and the hybrid gap-closure criterion is derived. Sec.~\ref{Sec_BC} is followed to study the BC, Chern-number
phase diagrams, and BCD in all three scenarios. Subsequently, in Sec.~\ref{Sec_transport}, the behavior
of three types of anomalous transport coefficients is presented for these scenarios. Finally, we summarize
our main results and discuss experimental implications in Sec.~\ref{Sec_summary}.

\section{Free model and band structure}\label{Sec_free_model}

We begin with introducing the free model for two-dimensional (2D) type-II SDSMs, whose
low-energy effective Hamiltonian can be written as~\cite{Huang2015PRB,Chen2022PRB,Xiong2023PRB,Shao2024PRX,Vargiamidis2025PRB,Liao2025Small,Elsayed2026PRR} 
\begin{equation}
\mathcal{H}(\mathbf{k})=(g_1k_x^2-vk_y)\sigma_x+g_2k_xk_y\sigma_y,\label{eq:Hamiltonian}
\end{equation}
which is generally expressed in a compact form
\begin{equation}
\mathcal{H}(\mathbf{k})=\mathbf{d}(\mathbf{k})\cdot \bm{\sigma}
\end{equation}
with the pseudomagnetic vector
\begin{equation}
\mathbf{d}(\mathbf{k})=\big(g_{1} k_{x}^{2}-v k_{y},\ g_{2} k_{x} k_{y},\ 0\big),\label{Eq_d_k}
\end{equation}
and $\sigma_{x,y,z}$ Pauli matrices
acting on the pseudospin degree of freedom of the low-energy two-band subspace.
Here, $g_1$ is the inverse effective mass governing the parabolic dispersion
along $k_x$, $v$ is the Fermi velocity for the linear $k_y$ dispersion,
and $g_2$ is the cross-anisotropy coefficient responsible for the characteristic type-II
correction of semi-Dirac fermions.

At first, it is of significance to emphasize that the type-II SDSMs are featured by an off-diagonal cross term ($\sim k_xk_y$) that
hybridizes the two momentum components, which is distinct from their type-I counterparts~\cite{Huang2015PRB}
whose energy spectrum is characterized by a quadratic dispersion along one momentum axis and a linear Dirac-like
dispersion along the perpendicular axis. This very term fundamentally modifies the Fermi-surface topology, breaks
certain intrinsic lattice symmetries, and consequently induces nonreciprocal band characteristics.

To proceed, we move to examine the symmetry properties of low-energy effective Hamiltonian~(\ref{eq:Hamiltonian}).
Considering the spatial inversion $\mathcal{P}$, under which momentum flips sign, namely,   $\mathcal{P}\mathbf{k}\mathcal{P}^{-1}=-\mathbf{k}$. Applying this transformation gives
\begin{equation}
\mathcal{P}\mathcal{H}(\boldsymbol{k})\mathcal{P}^{-1}=\big(g_1 k_x^2 +v k_y\big)\sigma_x + g_2
k_x k_y \sigma_y \neq \mathcal{H}(\boldsymbol{k}),
\end{equation}
which explicitly demonstrates that inversion symmetry is intrinsically broken in the free Hamiltonian of type-II SDSM.
As for the time-reversal operation $\mathcal{T}$, defined by $\mathcal{T}\mathbf{k}\mathcal{T}^{-1}=-\mathbf{k}$,
$\mathcal{T}\sigma_{x}\mathcal{T}^{-1}=\sigma_{x}$, and $\mathcal{T}\sigma_{y}\mathcal{T}^{-1}=-\sigma_{y}$.
Its action on the free Hamiltonian yields
\begin{eqnarray}
\mathcal{T}\mathcal{H}(\mathbf{k})\mathcal{T}^{-1}\!=\!\!\big(g_1 k_x^2 \!+\!v k_y\big)\sigma_x
\!-\!g_2 k_x k_y \sigma_y \!\neq \!\!\mathcal{H}(-\mathbf{k}),
\end{eqnarray}
thus confirming that time-reversal symmetry is likewise violated within this free model.

\begin{figure}
\centering
\includegraphics[width=2.5in]{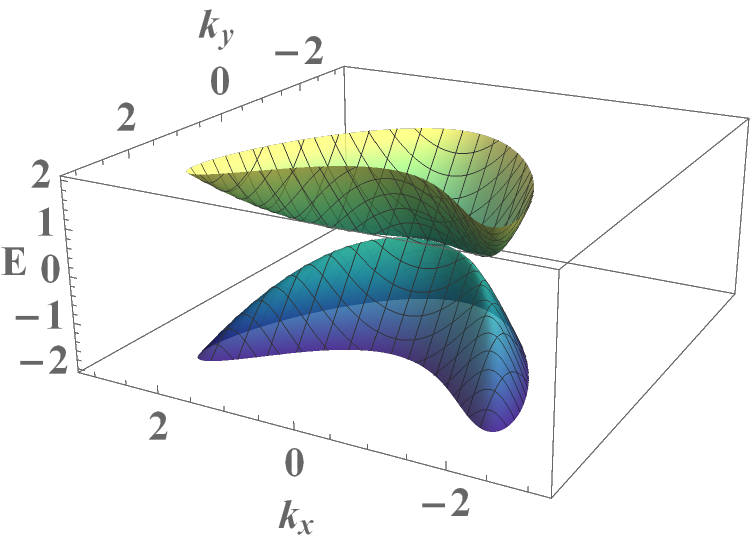}
\vspace{-0.2cm}
\caption{(Color online) Low-energy dispersion of the free type-II SDSM at $g_2/g_1=0.41$ and $v=2$.
The nonzero cross-coupling violates $\mathcal{P}$ and $\mathcal{T}$, giving rise to asymmetric nonreciprocal bands.}
\label{fig:E_k}
\end{figure}

The concomitant breaking of both symmetries inevitably renders the effective Hamiltonian intrinsically
nonreciprocal~\cite{Wakatsuki2017SA,Stavric2025PRL}, which induces nonreciprocity in the
energy dispersion as shown in Fig.~\ref{fig:E_k}. It is worth highlighting that this symmetry breaking
does not rely on external probes, such as external magnetic or electric fields~\cite{Tokura2018NC,Zhang2021PNASUSA},
but instead it is deeply rooted in the microscopic details, including the
lattice distortions, orbital hybridization, or crystalline point-group constraints
of realistic type-II SDSM candidates~\cite{Schoop2016NC}. As we address later, these broken
symmetries ultimately serve as a foundation for engineering topological responses via static
perturbations and periodic Floquet protocols in the remainder of this work.

Moreover, let us consider the band structure of free model.
This minimal two-band model of type-II SDSMs~(\ref{eq:Hamiltonian}) captures the
key low-energy excitations near the $\Gamma$ point,
neglecting high-energy bands far from the Fermi level.

Diagonalizing the free Hamiltonian~(\ref{eq:Hamiltonian}) gives particle-hole symmetric eigenvalues
$E(\mathbf{k})=\pm |\mathbf{d}(\mathbf{k})|$, which, after inserting the explicit components of
$\mathbf{d}(\mathbf{k})$~(\ref{Eq_d_k}), become
\begin{eqnarray}
E_\pm(\mathbf{k})=\pm\sqrt{(g_1k_x^2-vk_y)^2+g_2^2k_x^2k_y^2}.
\end{eqnarray}

These two bands touch at zero energy precisely at the $\Gamma$ point ($\mathbf{k}=0$), forming
a band-touching nodal point. The resulting band structure is strongly anisotropic and the energy dispersion
displays a strong directional dependence as depicted in Fig.~\ref{fig:E_k}. Restricting to the $k_x$-axis ($k_y=0$),
we recover a pure parabolic dispersion $E_\pm=\pm g_1 k_x^2$. In contrast, for a fixed nonzero $k_x$, the
dispersion depends almost linearly on $k_y$.  However, the dimensionless ratio $g_2/g_1$ directly tunes the
cross-anisotropic distortion. As the $g_2$ increases with a fixed $g_1$, the Fermi surface undergoes increasing
warping, which is labeled by a warped low-energy band manifold as illustrated in Fig.~\ref{fig:E_k} for two
representative values with $g_{2}/g_{1}=0.41$ and $v=2$.

\begin{figure*}[htpb]
\centering
\subfigure[]
{\includegraphics[width=1.8in]{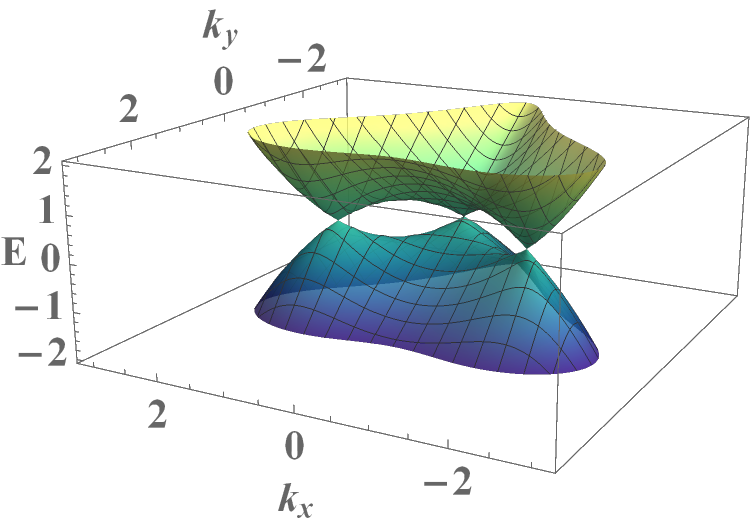}}\hspace{0.5cm}
\subfigure[]{\includegraphics[width=1.8in]{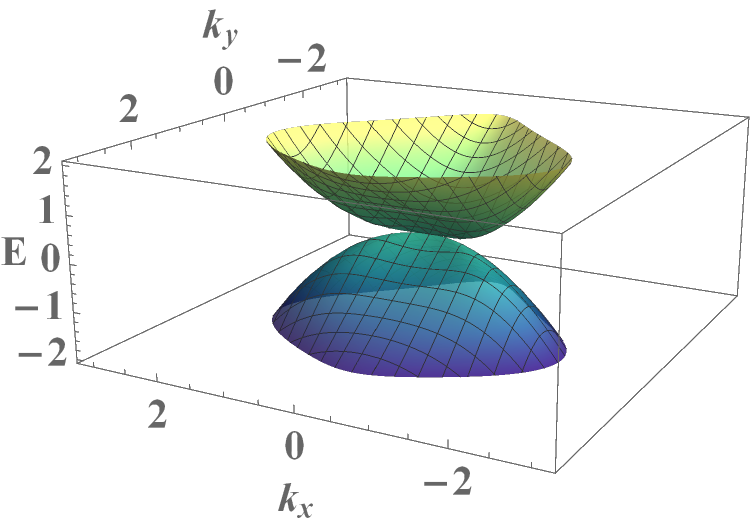}}\hspace{0.5cm}
\subfigure[]{\includegraphics[width=1.8in]{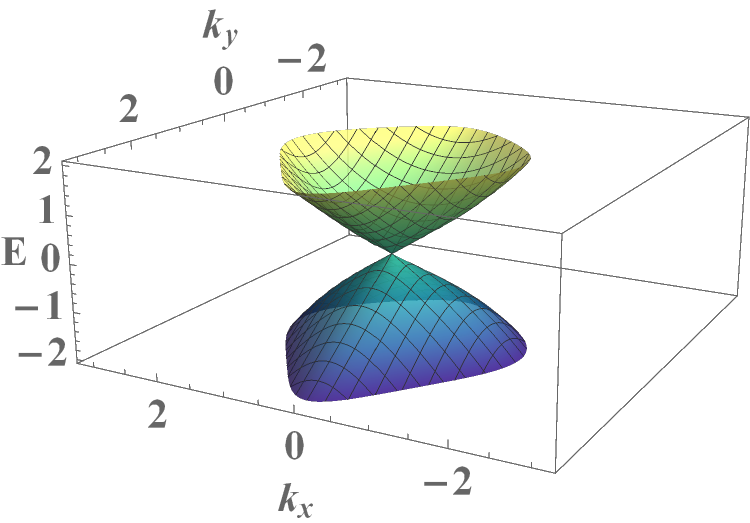}}
\vspace{-0.15cm}
\caption{(Color online) Low-energy electronic dispersion under a small perturbation $\Delta_x$ with $\Delta_y=m_z=0$
for (a) $\Delta_x<0$, (b) $\Delta_x=0$, and (c) $\Delta_x>0$. The $\Delta_x=0$ case preserves the gapless node,
while finite $\Delta_x$ displaces the band-touching point and remains symmetric about $k_x=0$.}
\label{fig:E_vs_k_Deltax}
\end{figure*}

Based on the above analysis of the free type-II SDSM Hamiltonian, we note that the
inherent breaking of both spatial-inversion and time-reversal symmetries makes this system an
excellent theoretical playground and a promising experimental platform for exploring symmetry-broken
topological band structures, Berry-phase responses, and nonreciprocal transport.
Nevertheless, the free Hamiltonian features only a nodal touching at $\Gamma$ without a full
gap across the whole Brillouin zone (BZ), which prevents the definition of robust Chern numbers.
To overcome this limitation, we introduce two complementary strategies~\cite{Saha2016PRB,Bian2026EPJP}, namely,
the static symmetry-breaking perturbations and time-periodic Floquet fields. Throughout the following sections,
we systematically examine their independent and cooperative influences
on the topological properties of type-II SDSMs.

\section{Modeling topological responses under static and periodically driven fields}\label{Sec_effectie_model}

Extending the free model from the previous section~\ref{Sec_free_model}, we within this section
develop a framework for the topological responses of type-II SDSMs in three distinct scenarios, namely,
(i) static symmetry-breaking perturbations that open gaps~\cite{Huang2015PRB,Elsayed2026PRR,Bian2026EPJP},
(ii) periodic Floquet driving via TFOPL~\cite{Kang2020PRA,Wang2023PRA,Maryam2026arXiv,Kukhar2022ZTF,Kanega2025PRB,Lesko2025arXiv},
and (iii) the cooperative interplay between these two strategies.
We will show that although neither static perturbations nor Floquet driving alone offers sufficient tunability,
their combination produces a qualitatively richer phase diagram with multiple gap-closing surfaces and
symmetry-breaking pathways that are absent when either is applied alone.

\subsection{Perturbative scenario}\label{subsection_pert}

At first, we introduce static perturbations into the free Hamiltonian~(\ref{eq:Hamiltonian}). Specifically, following Refs.~\cite{Huang2015PRB,Elsayed2026PRR,Bian2026EPJP}, we add two types of terms: (i) constant shifts in the
in-plane pseudospin components and (ii) a mass term arising from spin-orbit coupling. This yields the following
Hamiltonian density,
\begin{eqnarray}
\mathcal{H}(\mathbf{k})
&=&
(g_1k_x^2-vk_y+\Delta_x)\sigma_x
+(g_2k_xk_y+\Delta_y)\sigma_y\nonumber\\
&&+m_z\sigma_z,\label{eq:Hamiltonian_per}
\end{eqnarray}
which can be cast in vector form as $\mathcal{H}(\mathbf{k})\equiv\mathbf{h}(\mathbf{k})\cdot \bm{\sigma}$, where $\mathbf{h}(\mathbf{k})\equiv(h_x,h_y,h_z)$ is defined by
\begin{eqnarray}
\begin{cases}
h_x \equiv g_1 k_x^2 - v k_y+\Delta_x,\\
h_y \equiv g_2 k_x k_y+\Delta_y,\\
h_z \equiv m_z.\label{eq:vector_per}
\end{cases}
\end{eqnarray}

\begin{figure}
\centering
\subfigure[]
{\includegraphics[width=1.5in]{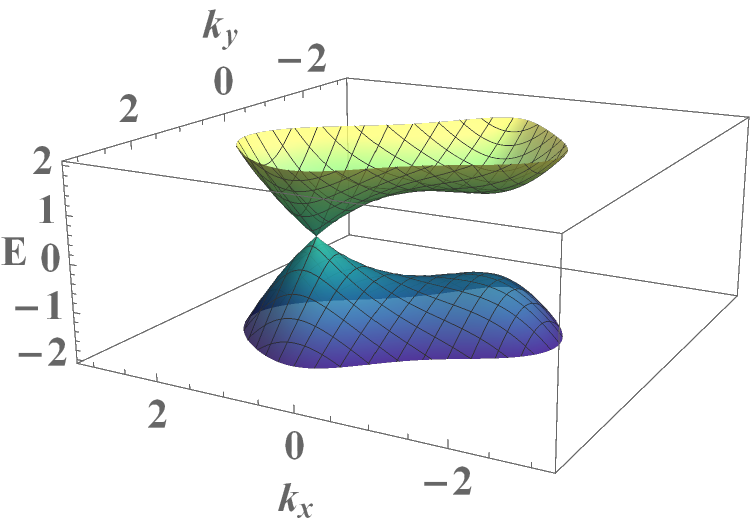}}
\subfigure[]{\includegraphics[width=1.5in]{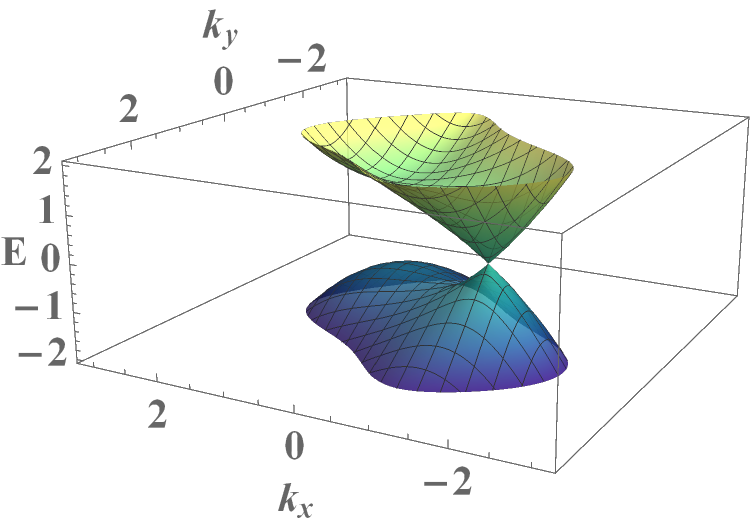}}
\vspace{-0.15cm}
\caption{(Color online) Low-energy electronic dispersion under a small perturbation $\Delta_y$ with $\Delta_x=m_z=0$
for (a) $\Delta_y<0$ and (b) $\Delta_y>0$. Finite $\Delta_y$ not only displaces the band-touching point, but also
breaks symmetry about $k_y=0$.}
\label{fig:E_vs_k_Deltay}
\end{figure}

After diagonalizing the Hamiltonian density in Eq.~\eqref{eq:Hamiltonian_per}, we obtain the band dispersion as follows,
\begin{eqnarray}
E_\pm &=& \pm\big[(g_1k_x^2+vk_y+\Delta_x)^2\nonumber\\
&&+(g_2k_xk_y+\Delta_y)^2+m_z^2\big]^{1/2}.\label{Eq_dispersion_per}
\end{eqnarray}
From the dispersion in Eq.~\eqref{Eq_dispersion_per}, we observe that the in-plane
parameters $\Delta_x$ and $\Delta_y$ primarily shift or split the nodal point(s) in momentum space without opening a gap,
as illustrated in Fig.~\ref{fig:E_vs_k_Deltax} and Fig.~\ref{fig:E_vs_k_Deltay}.

To clarify these effects, we set $m_z=0$. As shown in Fig.~\ref{fig:E_vs_k_Deltax}, for $\Delta_x<0$ the original semi-Dirac point splits into three Dirac cones located at $(0,\Delta_x/v)$ and $(\pm\sqrt{-\Delta_x/g_1},0)$. At $\Delta_x=0$, these three cones coalesce at the origin, recovering the type-II semi-Dirac point. For $\Delta_x>0$, only the cone at $(0,\Delta_x/v)$ remains. This indicates that the energy bands remain symmetric about $k_x=0$ when only $\Delta_x$ is present. However, $\Delta_y$ does break this mirror symmetry, as shown in Fig.~\ref{fig:E_vs_k_Deltay}. For $\Delta_y<0$, the Dirac point moves to
\begin{equation}
\mathbf{k}=\left(-\left(\frac{v \Delta_y}{g_1g_2}\right)^{1/3},
\left(\frac{g_1\Delta_y^2}{v g_2^2}\right)^{1/3}\right),
\end{equation}
and the bands lose mirror symmetry. For \(\Delta_y>0\), the cone shifts accordingly.

While the in-plane perturbations $\Delta_{x,y}$ merely shift or split the nodal points and leave the spectrum gapless
when $m_z=0$, the mass term $m_z\sigma_z$ plays a unique and decisive role. It alone can open a full bulk gap of size $2|m_z|$,
driving the system from a semi-Dirac semimetal into a Chern insulator~\cite{Huang2015PRB}.

\subsection{Floquet scenario}\label{subsection_Floquet}

Next, let us consider the periodic driving by TFOPL.
The corresponding vector potential is~\cite{Kang2020PRA,Wang2023PRA,Maryam2026arXiv,Kukhar2022ZTF,Kanega2025PRB,Lesko2025arXiv}
\begin{equation}
\mathbf{A}(t)=A_1\cos(\omega t)\hat{x}+A_2\cos(2 \omega t+\phi)\hat{y},\label{eq:light_vector}
\end{equation}
where $\phi$ is the relative phase and $A_{1,2}\equiv E_{1,2}/\omega_{1,2}$ are the amplitudes of
the two frequency components, with $\omega_1\equiv\omega$ and $\omega_2\equiv2\omega$. The quantities $E_{1,2}$
are the electric-field amplitudes, which satisfy $\mathbf{E}=-\partial \mathbf{A}/\partial t$.

We then couple the TFOPL~(\ref{eq:light_vector}) to the electrons governed by the 2D SDSM Hamiltonian~(\ref{eq:Hamiltonian}),
through the minimal substitution $\mathbf{k}\to\mathbf{k}+e\mathbf{A}(t)$. After performing a high-frequency
Floquet expansion to leading order in $1/\omega$ as detailed in Appendix~\ref{AppendixA}, the effective Hamiltonian
takes the form of
\begin{equation}
\mathcal{H}_{\rm eff}(\mathbf{k})
=(g_1k_x^2-v k_y+\delta_x)\sigma_x
+g_2k_xk_y\sigma_y+\gamma k_x\sigma_z,\label{eq:H_eff}
\end{equation}
where two coefficients $\delta_x$ and $\gamma$ are designated by
\begin{eqnarray}
\delta_x\equiv\frac{e^2g_1A_1^2}{2},\,\,\,\,\,\,\,\, \gamma\equiv\frac{5e^3g_1g_2A_1^2A_2\sin\phi}{4\omega}.\label{Eq_delta_gamma}
\end{eqnarray}

\begin{figure}
\centering
\subfigure[]
{\includegraphics[width=1.5in]{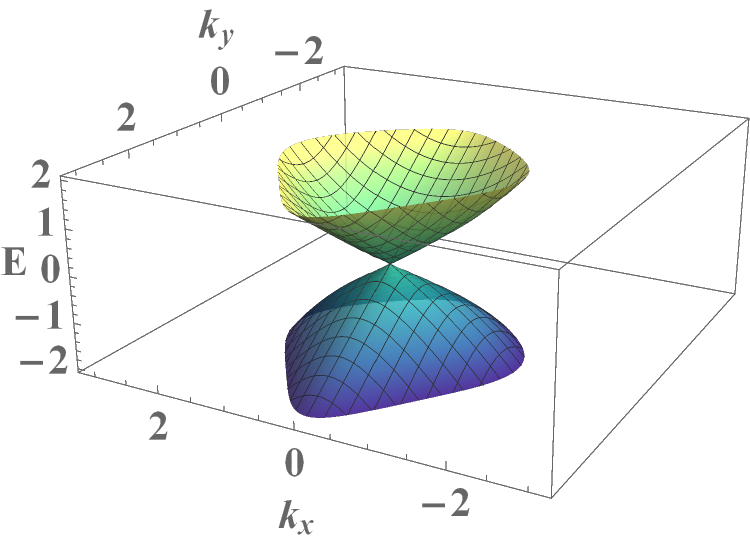}}\hspace{0.3cm}
\subfigure[]{\includegraphics[width=1.5in]{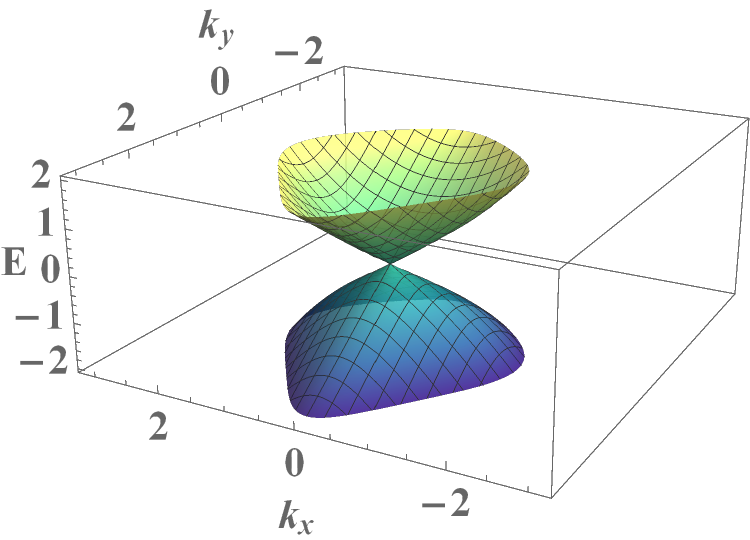}}
\vspace{-0.15cm}
\caption{(Color online) Low-energy electronic dispersion under the TFOPL driving for
(a) $\phi=0$ and (b) $\phi=\pi/3$ with $g_1/g_2=0.5$, $v=1$, $\omega=100$, $A_1=2$, and $A_2=1$.}
\label{fig:E_vs_k_Floquet}
\end{figure}

Following the analysis of the perturbative scenario in Sec.~\ref{subsection_pert}, we highlight two key
features of the effective Hamiltonian obtained from the driving field in Eq.~\eqref{eq:light_vector}.
First, for $\phi=0$ or $\pi$, the light only shifts the Dirac point by $\delta_x$ and does not open a gap.
Second, for $\phi\neq0,\pi$, it induces a momentum-dependent Zeeman term $\gamma k_x\sigma_z$~\cite{Oka-Aoki2009PRB,Zhu-Liang2020PRR},
but the spectrum remains gapless at $\mathbf{k}=(0,\delta_x/v)$. These behaviors are illustrated in
Fig.~\ref{fig:E_vs_k_Floquet}. Fig.~\ref{fig:E_vs_k_Floquet}(a) corresponds to $\phi=0$, where only
a shift is observed, and Fig.~\ref{fig:E_vs_k_Floquet}(b) is for $\phi=\pi/3$, where the band structure
is warped yet still gapless. Therefore, at first order in Floquet theory~\cite{Blanes-Ros2009PR,Haga2019PRE}, the TFOPL
alone cannot open a full band gap or drive a Chern-insulating phase. This motivates the hybrid scenario
presented below.

\subsection{Perturbative-Floquet hybrid scenario}\label{subsec_Per_Floquet_scenario}

Subsequently, we turn to the cooperative interplay between perturbative and Floquet scenario.
To this end, the static perturbations and the Floquet field act simultaneously on the system.
The total effective Hamiltonian is then obtained by combining Eq.~(\ref{eq:Hamiltonian_per}) and Eq.~(\ref{eq:light_vector}),
\begin{eqnarray}
\mathcal{H}_{\rm hyb}(\mathbf{k})=\mathbf{h}^{\rm hyb}(\mathbf{k})\cdot \bm{\sigma},\label{eq:H_tot}
\end{eqnarray}
where the pseudospin vector field $\mathbf{h}^{\rm hyb}(\mathbf{k})$ are designated as
\begin{eqnarray}
\begin{cases}
h^{\rm hyb}_x\equiv g_1k_x^2-vk_y+\tilde{\Delta}_x,\\
h^{\rm hyb}_y\equiv g_2k_xk_y+\Delta_y,\\
h^{\rm hyb}_z\equiv m_z+\gamma k_x,
\end{cases} \label{Eq_h_xyz}
\end{eqnarray}
with $\tilde{\Delta}_x\equiv \Delta_x+\delta_x$.

It is particularly important to note that $h^{\rm hyb}_z$ reveals a direct competition between the TFOPL-induced
momentum-dependent Zeeman term $\gamma k_x\sigma_z$ and the static mass $m_z\sigma_z$. As demonstrated below,
this competition produces a phase diagram qualitatively richer than that of either the perturbation-only or Floquet-only scenario.

\subsubsection{Gap-closure criterion and symmetry breaking}

To investigate how time-periodic driving cooperates with static perturbations to modulate
the topological properties, we focus on the hybrid Hamiltonian $\mathcal{H}_{\rm hyb}$ in Eq.~(\ref{eq:H_tot}).
Following the analysis in Secs.~\ref{subsection_pert} and~\ref{subsection_Floquet}, we find that the energy gap
closes only when all three components defined in Eq.~(\ref{Eq_h_xyz}) vanish simultaneously.
At a generic point in the BZ, this condition reads $h^{\rm hyb}_x=h^{\rm hyb}_y=h^{\rm hyb}_z=0$. By eliminating $k_x$
and $k_y$ from these three algebraic equations, we obtain the central criterion (i.e, the gap-closure equation),
\begin{equation}
g_1g_2m_z^3+\tilde{\Delta}_xg_2\gamma^2m_z-v\gamma^3\Delta_y=0,\label{eq:condition}
\end{equation}
which defines a surface in the three-dimensional parameter space $(\tilde{\Delta}_x,\Delta_y,m_z)$ for fixed $(g_1,g_2,v,\gamma)$.

To proceed, it is instructive to identify the physical origin of each term in Eq.~(\ref{eq:condition}).
The first term, $g_1g_2m_z^3$, is purely intrinsic to the static perturbation, which stems from the mass
term $m_z\sigma_z$ and survives even at $\gamma=0$. The second term, \(\tilde{\Delta}_xg_2\gamma^2m_z\),
is a cross term that requires both the static $x$-component shift $\tilde{\Delta}_x$ renormalized by the
light via $\delta_x$ and the Floquet-induced $z$-component measured by $\gamma$. Physically, it captures
how the light modifies the effective mass anisotropy. The third term, $-v \gamma^3 \Delta_y$, is purely
of hybrid origin, which vanishes if either $\gamma=0$ (the absence of TFOPL) or $\Delta_y$ is zero (no static $y$-perturbation).
In particular, this term is odd in both $\gamma$ and $\Delta_y$ and has no analogue in either
individual scenario, which directly reflects the cooperative interplay between the two distinct mechanisms.

In addition, Eq.~(\ref{eq:condition}) is mathematically cubic in $m_z$ but only linear
in $\tilde{\Delta}_x$ and $\Delta_y$. Owing to this structure, multiple solutions for $m_z$
can emerge across a wide parameter range, leading to multiple gap-closing points. This feature
has no counterpart in the perturbation-only scenario, where the gap closes only at $m_z=0$.
Fig.~\ref{fig:condition} illustrates this behavior through representative solutions,
with the curves marking the phase boundaries between the insulating and gapless semimetallic phases.

Furthermore, Eq.~(\ref{eq:condition}) uncovers an important symmetry-breaking effect. The presence of
$-v\gamma^3\Delta_y$ breaks the symmetry of gap-closure position under $\Delta_y\rightarrow-\Delta_y$. In other words,
it renders the phase diagram asymmetric in the sign of $\Delta_y$ by changing the gap-closure condition
unless $\gamma$ flips sign simultaneously (i.e., $\phi\rightarrow\phi+\pi$). This asymmetry originates
from the breaking of inversion symmetry by the Floquet field in
the perturbative-Floquet hybrid scenario. In the perturbation-only scenario with $\gamma=0$,
by contrast, the phase boundaries are symmetric under $\Delta_y\rightarrow-\Delta_y$, as the gap condition
is independent of the sign of $\Delta_y$ presented in Sec.~\ref{subsection_pert}.
Consequently, the TFOPL introduces a preferred chirality in momentum space, which produces the nontrivial
behaviors for both the Berry curvature dipole and the nonlinear Hall response as will be studied in Sec.~\ref{subsec_BCD}.

\subsubsection{Cooperative effects between two scenarios}\label{subsub_coop_effects_three_roots}

To elucidate the cooperative effects in the hybrid scenario, let us focus on several
situations in which Eq.~(\ref{eq:condition}) exhibits its cooperative nature most clearly,
and which are strictly forbidden in the individual scenarios.

At first, we find that a band gap can be generated in the absence of a static mass.
Specifically, as presented in Sec.~\ref{subsection_pert}, when $m_z=0$, the perturbation-only scenario is always gapless,
independent of $\Delta_y$. In the hybrid scenario, however, Eq.~(\ref{eq:condition}) with $m_z=0$ reduces to
\begin{equation}
-v\gamma^3\Delta_y=0.
\end{equation}
Since this condition cannot be satisfied for $\gamma\neq 0$ and $\Delta_y\neq 0$, the system
acquires a finite gap even though the static mass is exactly zero.
Physically, this stems from the cooperative effects between two scenarios.  In other words,
the light-induced term $\gamma k_x\sigma_z$ cooperates with the static $\Delta_y\sigma_y$ term
to generate an effective mass $m_{\rm eff}\propto \gamma \Delta_y^{1/3}$.
This indicates that the TFOPL serves as a catalyst for mass generation
in the presence of an appropriate static in-plane perturbation.

Additionally, for $m_z\neq0$, an appropriate combination of $\tilde{\Delta}_x$, $\Delta_y$,
and $\gamma$ can still satisfy the gap-closure condition of Eq.~(\ref{eq:condition}). This implies
that even a finite static mass, which is sufficient to open a gap, can be counteracted by the TFOPL,
driving the system back to a gapless semimetallic phase. This behavior is also due to the cooperative
effects. The momentum-space shift induced by $\Delta_y$, together with the Zeeman-like term $\gamma k_x$,
displaces the band-touching point to a momentum at which the effective $z$-field $m_z+\gamma k_x$ vanishes.
At the same time, the in-plane components are driven to zero as well.

\begin{figure}
\centering
\subfigure[]
{\includegraphics[width=2.5in]{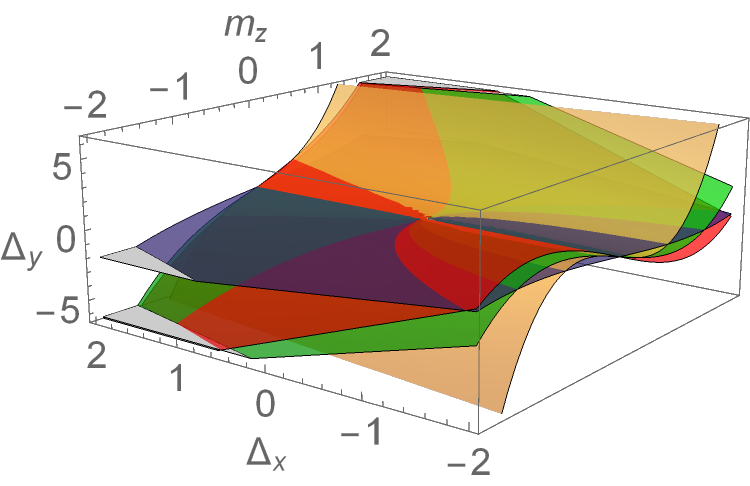}}
\subfigure[]
{\includegraphics[width=2.5in]{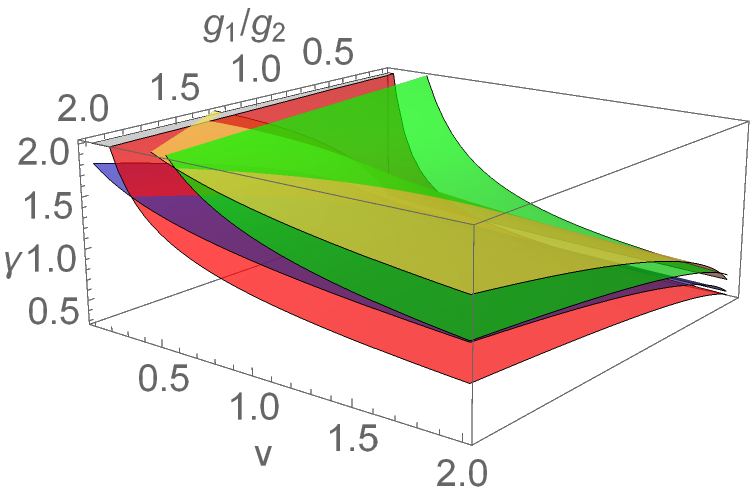}}
\vspace{-0.15cm}
\caption{(Color online) Surfaces of gap-closing solutions to the energy-gap
closure equation~(\ref{eq:condition}) in the (a) $(\Delta_x, \Delta_y, m_z)$
and (b) $(v, \gamma, g_1/g_2)$ parameter space. The four surfaces (red, green, blue, yellow)
correspond to the representative parameter sets $(g_1/g_2, v, \gamma)=
(0.5,1,1.5)$, $(1,0.8,1.2)$, $(1.5,1.2,2)$, $(2,0.9,1)$, and $(m_z,\Tilde{\Delta}_x,\Delta_y)
=(0.5,0.2,0.4)$, $(0.8,0.5,0.6)$, $(1,-0.5,0.6)$, $(1.5,-0.6,0.8)$, respectively.}
\label{fig:condition}
\end{figure}

Furthermore, since the gap-closure equation~(\ref{eq:condition}) is cubic in $m_z$, it admits up to three distinct
real roots for $m_z$ at fixed $\tilde{\Delta}_x$ and $\Delta_y$. A particularly instructive case is
$\Delta_y=0$, in which Eq.~(\ref{eq:condition}) factorizes into
\begin{equation}
m_z\left(g_1g_2m_z^2+\tilde{\Delta}_x g_2\gamma^2\right)=0.\label{Eq_factorized}
\end{equation}
This immediately yields one root at $m_z=0$, together with two additional roots
\begin{equation}
m_z=\pm\sqrt{-\frac{\tilde{\Delta}_x\gamma^2}{g_1}}.\label{Eq_roots}
\end{equation}
Since $g_1>0$ and $\gamma^2\ge 0$, the two additional roots in Eq.~(\ref{Eq_roots}) are real only
when $\tilde{\Delta}_x\le 0$.
For $\tilde{\Delta}_x>0$, only the root \(m_z=0\) remains real and the other two are purely imaginary and thus unphysical.
This shows that the number of gap-closing points is controlled directly by the sign of $\tilde{\Delta}_x$,
providing a simple tunable knob for engineering the topological phase diagram.

Concretely, this implies that, as $m_z$ is adiabatically varied from negative to positive values,
the system can undergo a sequence of phase transitions, namely, insulator $\rightarrow$ semimetal
$\rightarrow$ insulator $\rightarrow$ semimetal. Such a phenomenon does not occur in the
perturbation-only case, where a single critical point exists at $m_z=0$. The resulting controllability
of gap-closing points, along with the accompanying phase transitions, directly reflects the cooperative
interplay between the static perturbation and the Floquet field. This can also be inferred from the
nonmonotonic phase boundaries shown in Fig.~\ref{fig:condition}.

\subsubsection{Physical picture of the cooperative effects}

Physically, the physical origin of the cooperative effects can be understood as follows.
In the context of perturbative-Floquet hybrid scenario, the TFOPL field does not create
real excitations but instead dresses the band structure through virtual photon processes,
renormalizing the hopping amplitudes and generating a momentum-dependent
pseudospin field $\gamma k_x\sigma_z$. As shown in Eqs.~(\ref{eq:H_tot}) and~(\ref{Eq_h_xyz}),
this dynamical field competes directly with the static pseudospin fields
$\Delta_x\sigma_x$, $\Delta_y\sigma_y$, and $m_z\sigma_z$.
A band touching occurs whenever the total pseudospin vector $\mathbf{h}^{\rm hyb}(\mathbf{k})$~(\ref{eq:H_tot}) vanishes,
i.e., when the electron experiences zero net pseudospin splitting. Since the light-induced term is linear in $k_x$,
it can cancel the static $z$-component only at a specific $k_x$, while the in-plane components must vanish simultaneously.
This requirement is satisfied only for special parameter combinations, which is the origin of the nonlinear,
cooperative structure of Eq.~(\ref{eq:condition}).

To wrap up, this cooperative interplay is the central theme of this work.
It provides a powerful tuning knob, via the laser amplitude, frequency, phase ($\phi$),
and static gate voltages (tuning $\Delta_{x,y}$ and $m_z$), for engineering topological
phase transitions beyond the reach of either the perturbative or Floquet alone. Accordingly,
the subsequent calculations of the Berry curvature, Chern number, and transport signatures
will directly reflect the phase boundaries derived here.

\section{Berry curvature and its physical consequences}\label{Sec_BC}

In this section, we aim to characterize the band geometry and topological properties
of the perturbative, Floquet, and hybrid scenarios. To this end, we compute the Berry
curvature (BC) in each case, which allows us to map out the topological phase diagrams
via the Chern number and to quantify the asymmetry of the BC distribution through the
Berry curvature dipole (BCD) that governs the nonlinear Hall response. Crucially, due to
the cooperative effects, our analysis reveals that the hybrid scenario supports
a wider range of topological phases than either perturbative or Floquet scenario
can realize individually.

\begin{figure*}
\centering
\subfigure[]
{\includegraphics[width=1.7in]{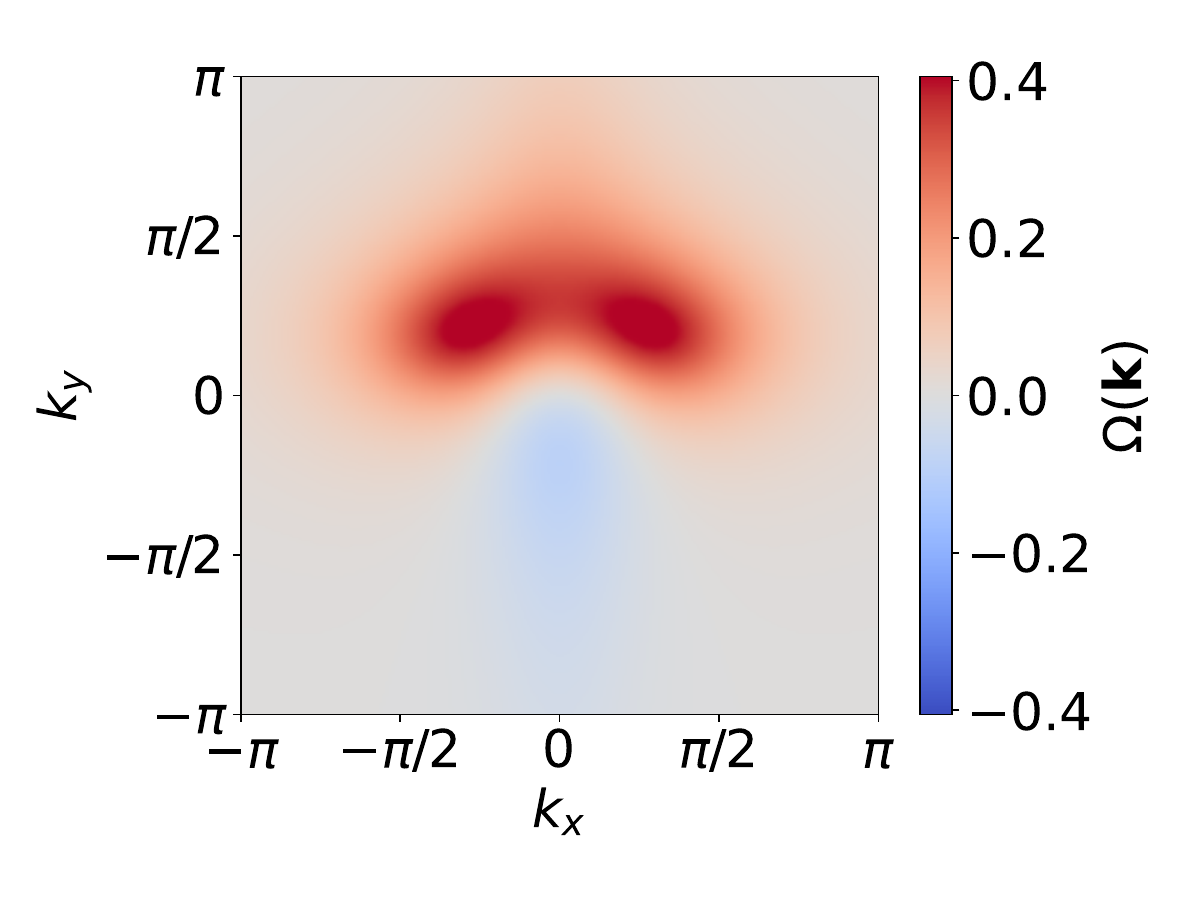}}
\subfigure[]
{\includegraphics[width=1.7in]{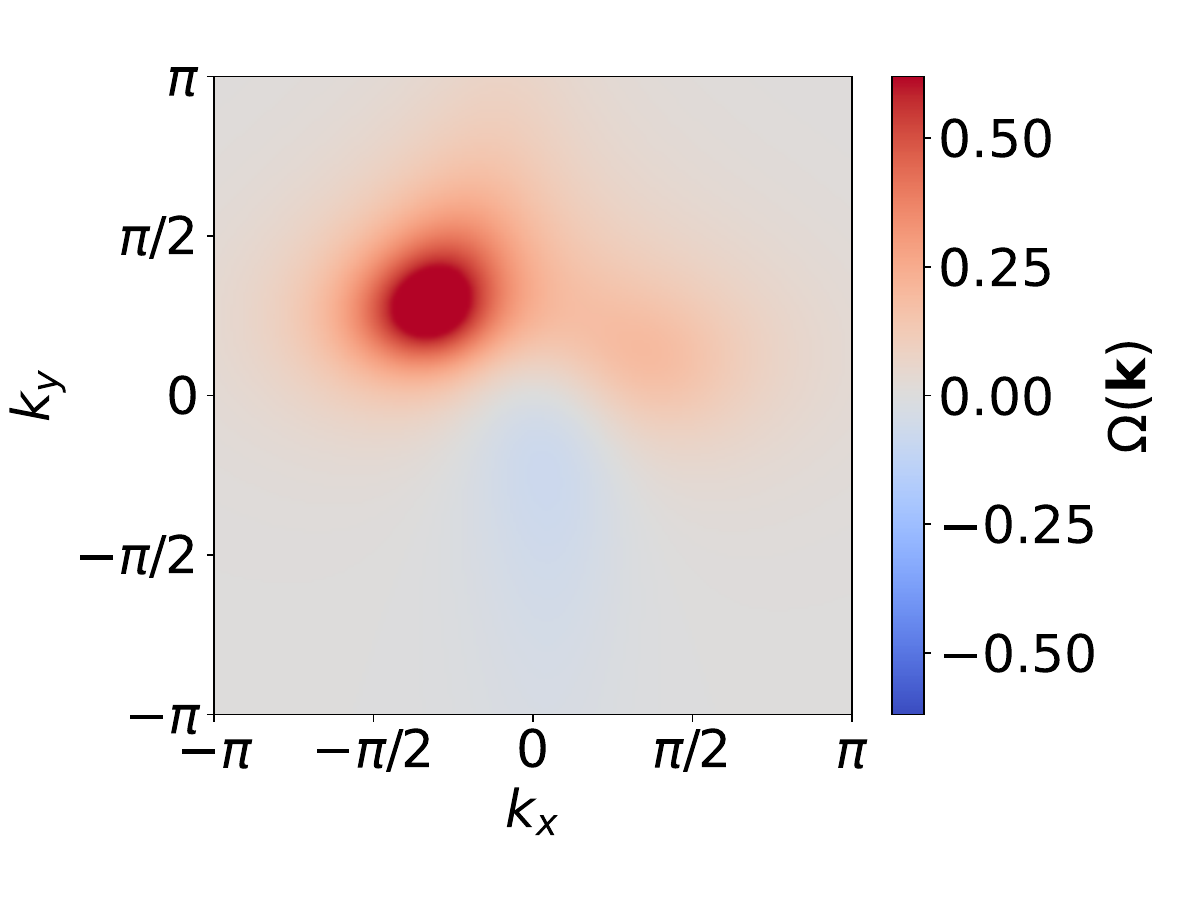}}
\subfigure[]
{\includegraphics[width=1.7in]{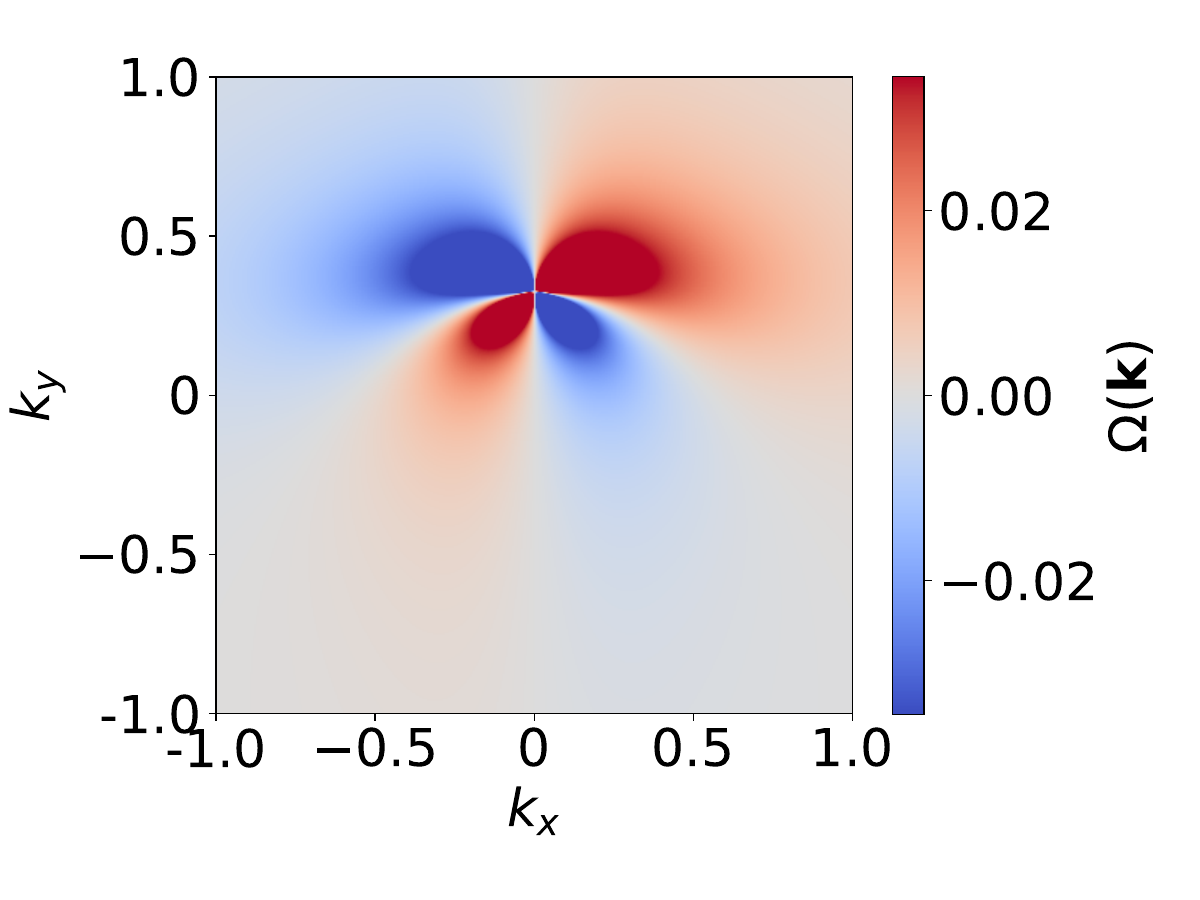}}
\subfigure[]
{\includegraphics[width=1.7in]{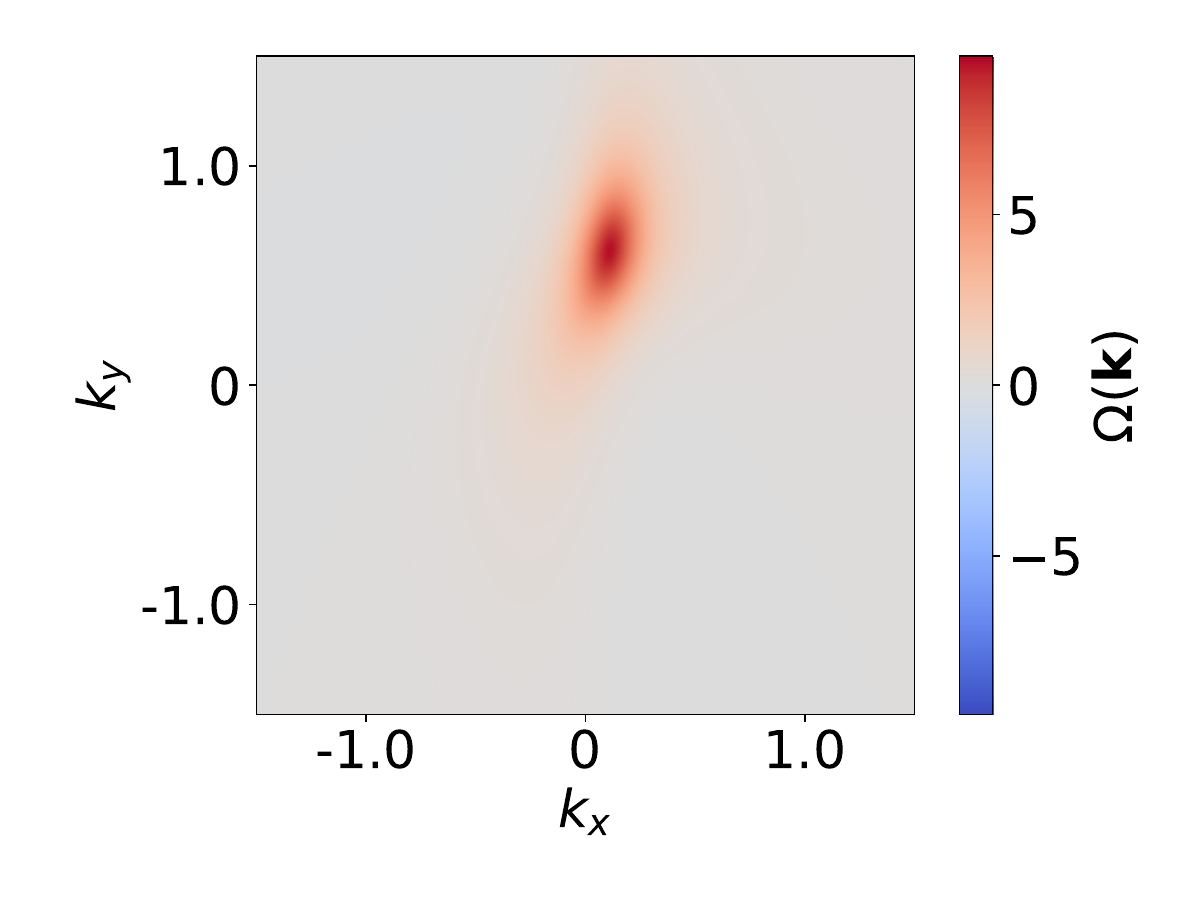}}\\
\subfigure[]
{\includegraphics[width=1.7in]{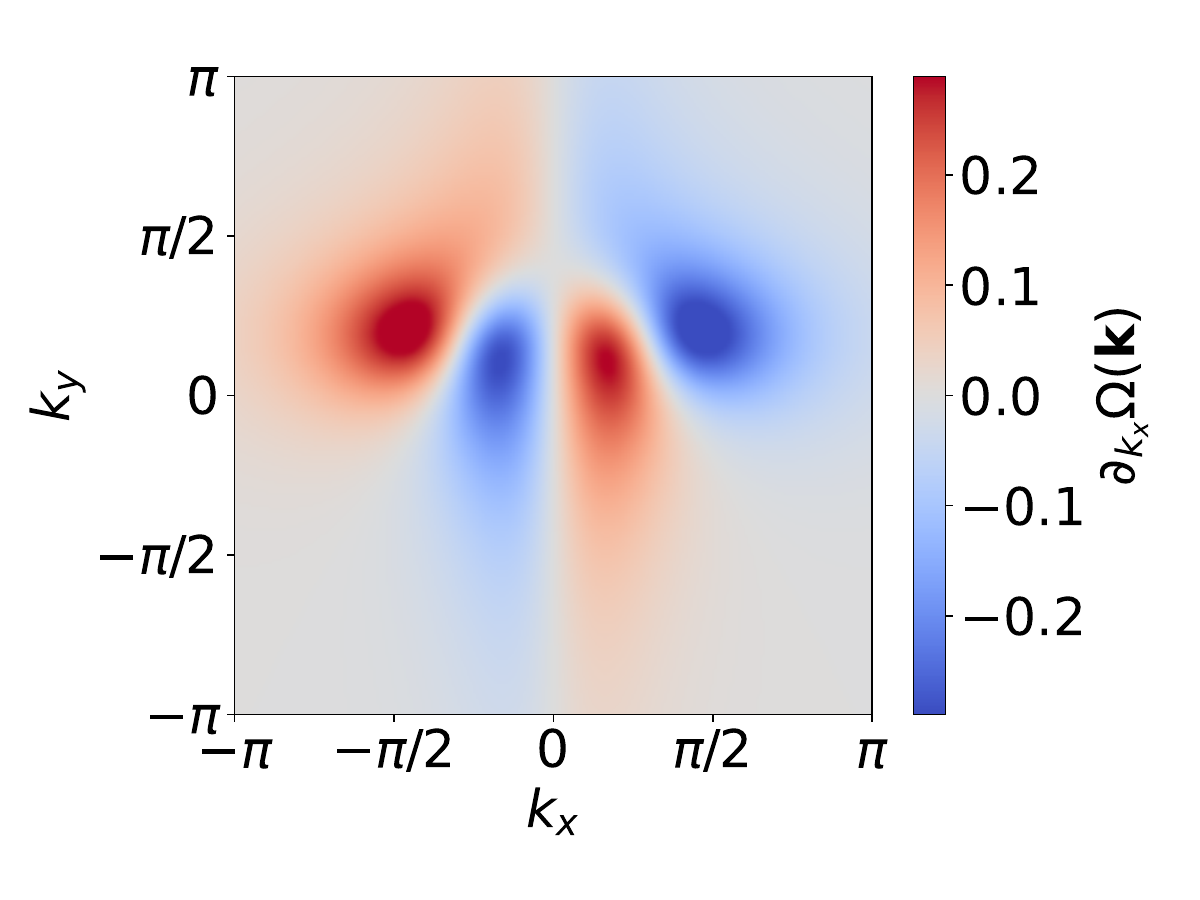}}
\subfigure[]
{\includegraphics[width=1.7in]{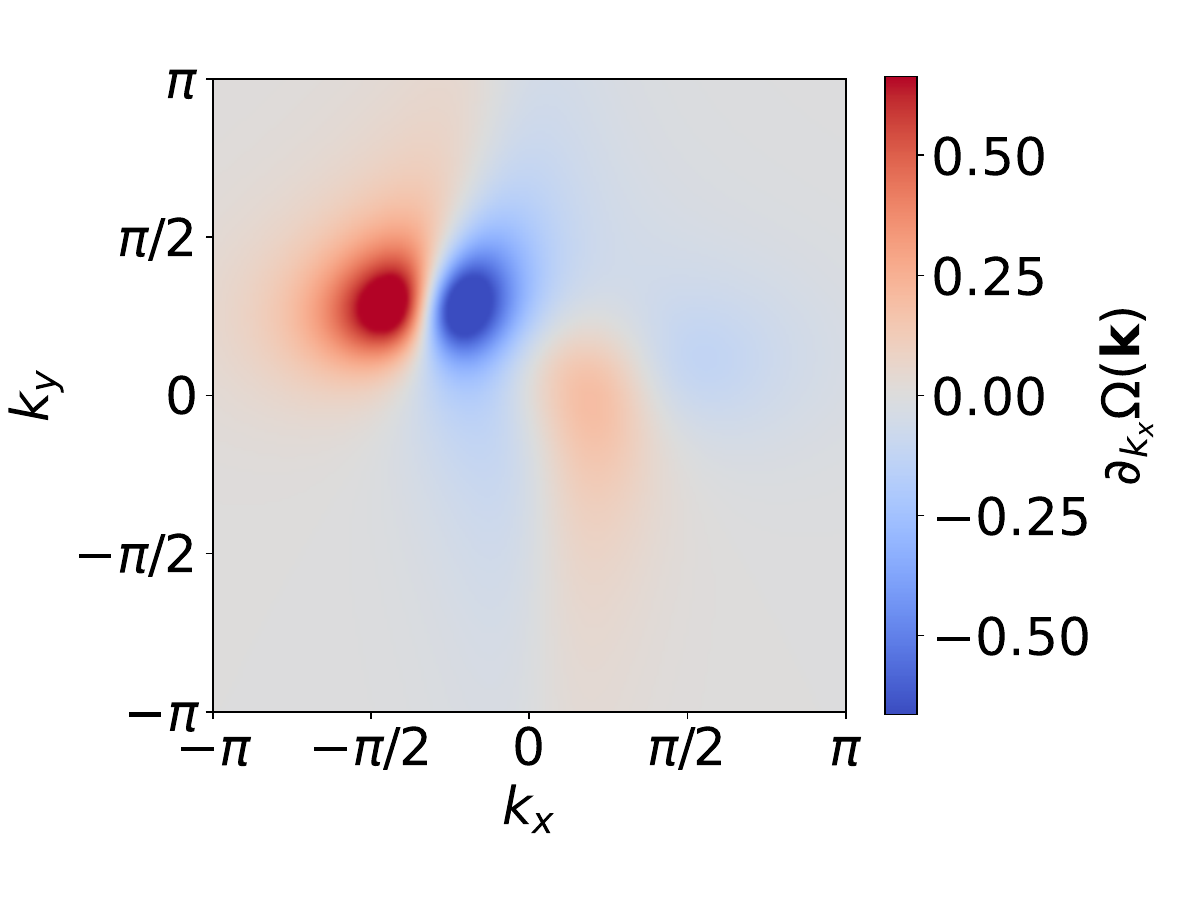}}
\subfigure[]
{\includegraphics[width=1.7in]{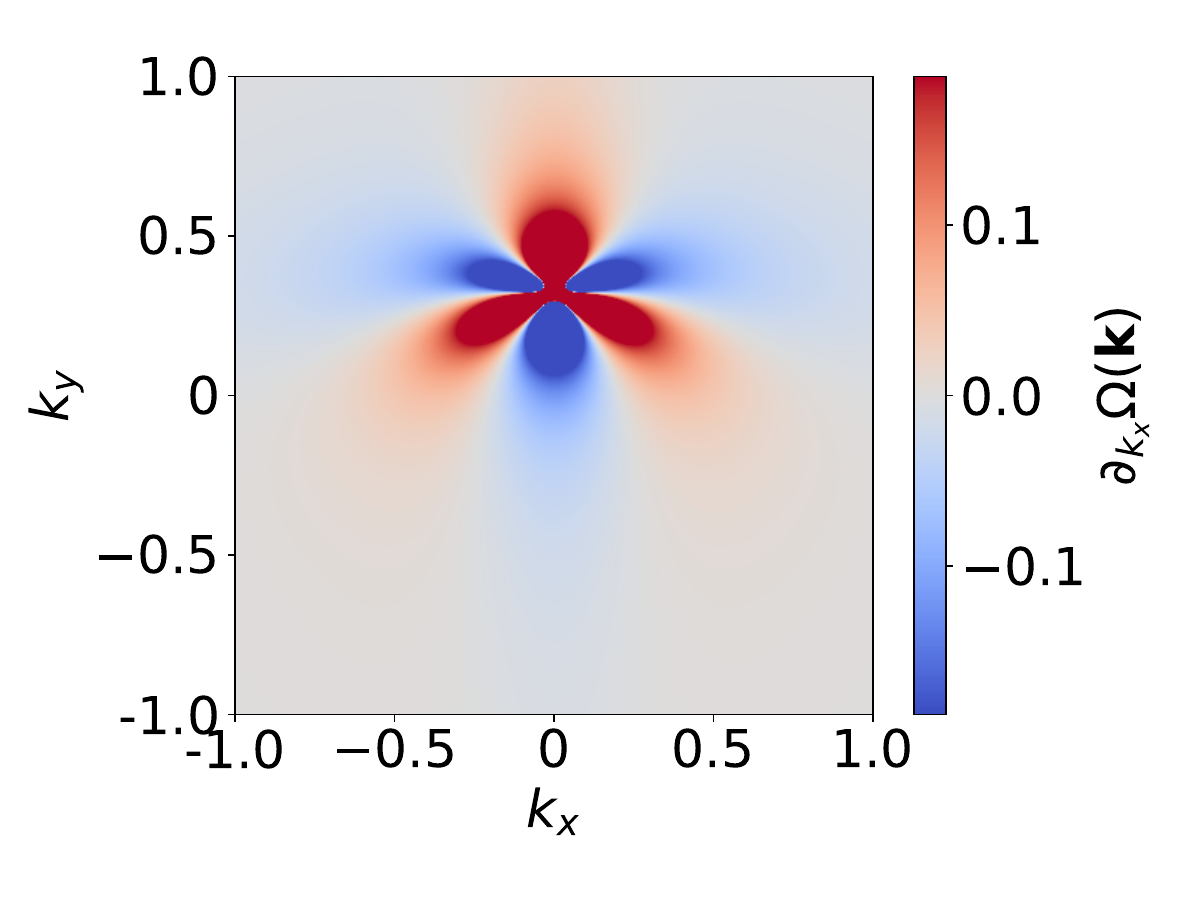}}
\subfigure[]
{\includegraphics[width=1.7in]{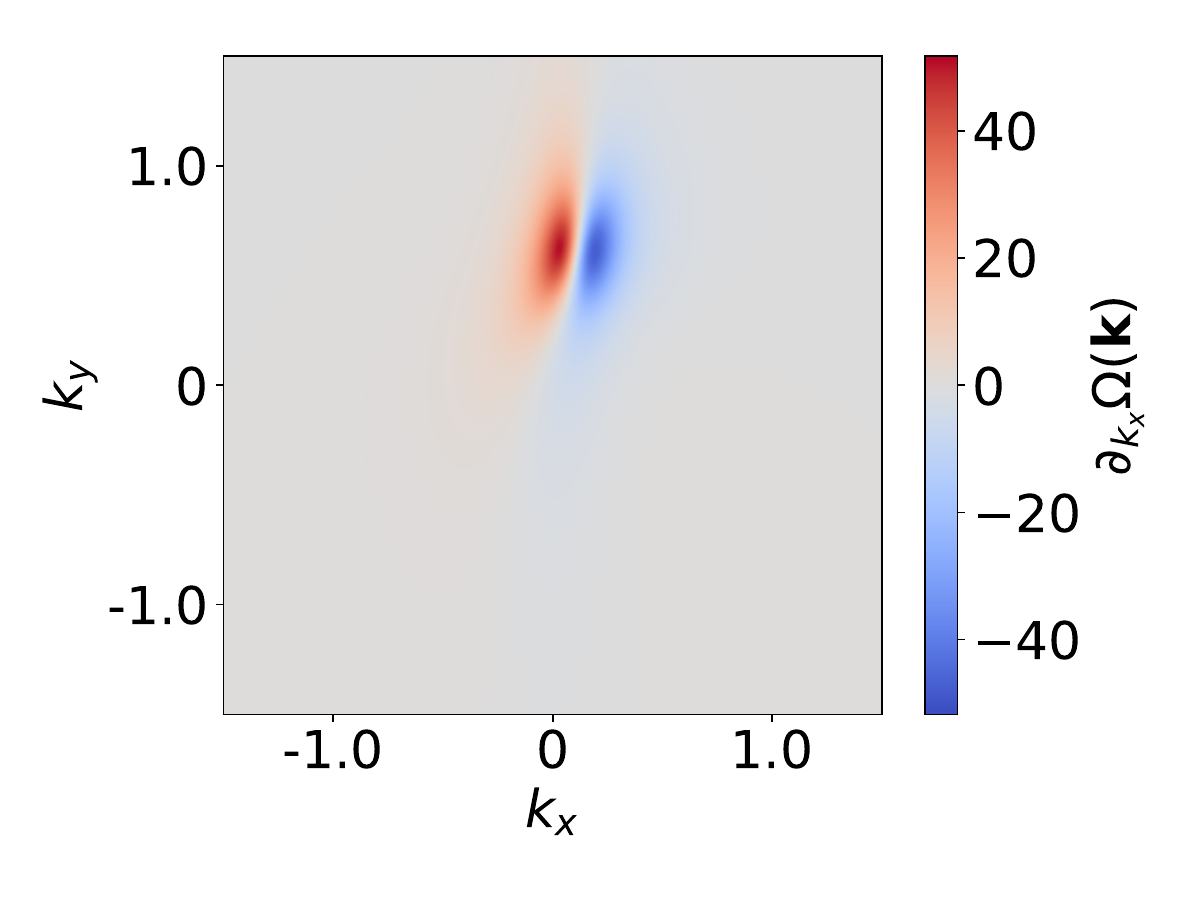}}\\
\subfigure[]
{\includegraphics[width=1.7in]{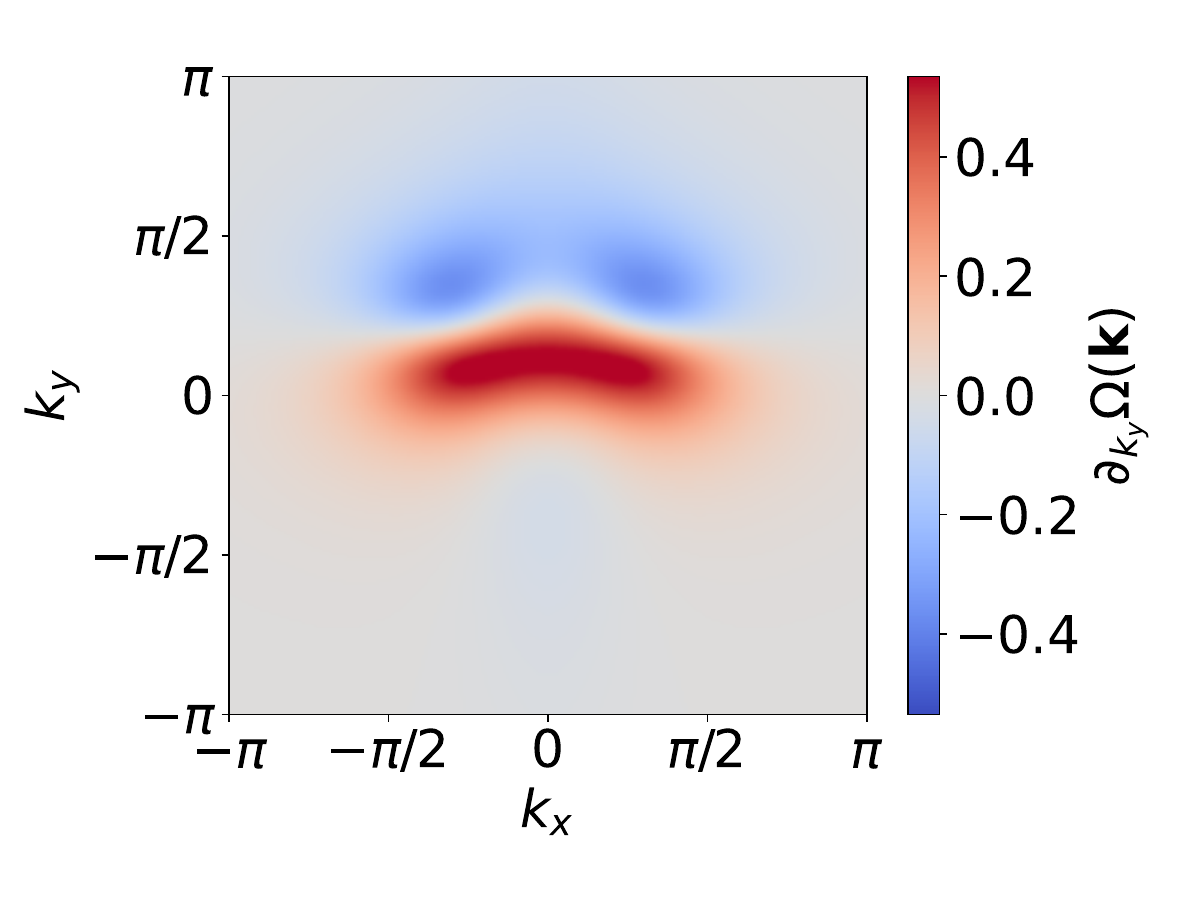}}
\subfigure[]
{\includegraphics[width=1.7in]{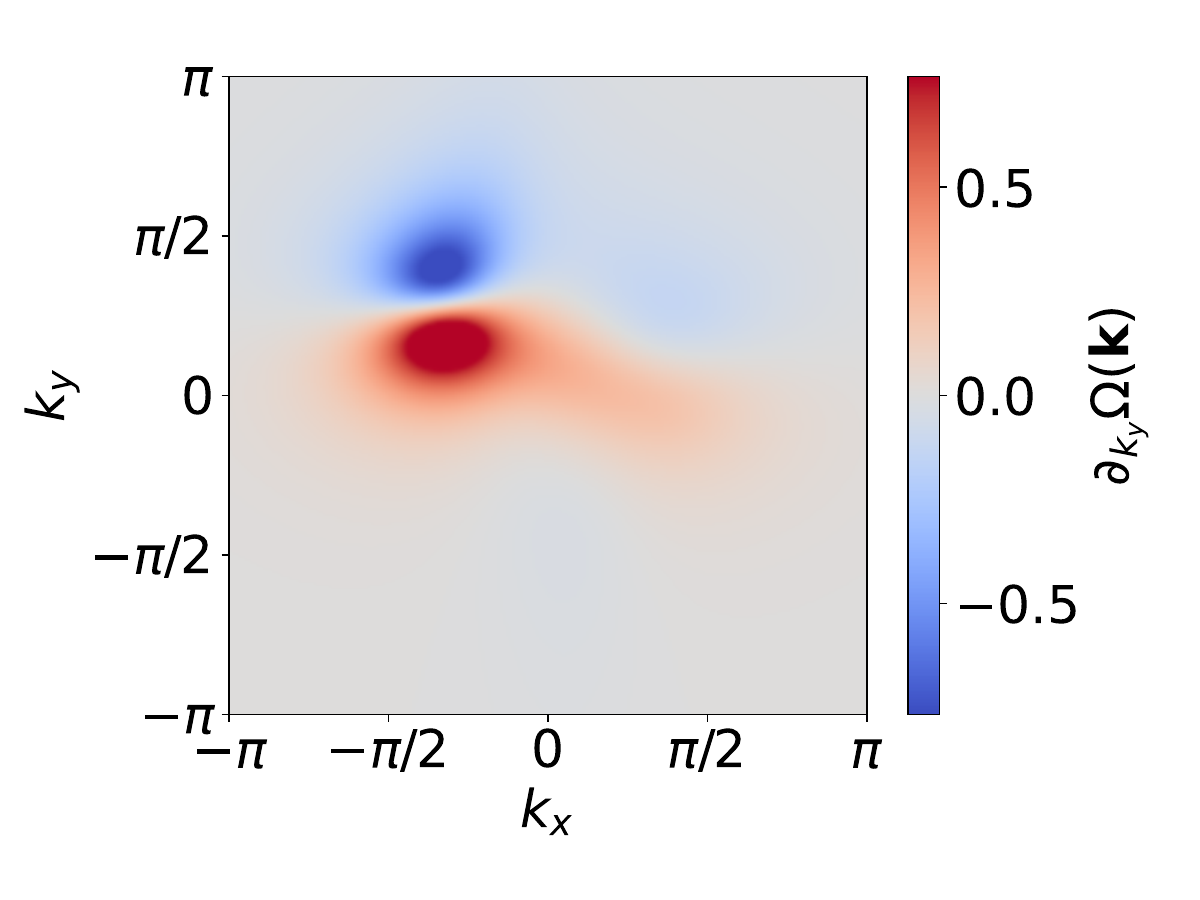}}
\subfigure[]
{\includegraphics[width=1.7in]{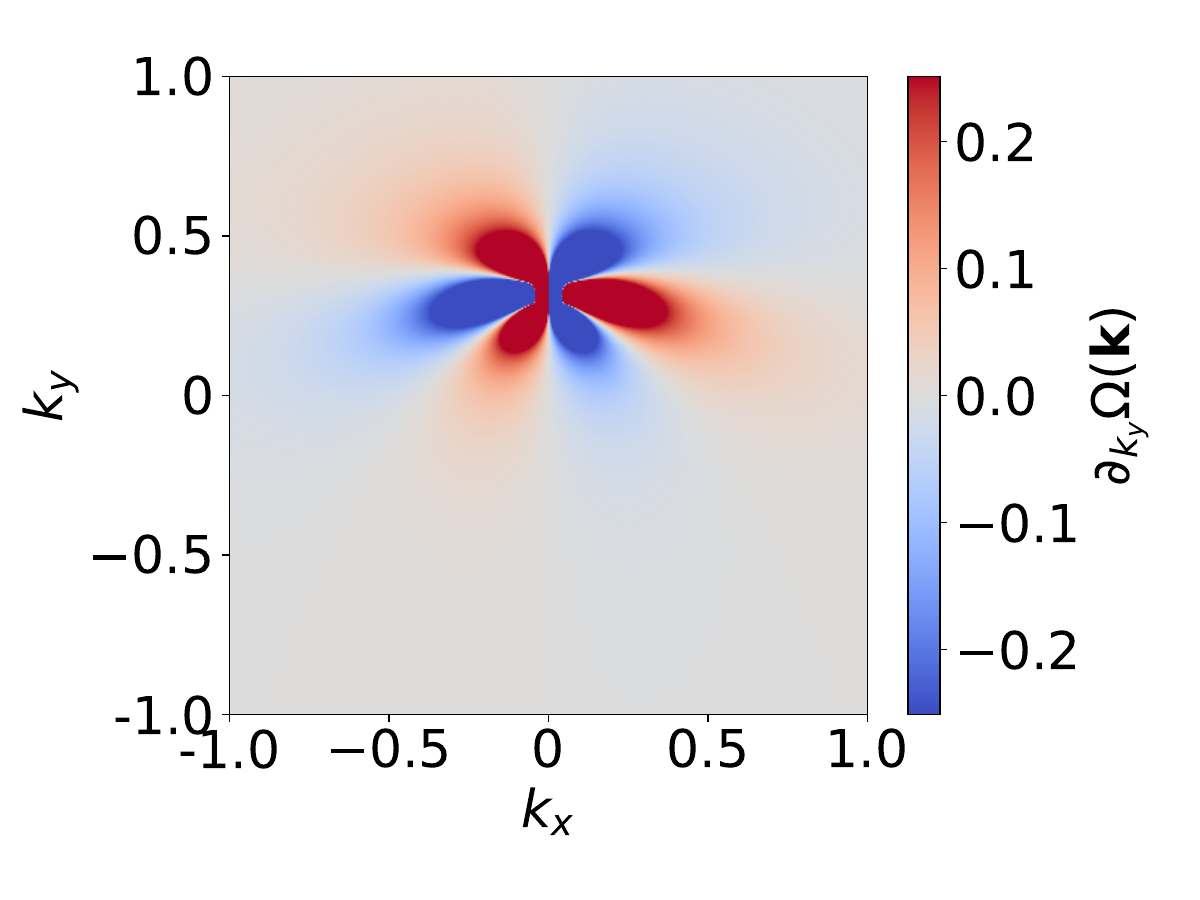}}
\subfigure[]
{\includegraphics[width=1.7in]{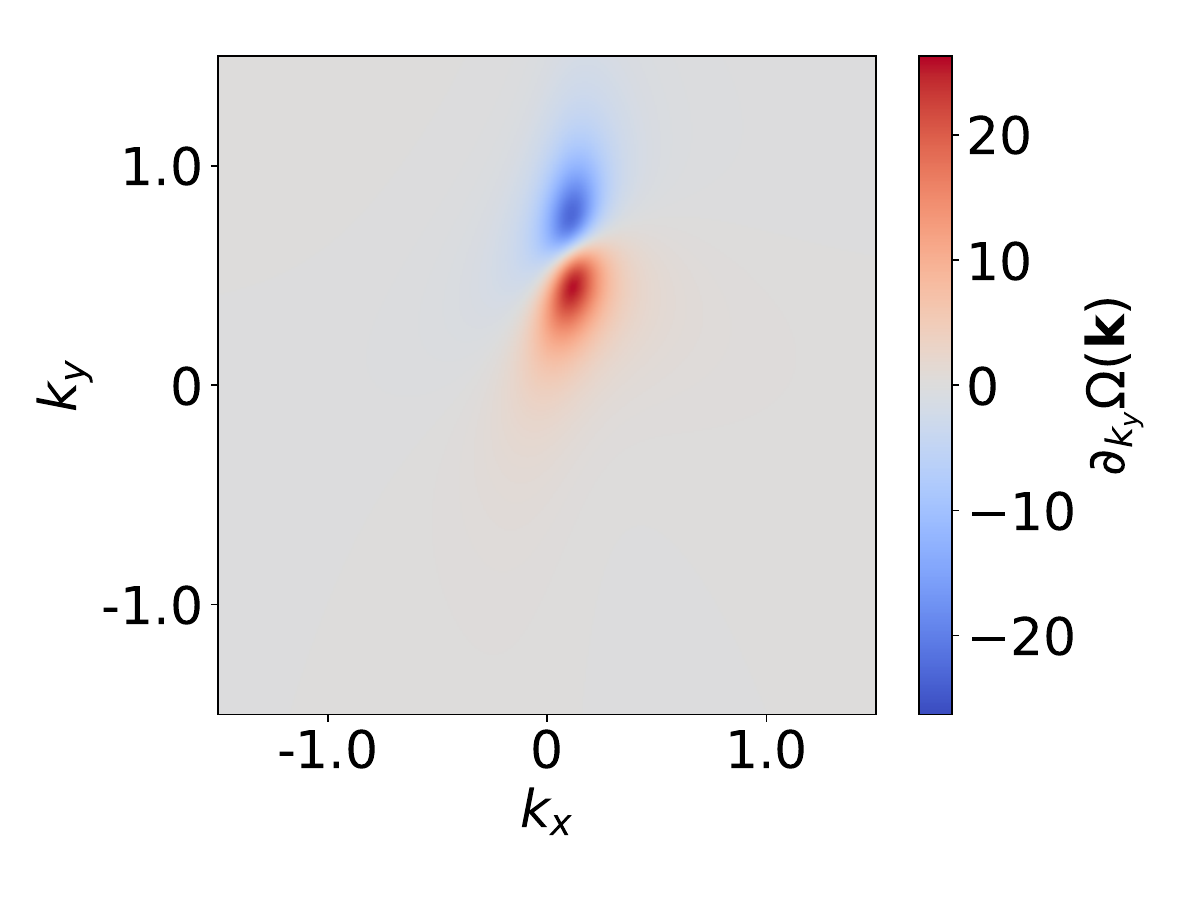}}
\vspace{-0.2cm}
\caption{(Color online) BC and its derivatives for three scenarios. The first row shows $\Omega(\mathbf{k})$ for
(a) the perturbative scenario with $\Delta_x=0.5$, $\Delta_y=0$, $m_z=1.0$, (b) the perturbative scenario
with $\Delta_x=0.5$, $\Delta_y=0.6$, $m_z=1.0$, (c) the Floquet scenario with $\phi=\pi/3$, and (d) the
hybrid scenario with $\Delta_y=-0.4$, $\phi=\pi/2$, $\Delta_x=0.2$, $m_z=0$. The second and third rows
display the density of BCD $\partial_x\Omega(k)$ and $\partial_y\Omega(k)$, respectively, for the same parameter sets.
These derivatives determine the BCD. Parameters common to all related panels $A_1=0.8$, $A_2=0.9$, $\omega=100$, $g_1/g_2=0.5$, and $v=1.0$.}
\label{fig:bc_bcd}
\end{figure*}

\subsection{Berry curvature}\label{subsec_BC}

For a two-band model described by the Hamiltonian $\mathcal{H}(\mathbf{k})=\mathbf{h}(\mathbf{k})\cdot\bm{\sigma}$,
the Berry curvature (BC) is exclusively $z$-polarized and is given by~\cite{Bian2026EPJP,Saha2016PRB,Oliveira2024arXiv,Kim2024arXiv}
\begin{eqnarray}
\Omega_z(\mathbf{k})
=\frac{1}{2|\mathbf{h}(\mathbf{k})|^3}\mathbf{h}(\mathbf{k})\cdot \Bigl(\partial_{k_x}\mathbf{h}(\mathbf{k})\times\partial_{k_y}\mathbf{h}(\mathbf{k})\Bigr),\label{eq:BC_Formula}
\end{eqnarray}
with the vector field $\mathbf{h}(\mathbf{k})\equiv(h_x,h_y,h_z)$ and
$|\mathbf{h}(\mathbf{k})|=\sqrt{h_x^2+h_y^2+h_z^2}$. Physically, the BC encodes the local geometric curvature
of the pseudospin texture on the momentum-space manifold.

\subsubsection{Perturbative scenario}\label{subsubsec_BC}

For the perturbative Hamiltonian in Eq.~(\ref{eq:Hamiltonian_per}), the corresponding vector field $\mathbf{h}(\mathbf{k})$
is given by Eq.~(\ref{eq:vector_per}). Substituting this into Eq.~(\ref{eq:BC_Formula}) yields
\begin{eqnarray}
\Omega_z^{\mathrm{per}}(\mathbf{k})
=\frac{m_zg_2(2g_1k_x^2+vk_y)}{2|\mathbf{h}|^3}.\label{eq:bc_per}
\end{eqnarray}
Two features are immediately apparent. On one hand, the sign of $\Omega_z^{\mathrm{per}}$ is determined entirely by $m_z$,
which enters as a multiplicative prefactor. On the other hand, the numerator contains the term $2g_1k_x^2+vk_y$, which depends quadratically on $k_x$ and linearly on $k_y$. This asymmetry in $k_y$ directly reflects the broken inversion symmetry
of the underlying semi-Dirac model.

Figs.~\ref{fig:bc_bcd}(a) and (b) present representative BC distributions
for the perturbative scenario. In the presence of $\Delta_x$ alone, the distribution
respects the mirror symmetry $\Omega_z(k_x,k_y)=\Omega_z(-k_x,k_y)$, which is a direct consequence
of the invariance of the Hamiltonian in Eq.~(\ref{eq:Hamiltonian_per}) under $k_x\rightarrow-k_x$.
The BC then exhibits two lobes of opposite sign centered at the shifted Dirac points, whose magnitudes
are governed by $m_z$.

When a nonzero $\Delta_y$ is also introduced, as shown in Fig.~\ref{fig:bc_bcd}(b),
this mirror symmetry is entirely broken and the BC develops a skewed,
asymmetric profile in momentum space. The $\Delta_y$ term tilts the band structure and moves
the Berry-curvature hot spots off the high-symmetry lines, producing a non-centrosymmetric distribution
with important implications for the BCD (see Sec.~\ref{subsec_BCD}).

\subsubsection{Floquet scenario}\label{subsubsec_BC_Floquet}

Paralleling the derivation for the perturbative scenario in Sec.~\ref{subsubsec_BC}, the BC for the Floquet scenario
can be obtained from its effective Hamiltonian in Eq.~(\ref{eq:H_eff}) as
\begin{eqnarray}
\Omega_z^{\mathrm{Flo}}(\mathbf{k})=\frac{\gamma g_2 k_x(g_1k_x^2+vk_y-\delta_x)}{2|\mathbf{h}^{\rm Flo}(\mathbf{k})|^3},\label{eq:bc_Flo}
\end{eqnarray}
with $\mathbf{h}^{\rm Flo}(\mathbf{k})\equiv(g_1k_x^2-v k_y+\delta_x,g_2k_xk_y,\gamma k_x)$.

It is apparent that the BC is an odd function of $k_x$, satisfying $\Omega_z^{\mathrm{Flo}}(k_x,k_y)
=-\Omega_z^{\mathrm{Flo}}(-k_x,k_y)$. This properties arises directly from the $\gamma k_x\sigma_z$ term
with a nonzero $\gamma$ in the Hamiltonian~(\ref{eq:H_eff}). Fig.~\ref{fig:bc_bcd}(c) presents
$\Omega_z^{\mathrm{Flo}}$ for $\phi=\pi/3$. We observe that the Berry curvature is concentrated
near the Dirac point $\mathbf{k}=(0,\delta_x/v)$, while its positive and negative lobes are
distributed antisymmetrically about $k_x=0$, with the values scaling as $\gamma\propto\sin\phi$.
In particular, the BC clearly possesses odd parity under $k_x\rightarrow-k_x$.

In the opposite limit, $\phi=0,\pi$
implying $\gamma=0$, the BC vanishes identically, which is consistent with the fact that the Floquet
field only displaces the Dirac point without creating any pseudospin texture. It is worth emphasizing
that the system remains gapless in the Floquet scenario, so the absence of a global gap implies that
no well-defined Chern number exists (see Sec.~\ref{subsec_Chern_number}).

\begin{figure*}
\centering
\subfigure[]
{\includegraphics[width=2.2in]{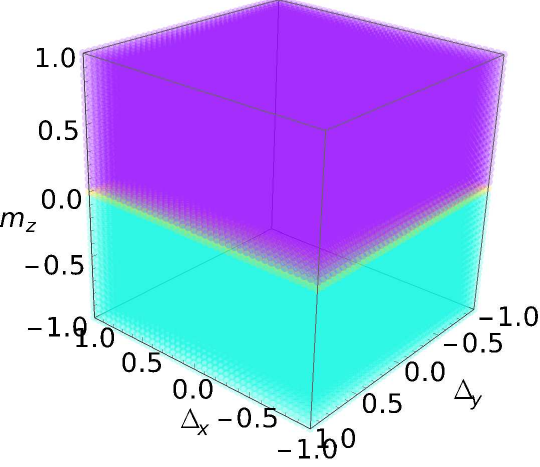}}\hspace{0.5cm}
\subfigure[]{\includegraphics[width=2in]{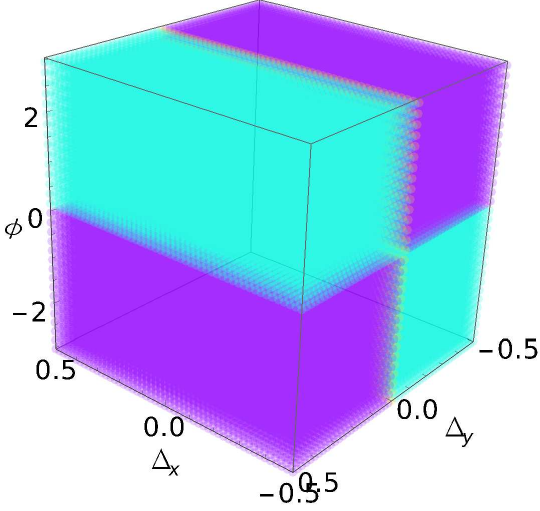}}\hspace{0.5cm}
\subfigure[]{\includegraphics[width=2in]{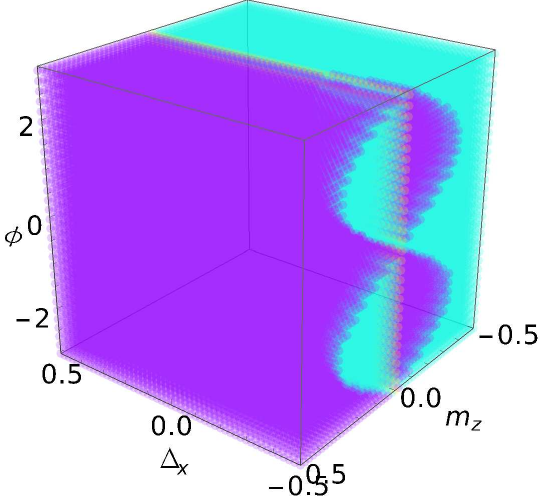}}
\vspace{-0.2cm}
\caption{(Color online) Topological phase diagrams showing the Chern number for (a) the perturbative
scenario and the hybrid scenario with (b) $m_z=0$ and (c) $\Delta_y=0$. The colors purple, yellow, and blue regions
correspond to $C^{\mathrm{per}/\mathrm{hyb}}=+2$, ill-defined, and $C^{\mathrm{per}/\mathrm{hyb}}=-2$, respectively.}
\label{fig:Cnumber}
\end{figure*}

\subsubsection{Hybrid scenario}\label{subsubsec_BC_hybrid}

Applying the same BC formula to the hybrid effective Hamiltonian in Eq.~(\ref{eq:H_tot}),
in analogy with the perturbative scenario of Sec.~\ref{subsubsec_BC}, yields
\begin{eqnarray}
\Omega_z^{\mathrm{hyb}}(\mathbf{k})
&=&
\frac{1}{2|\mathbf{h}^{\rm hyb}(\mathbf{k})|^3}\big\{g_1g_2(\gamma k_x^3+2m_z k_x^2)-v\gamma \Delta_y\nonumber\\
&&
+g_2\big[\gamma(vk_y-\Tilde{\Delta}_x)k_x+v m_zk_y\big]\big\},\label{eq:bc_hyb}
\end{eqnarray}
where the vector field $\mathbf{h}^{\mathrm{hyb}}(\mathbf{k})$ is defined in Eq.~(\ref{Eq_h_xyz}).

Hereby, the hybrid $\Omega_z^{\mathrm{hyb}}(\mathbf{k})$ receives contributions
from both the static perturbation parameters $(\tilde{\Delta}_x,\Delta_y,m_z)$ and the
Floquet parameter $\gamma$. Compared with the perturbative and Floquet scenarios,
Eq.~(\ref{eq:bc_hyb}) exhibits a significantly more nontrivial structure.
When $\phi=0,\pi$ (i.e., $\gamma=0$), the TFOPL amplitude alone plays a significant role,
as captured by $\delta_x$ in Eq.~(\ref{Eq_delta_gamma}). As a result, Eq.~(\ref{eq:bc_hyb})
takes the same form as Eq.~(\ref{eq:bc_per}), with $\Delta_x$ replaced by $\tilde{\Delta}_x=\Delta_x+\delta_x$.
In this limit, the BC inherits the symmetry properties of the perturbative scenario.
In sharp contrast, for general $\phi\neq0,\pi$, both the amplitude and the phase contribute to the BC.
This generates a combination of terms $\gamma g_1g_2 k_x^3$, $\gamma g_2(vk_y-\tilde{\Delta}_x)k_x$,
and $-v\gamma\Delta_y$, which break all reflection symmetries in momentum space. As a result,
$\Omega_z^{\mathrm{hyb}}(\mathbf{k})$ satisfies neither even nor odd symmetry under
$k_x\rightarrow-k_x$ or $k_y\rightarrow-k_y$, yielding a fully asymmetric BC distribution.

Fig.~\ref{fig:bc_bcd}(d) displays $\Omega_z^{\mathrm{hyb}}$ for a representative parameter set.
The positive and negative lobes are neither mirror-symmetric nor point-symmetric, resulting in a
pronounced asymmetry. This complete symmetry breaking originates from the cooperative interplay
between the static $\Delta_y\sigma_y$ term and the Floquet-induced $\gamma k_x\sigma_z$ term.
Two limiting cases are particularly interesting. When $m_z=0$, the sign of the overall BC is
controlled by the product $\phi\Delta_y$. When $\Delta_y=0$, the distribution remains weak asymmetric
because the light-induced $\gamma$ term generates an asymmetric $k_x$-dependence, making
the profile qualitatively distinct from the perturbative case.

Physically, the asymmetric Berry curvature in the hybrid scenario implies that
a finite BCD can arise even when the Chern number is zero (see Sec.~\ref{subsec_BCD}),
which in turn generates a nonlinear Hall response. Because $\Omega_z^{\mathrm{hyb}}$ is
fully tunable through $\phi$, $A_{1,2}$, and the static gate voltages, these parameters
provide a versatile experimental handle for engineering the band geometry.

\subsection{Chern number and topological phase transition} \label{subsec_Chern_number}

With the BC in hand, we are now in a suitable position to examine
the Chern number and the associated topological phase transitions.
The Chern number is generally defined as the integral of the BC over the entire BZ~\cite{Bian2026EPJP,Saha2016PRB,Oliveira2024arXiv,Kim2024arXiv,Hasan2010RMP,
Xiao2010RMP,Mukherjee2024arXiv,Biscocho2025PRB},
\begin{eqnarray}
C\equiv\frac{1}{2\pi}\int_{\rm BZ}d^2\mathbf{k}\,\,\Omega_z(\mathbf{k}),\label{eq:C_formula}
\end{eqnarray}
which is quantized to an integer for gapped systems and characterizes the global topology of the insulating state.
Since the Floquet scenario is gapless and therefore has no well-defined Chern number as discussed in
Sec.~\ref{subsubsec_BC_Floquet}, we put our focus on the perturbative and hybrid scenarios in the following analysis.

\begin{table*}[t]
\centering
\caption{Semimetallic (SM) phase conditions and Chern numbers in different scenarios and parameter regimes.}
\label{tab:Cnumber}
\vspace{0.2cm}
\begin{tabular}{>{\centering\arraybackslash}p{2.3cm}>{\centering\arraybackslash}p{2.3cm}>{\centering\arraybackslash}p{3.15cm}
>{\centering\arraybackslash}p{3.35cm}>{\centering\arraybackslash}p{4.5cm}}
\hline\hline
Scenario & Setup& Parameters & SM phase & Chern number \\
\hline
Perturbative & --- & --- & $m_z=0$ & $\pm 2\,\mathrm{sgn}(m_z)$ \\
Floquet & --- & --- & Always (gapless) & --- \\
Hybrid & $\phi=0,\pi$ & --- & $m_z=0$ & $\pm 2\,\mathrm{sgn}(m_z)$ \\
Hybrid & $\phi\neq0,\pi$ & $m_z=0$ & $\Delta_y=0$ & $\mp 2\,\mathrm{sgn}(\phi\Delta_y)$ \\
Hybrid & $\phi\neq0,\pi$ & $\Delta_y=0,\ \tilde{\Delta}_x\geq 0$ & $m_z=0$ & $\pm 2\,\mathrm{sgn}(m_z)$ \\
Hybrid & $\phi\neq0,\pi$ & $\Delta_y=0,\ \tilde{\Delta}_x<0$ & $m_z=0,\ \pm\sqrt{-\tilde{\Delta}_x\gamma^2/g_1}$ & Depends on $(\tilde{\Delta}_x,m_z,\phi)$ \\
\hline\hline
\end{tabular}
\end{table*}

\subsubsection{Perturbative scenario}

For the perturbative Hamiltonian in Eq.~(\ref{eq:Hamiltonian_per}),
we substitute Eq.~(\ref{eq:bc_per}) into Eq.~(\ref{eq:C_formula}) and integrate
over the BZ, obtaining
\begin{eqnarray}
C^{\mathrm{per}}=\pm 2~{\rm sgn} (m_z),\label{eq:C_per}
\end{eqnarray}
where the sign reflects the choice of the upper or lower band. The factor of two stems from the presence
of four symmetry-related semi-Dirac points in the extended BZ~\cite{Huang2015PRB}.

It can be found from Eq.~(\ref{eq:C_per}) that the sign of $C^{\mathrm{per}}$ is determined solely by the sign of $m_z$.
This indicates that reversing the mass term changes the Chern number from $\pm2$ to $\mp2$, therefore driving
a topological phase transition. Fig.~\ref{fig:Cnumber}(a) presents the phase diagram for the perturbative scenario.
The Chern number takes the value $C^{\mathrm{per}}=+2$ in the region $m_z>0$ and $C^{\mathrm{per}}=-2$ in the region $m_z<0$,
while the system is a gapless semimetal at $m_z=0$. The phase boundary therefore consists of a
single critical point at $m_z=0$, which is in well agreement with Eq.~(\ref{eq:C_per}).

In contrast, $\Delta_x$ and $\Delta_y$ are absent in Eq.~(\ref{eq:C_per}) and thus do not alter the quantized value of $C^{\mathrm{per}}$, even though they shift the Dirac points and break symmetries. This robustness is a direct consequence of topological
protection. In other words, the total Berry-curvature integral is a topological invariant protected by
the gap opened by $m_z$.

\subsubsection{Hybrid scenario}

With respect to the hybrid scenario, the Chern number is obtained by integrating Eq.~(\ref{eq:bc_hyb})
over the BZ. Its value depends nontrivially on the interplay between the static parameters and the Floquet field.
We summarize the main cases below, with the corresponding full phase diagrams presented in Fig.~\ref{fig:Cnumber}(b)
and Fig.~\ref{fig:Cnumber}(b)(c).

At first, let us consider the case $\phi=0$ or $\pi$. In this limit, the light-induced
term $\gamma k_x\sigma_z$ vanishes (i.e., $\gamma=0$), so only the TFOPL amplitude remains
relevant, which is captured by $\delta_x$ in Eq.~(\ref{Eq_delta_gamma}). As shown in
Sec.~\ref{subsubsec_BC_hybrid}, the BC then reduces to the perturbative form with $\Delta_x$
replaced by $\tilde{\Delta}_x=\Delta_x+\delta_x$. Consequently, the Chern number is
\begin{eqnarray}
C^{\mathrm{hyb}}=\pm 2~{\rm sgn}(m_z),
\end{eqnarray}
and the phase diagram is identical to that in Fig.~\ref{fig:Cnumber}(a).
Thus, for $\phi=0,\pi$, the light merely renormalizes the effective $x$-component
of the pseudospin field without altering the topological classification.

\begin{figure*}
\centering
\subfigure[]
{\includegraphics[width=2.2in]{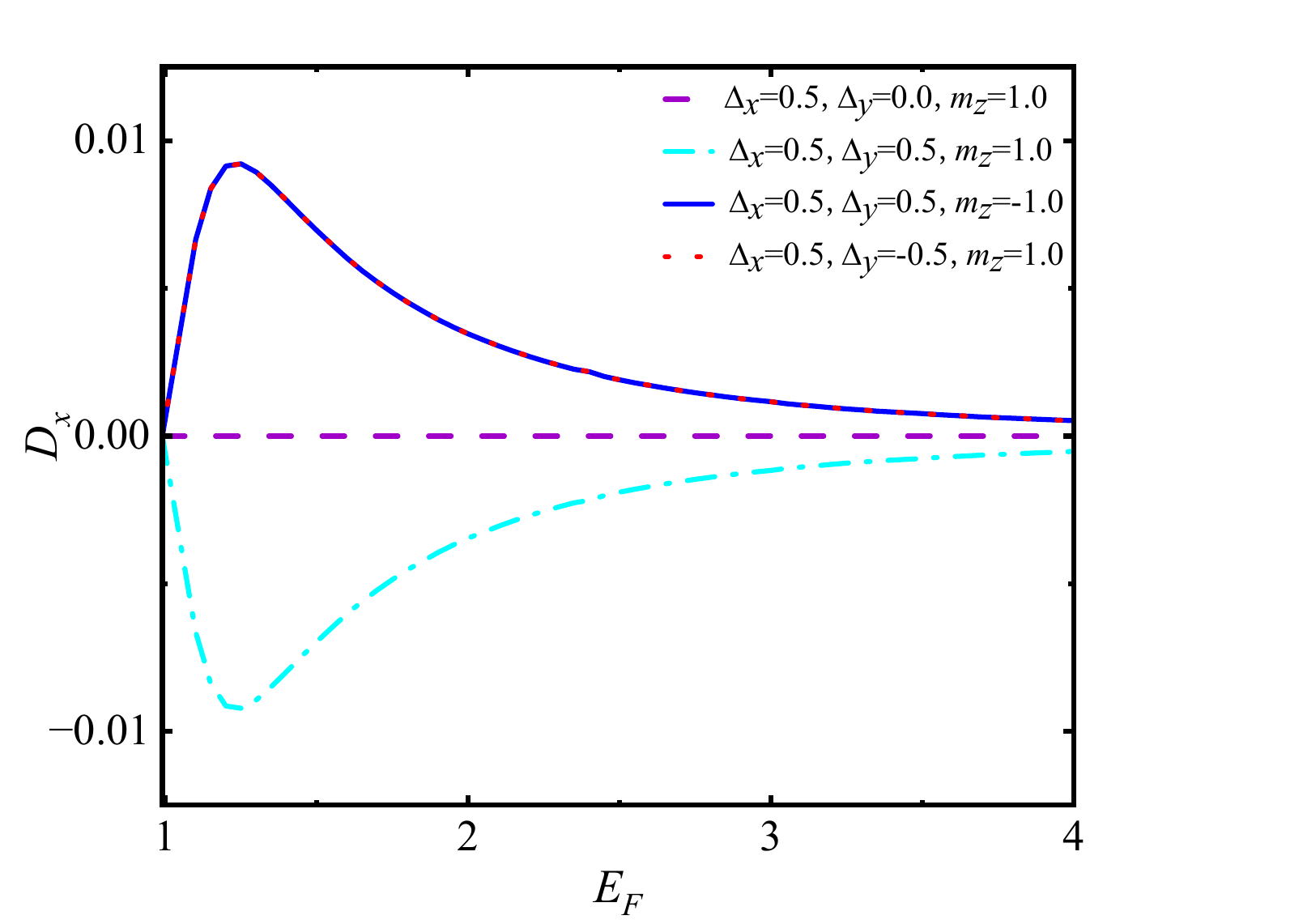}}
\subfigure[]{\includegraphics[width=2.2in]{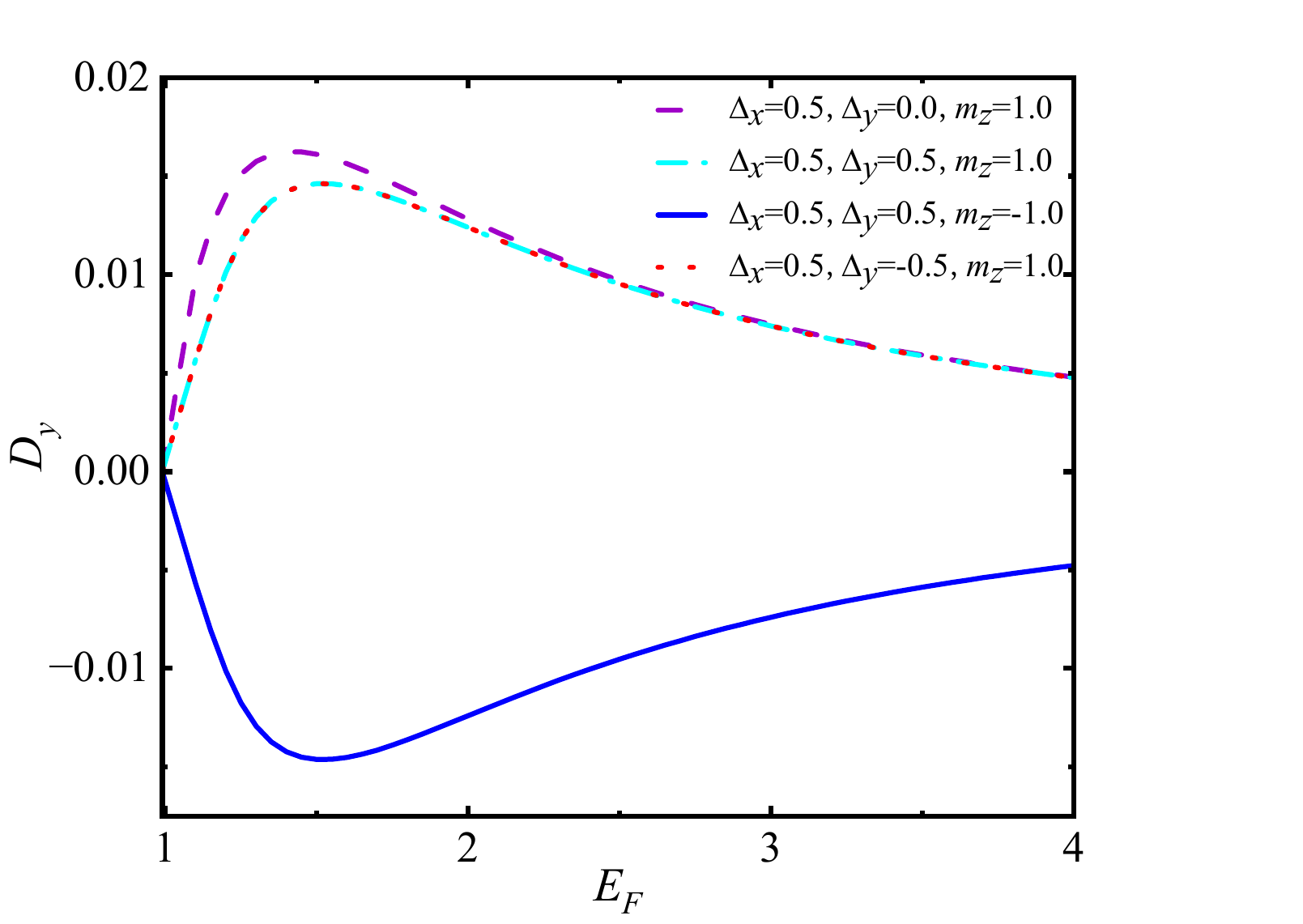}}
\subfigure[]{\includegraphics[width=2.2in]{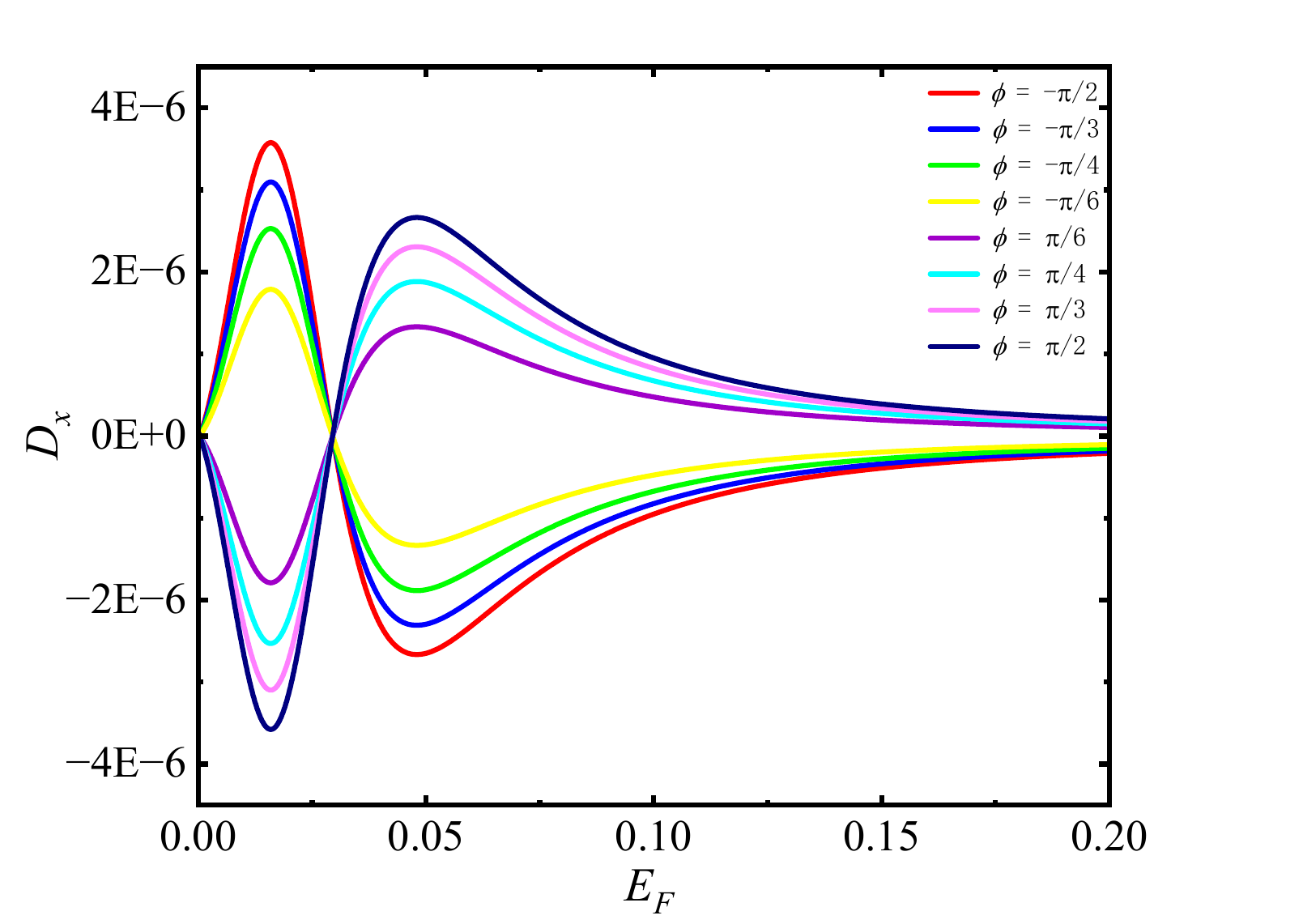}}
\subfigure[]{\includegraphics[width=2.2in]{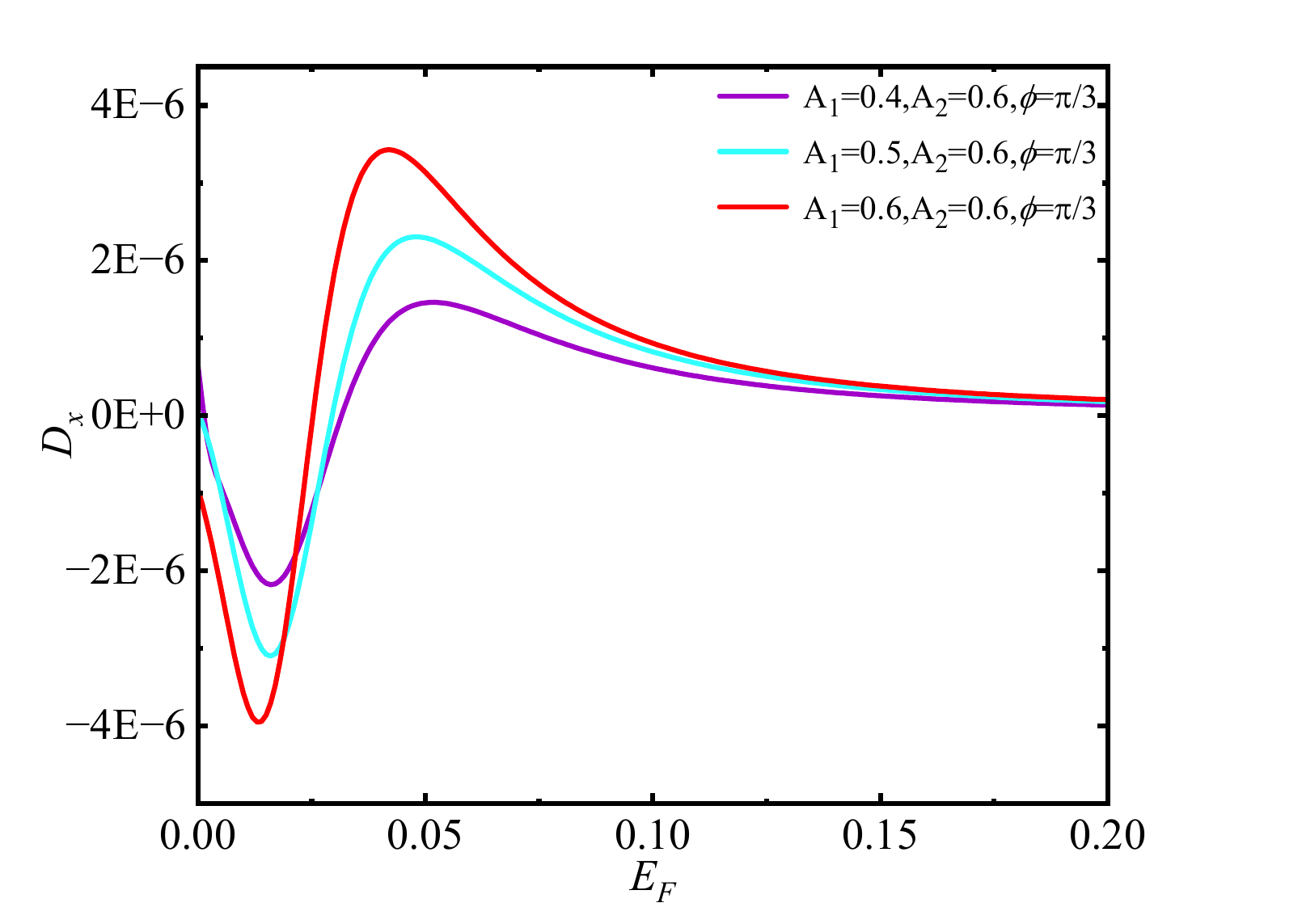}}
\subfigure[]{\includegraphics[width=2.2in]{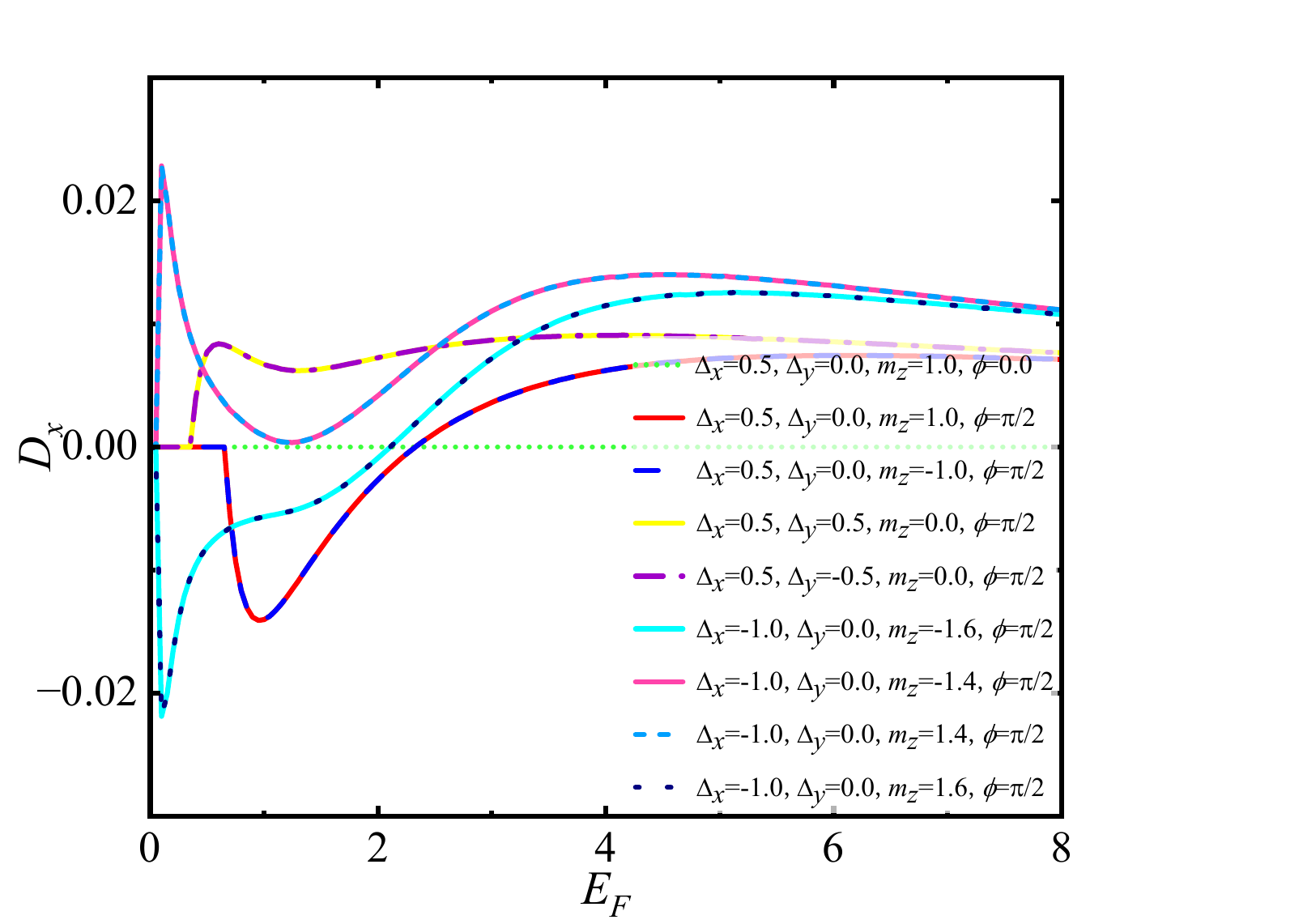}}
\subfigure[]{\includegraphics[width=2.2in]{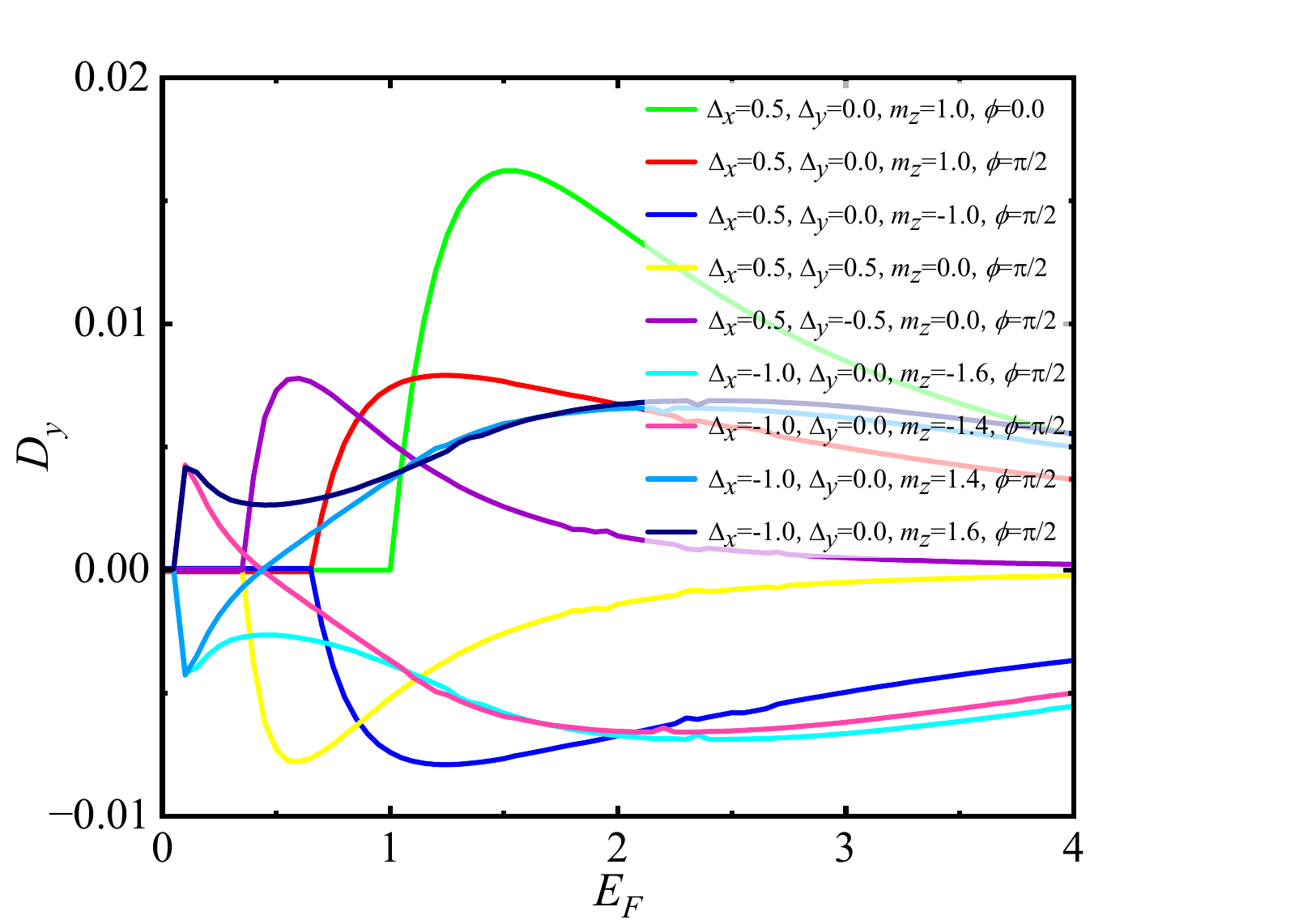}}
\vspace{-0.2cm}
\caption{(Color online) BCD as a function of the Fermi energy $E_F$ for three scenarios.
Perturbative scenario: (a) $D_x$ and (b) $D_y$ for different $\Delta_x,\Delta_y,m_z$.
Floquet scenario: (c) $D_x$ and (d) $D_x$ for different $\phi$ and $A_1$, respectively.
Hybrid scenario: (e) $D_x$ and (f) $D_y$ with $g_1=0.5$, $g_2=1.0$, $A_1=0.5$, $A_2=0.6$, and $\omega=10^5$.}
\label{fig:BCD}
\end{figure*}

Next, we move to the case with $m_z=0$. In the perturbative scenario, the system is gapless for all $\Delta_y$.
In the hybrid scenario, however, the combination $\phi\neq0,\pi$ with $\Delta_y\neq0$ opens a finite gap (see Sec.~\ref{subsec_Per_Floquet_scenario}), and the Chern number becomes
\begin{eqnarray}
C^{\mathrm{hyb}}=\mp 2~{\rm sgn}(\phi\Delta_y),\label{Eq_hyb_C}
\end{eqnarray}
where the upper (lower) sign corresponds to the upper (lower) band and the factor of two again
reflects the four semi-Dirac points in the BZ.

This result is of particular significance. The sign of $C^{\mathrm{hyb}}$ is controlled by the
product $\phi\Delta_y$. Reversing either the light's phase difference or the sign of $\Delta_y$ therefore
flips the Chern number from $+2$ to $-2$ or vice versa, driving a topological phase transition.
This behavior is captured by the checkerboard phase diagram in Fig.~\ref{fig:Cnumber}(b), where the Chern number
changes its sign as $\phi\Delta_y$ crosses zero. Accordingly, the cooperative effects is able to generate a
topological Chern-insulator phase from a system that would otherwise be a gapless semimetal.

At last, we turn to the case without the $y$-component perturbation, i.e., $\Delta_y=0$.
In this sense, the gap-closure criterion in Eq.~(\ref{eq:condition}) then reduces to
\begin{equation}
 m_z\left(g_1g_2m_z^2+\tilde{\Delta}_xg_2\gamma^2\right)=0,\label{Eq_CN_gap_closure}
\end{equation}
and the behavior of the Chern number depends on the sign of $\tilde{\Delta}_x$.

For $\tilde{\Delta}_x\geq0$, the equation admits only the real solution $m_z=0$. This implies
that the system is gapped for all $m_z\neq0$, with
\begin{equation}
C^{\mathrm{hyb}}=\pm 2\,\mathrm{sgn}(m_z),
\end{equation}
which identical to the perturbative scenario as shown in Fig.~\ref{fig:Cnumber}(a).

By contrast, for $\tilde{\Delta}_x<0$, the equation~(\ref{Eq_CN_gap_closure}) allows three distinct real solutions:
$m_z=0$ and $m_z=\pm\sqrt{-\tilde{\Delta}_x\gamma^2/g_1}$ as detailed in Sec.~\ref{subsub_coop_effects_three_roots}.
At each of these critical values, the gap closes and the system becomes semimetallic. Consequently,
as $m_z$ is tuned from negative to positive values, the system undergoes three
successive insulator¨Csemimetal transitions, with the Chern number alternating between $\pm2$ and $\mp2$.
This multiple-critical-point behavior is illustrated by the intricate phase diagram in Fig.~\ref{fig:Cnumber}(c),
where $C^{\mathrm{hyb}}$ depends nontrivially on $\tilde{\Delta}_x$, $m_z$, and $\phi$.

To be brief, unlike the perturbative scenario, where topological phase transitions are driven
solely by $m_z$, the hybrid scenario supports topological phase transitions driven not only by
$m_z$ but also by $\phi$, $\Delta_y$, and $\tilde{\Delta}_x$. Table~\ref{tab:Cnumber} summarizes
the Chern number results for all scenarios. These offer multiple experimental handles for
controlling the topological state.

\subsection{Berry curvature dipole} \label{subsec_BCD}

While the Chern number is a global topological invariant obtained by integrating the BC over the BZ,
it does not capture the momentum-space distribution of the BC. To characterize the latter,
we introduce the Berry curvature dipole (BCD), defined as the Fermi-surface-weighted first moment of the Berry-curvature gradient~\cite{Battilomo2019PRL,Sodemann2015PRL,Ortix2021AQT},
\begin{eqnarray}
D_a\equiv\int_{\rm BZ} \frac{d^2\mathbf{k}}{(2\pi)^2}f_0(\mathbf{k})\,\,\partial_{k_a}\Omega_z(\mathbf{k}),\label{eq:BCD_formula}
\end{eqnarray}
where $a=x,y$ and $f_0(\mathbf{k})=f(E(\mathbf{k})-E_F)$ is the Fermi-Dirac distribution function, with $E_F$ the Fermi energy.

Since the BCD is a pseudovector, it directly governs the nonlinear Hall conductivity via $\chi_{abc}\propto\varepsilon_{abd}D_d\delta_{bc}$~\cite{Sodemann2015PRL,Ortix2021AQT}.
At zero temperature, its defining integral is restricted to the Fermi surface, making it a sensitive probe
of low-energy band geometry. Below, we analyze its symmetry constraints and tunable behavior across the perturbative,
Floquet-only, and hybrid scenarios

\subsubsection{Perturbative scenario}\label{subsubsec_BCD_perp}

We begin by analyzing the BCD in the perturbative scenario. Substituting
Eq.~(\ref{eq:bc_per}) into Eq.~(\ref{eq:BCD_formula}) shows that the symmetry
properties of $D_{x,y}$ are controlled by the perturbation parameters.

When $\Delta_y=0$, the system respects the $k_x\rightarrow -k_x$ mirror symmetry, which enforces $\Omega_z^{\text{per}}(k_x,k_y)=\Omega_z^{\text{per}}(-k_x,k_y)$. This symmetry has two immediate
consequences for the density of BC (derivatives), namely, $\partial_{k_x}\Omega_z^{\text{per}}$ is odd in $k_x$, but $\partial_{k_y}\Omega_z^{\text{per}}$ is even, as clearly shown in Fig.~\ref{fig:bc_bcd}(e) and (i).
Consequently, $D_x=0$, while $D_y$ acquires a finite value. In particular, the sign of $D_y$ is
determined by $m_z$, and its magnitude depends on the position of the Fermi level relative to
the hot spots of the Berry curvature. In sharp contrast, a finite $\Delta_y$ breaks the $k_x\rightarrow-k_x$
mirror symmetry of the BC. As a result, the density of BCD exhibits no residual symmetry
as shown in Figs.~\ref{fig:bc_bcd}(f) and (j), and both $D_x$ and $D_y$ become generically nonzero.

Fig.~\ref{fig:BCD}(a) and Fig.~\ref{fig:BCD}(b) show these behavior of $D_x$ and $D_y$
for several representative parameter sets. In Fig. 8(a), when $\Delta_y=0$, the
$k_x\rightarrow -k_x$ mirror symmetry enforces $D_x=0$ over the entire $E_F$ range,
as indicated by the flat purple dashed line. Once $\Delta_y\neq0$, this mirror symmetry
is broken and $D_x$ becomes nonzero, developing pronounced peaks near the band edges.
The sign and magnitude of these peaks depend sensitively on $\Delta_x$, $\Delta_y$, and $m_z$.
In Fig. 8(b), $D_y$ remains finite even at $\Delta_y=0$ and varies continuously with $E_F$.
For $\Delta_y\neq0$, the sign and line shape of $D_y$ are modified. In both panels, $D_x$ and $D_y$
are strongly enhanced near the band edges and decay at larger $E_F$.

These results suggest that a nonzero $D_{x}$ therefore serves as a direct experimental
signature of finite $\Delta_{y}$. In addition, reversing the sign of $m_z$ flips the overall
sign of $\Omega_z^{\text{per}}$, which in turn reverses both $D_x$ and $D_y$ without changing
their dependence on $E_{\mathrm{F}}$. This leads to a testable prediction, i.e., reversing the
static mass term (e.g., through strain or an external magnetic field)~\cite{You-Low2018PRB,Roy-Narayan2022JPCM}
reverses the full nonlinear Hall response while preserving its gate-voltage line shape.

\subsubsection{Floquet scenario}

We next turn to the Floquet scenario. In this circumstance, the light-induced $\gamma k_x\sigma_z$ term
imposes the odd parity $\Omega_z^{\text{Flo}}(k_x,k_y)=-\Omega_z^{\text{Flo}}(-k_x,k_y)$. As shown
in Fig.~\ref{fig:bc_bcd}(g) and Fig.~\ref{fig:bc_bcd}(k), this forces $\partial_{k_y}\Omega_z^{\text{Flo}}$
to be odd in $k_x$, so that $D_y=0$. In comparison,  $\partial_{k_x}\Omega_z^{\text{Flo}}$ is even
under $k_x\rightarrow -k_x$ and yields a finite $D_x$. The phase dependence of $D_x$ follows directly
from its proportionality to $\gamma$. Since $\gamma\propto\sin\phi$, it is maximal near $\phi=\pm\pi/2$
and vanishes at $\phi=0,\pi$.

Fig.~\ref{fig:BCD}(c) and Fig.~\ref{fig:BCD}(d) present the main results of the Floquet BCD
component $D_x$ versus $E_F$ for the Floquet scenario. Specifically, Fig.~\ref{fig:BCD}(c) shows
that $D_x$ depends strongly on the phase $\phi$, with multiple peaks appearing near the band edge
and both its sign and spectral shape changing markedly as $\phi$ varies. In Fig. 8(d), by contrast,
$\phi$ is fixed at $\pi/3$ and the TFOPL amplitude $A_1$ is varied. Increasing $A_1$ enhances the
peak magnitude of $D_x$ monotonically, while the overall line shape and peak positions are
nearly unchanged. In particular, Fig.~\ref{fig:BCD}(c) and (d) demonstrate that $D_x$ undergoes
sign change at a certain $E^\star_F$ that is associated with the gapless Dirac node at $\mathbf{k}=(0,\delta_x/v)$
given in Sec.~\ref{subsection_Floquet}. This zero-crossing
energy $E^\star_F$ can be continuously shifted by tuning the laser intensity, which renormalizes
$\delta_x$~(\ref{Eq_delta_gamma}) and thereby displaces the Dirac point in momentum space.

In short, $\phi$ controls the sign and profile of $D_x$, while $A_1$ governs its amplitude.
Tuning the light intensity shifts the Dirac point and moves the $D_x$ zero-crossing energy to a different Fermi energy.
These make the BCD a direct experimental clue that uniquely identifies the Floquet scenario.

\subsubsection{Hybrid scenario}

Finally, we consider the BCD in the perturbative-Floquet hybrid scenario,
which produces the richest BCD phenomenology among all cases studied in Fig.~\ref{fig:BCD}.

At the limit case with $\phi=0,\pi$, the behavior is analogous to its perturbative counterpart provided in
Sec.~\ref{subsubsec_BCD_perp}. However, for generic $\phi\neq0,\pi$, all reflection symmetries of
the BCD distribution in momentum space are broken as depicted in Fig.~\ref{fig:bc_bcd}(h) and Fig.~\ref{fig:bc_bcd}(l),
rendering both $D_x$ and $D_y$ generically nonzero. Their amplitudes, relative magnitudes,
and signs can be continuously tuned via the TFOPL parameters $(A_1,A_2,\phi,\omega)$
and the gate-controlled static perturbations $(\Delta_x,\Delta_y,m_z)$.
Fig.~\ref{fig:BCD}(e)-(f) specifically present our central numerical results.

At the outset, we can find from Fig.~\ref{fig:BCD}(e)-(f) that the most
distinctive feature of this hybrid regime is the dramatic contrast between
the BCD near and away from topological phase boundaries. Close to the gap-closing manifolds,
the BC develops certain sharp momentum-space hot spots that greatly amplify its gradient
$\partial_a\Omega_z$ and hence the BCD.  In contrast, away from the hot spots and deep
within the gapped insulating phases, the Berry-curvature profile varies smoothly and
the BCD is heavily suppressed. Consequently, by sweeping system parameters across a phase boundary,
one can toggle the nonlinear Hall response~\cite{Sodemann2015PRL,Ortix2021AQT} on and off.
This switching effect is unique to the hybrid scenario and is absent in both the purely
perturbative and purely Floquet-driven scenarios.

Additionally, our numerical calculations reveal a characteristic double-peak structure in both $D_x$ and $D_y$
clearly shown in Fig.~\ref{fig:BCD}(e)-(f). The sharp peak near the renormalized Dirac-point energy $E_F\approx0$
and a broader secondary peak at higher Fermi energies. These two peaks have distinct physical origins. The former
arises from proximity to gap-closing conditions, where rapid pseudospin reorientation amplifies $\partial_a\Omega_z$,
making it highly sensitive to $\gamma$, $\Delta_y$, and $\tilde{\Delta}_x$. The latter originates from extended
momentum-space regions with moderate Berry-curvature gradients, rendering it comparatively robust against parameter
variations. This two-peak feature constitutes another distinct experimental fingerprint of the cooperative interplay
between the static and Floquet scenarios, and it is absent in either individual scenario.

Moreover, the full BCD vector ($\mathbf{D}=D_x\mathbf{e}_x+D_y\mathbf{e}_y$) or its $y$-component ($D_y$)
can undergo a synchronized sign reversal by tuning a single external parameter. Since $\gamma\propto\sin\phi$,
sweeping $\phi$ across zero flips the sign of $\gamma$, which reverses all $\gamma$-odd contributions to
$\Omega_z^{\text{hyb}}$ and thereby flips $D_x$ and $D_y$ simultaneously. Likewise, a sign change in either
$\Delta_y$ or $m_z$ can reverse $y$-component $D_y$ as shown in Fig.~\ref{fig:BCD}(f),
which stems from the odd parity of $D_y$ in both parameters. In both cases, the joint sign reversal
reflects a global inversion of the pseudospin texture, which is induced
by broken inversion symmetry in the hybrid scenario. Experimentally, the corresponding zero
crossings of the BCD and associated nonlinear Hall response~\cite{Sodemann2015PRL,Ortix2021AQT}
can be observed by tuning the laser phase or the in-plane static perturbations,
providing unambiguous evidence of cooperative effects between two distinct scenarios.

To be brief, we examine the behavior of BCD in three scenarios.
In the perturbative scenario, $k_x\to-k_x$ mirror symmetry enforces $D_x=0$ and $D_y\neq0$ for $\Delta_y=0$, while
finite $\Delta_y$ makes both nonzero. In the Floquet scenario, the $\gamma k_x\sigma_z$ term imposes odd parity,
giving $D_y=0$ and $D_x\propto\sin\phi$. The hybrid scenario breaks all reflection symmetries at $\phi\neq0,\pi$,
producing tunable $D_x$ and $D_y$, a double-peak BCD structure, enhancement near gap-closing manifolds,
and topological switching across phase boundaries. Besides, synchronized sign reversal of the full BCD
vector or its $y$-component provides a distinctive experimental fingerprint of cooperative effects.

\section{Anomalous transport properties}\label{Sec_transport}

We are now in a suitable position to investigate the anomalous transport coefficients of the system,
focusing on the anomalous Hall conductivity (AHC)~\cite{Qi2006PRB}, anomalous Nernst
conductivity (ANC)~\cite{Xiao2006PRL}, and anomalous
thermal Hall conductivity (ATHC)~\cite{Luttinger1964PR}. As established in the previous section~\ref{Sec_BC},
the BC in the Floquet scenario is an odd function of $k_x$, so its BZ integral vanishes identically, yielding no net
contribution to these quantities. We therefore restrict our analysis to the perturbative and hybrid scenarios,
in which the broken symmetries and finite BC gradients give rise to nontrivial transport coefficients. The
following subsections detail the dependence of these quantities on the Fermi energy, temperature, and
system parameters.

\subsection{Anomalous Hall conductivity}\label{subsec_AHC}

The AHC is one of the most direct transport manifestations of band topology, originating from the BC of
the occupied electronic states~\cite{Bian2026EPJP,Saha2016PRB,Qi2006PRB,
Pradhan2024arXiv,Habe2025PRB,Karplus2954PR,Streda2010PRB,Baidya2025arXiv}.
It is given by the BC integrated over the BZ and weighted by the Fermi¨CDirac distribution $f(E^n(\mathbf{k})-E_F)$,
\begin{eqnarray}
\sigma_{xy}\equiv\frac{e^2}{\hbar}\sum_n\int_{\mathrm{BZ}}\frac{d^2\mathbf{k}}{(2\pi)^2}
f\bigl(E^n(\mathbf{k})-E_F\bigr)\Omega_z^n(\mathbf{k}),
\end{eqnarray}
where the summation $n$ spans all energy bands.

When the Fermi level lies inside the bulk gap of a fully gapped Chern insulator, $\sigma_{xy}$
becomes quantized in plateaus following the Thouless-Kohmoto-Nightingale-den Nijs (TKNN)
relation $\sigma_{xy}=e^2C/h$, where $C$ is the Chern number~\cite{Thouless1982PRL}.
This quantization is a defining feature of  a topologically nontrivial state and provides a direct,
experimentally accessible signature of the Chern number.

\begin{figure*}
\centering
\subfigure[]
{\includegraphics[width=2.8in]{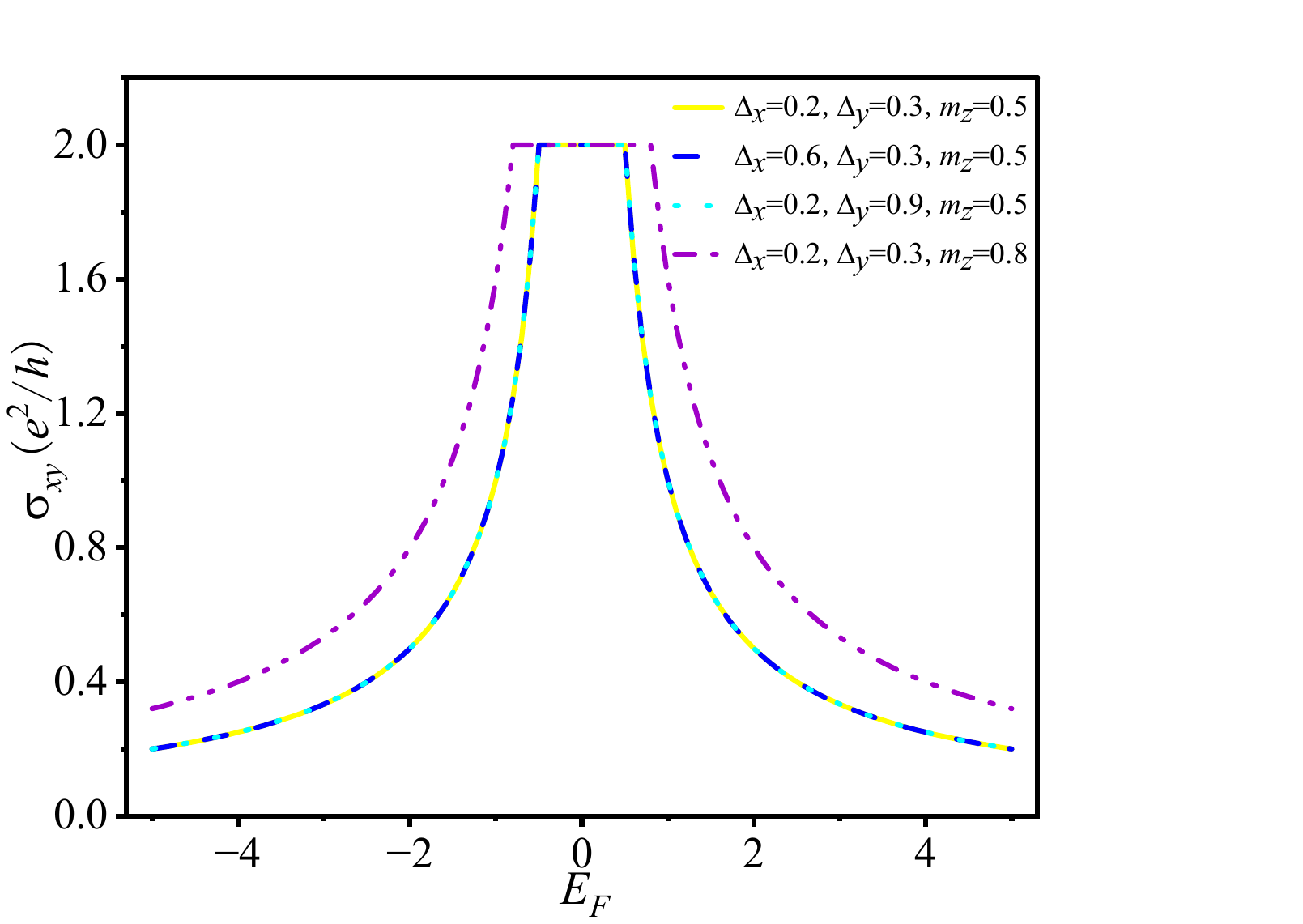}}
\hspace{-0.1cm}
\subfigure[]
{\includegraphics[width=2.8in]{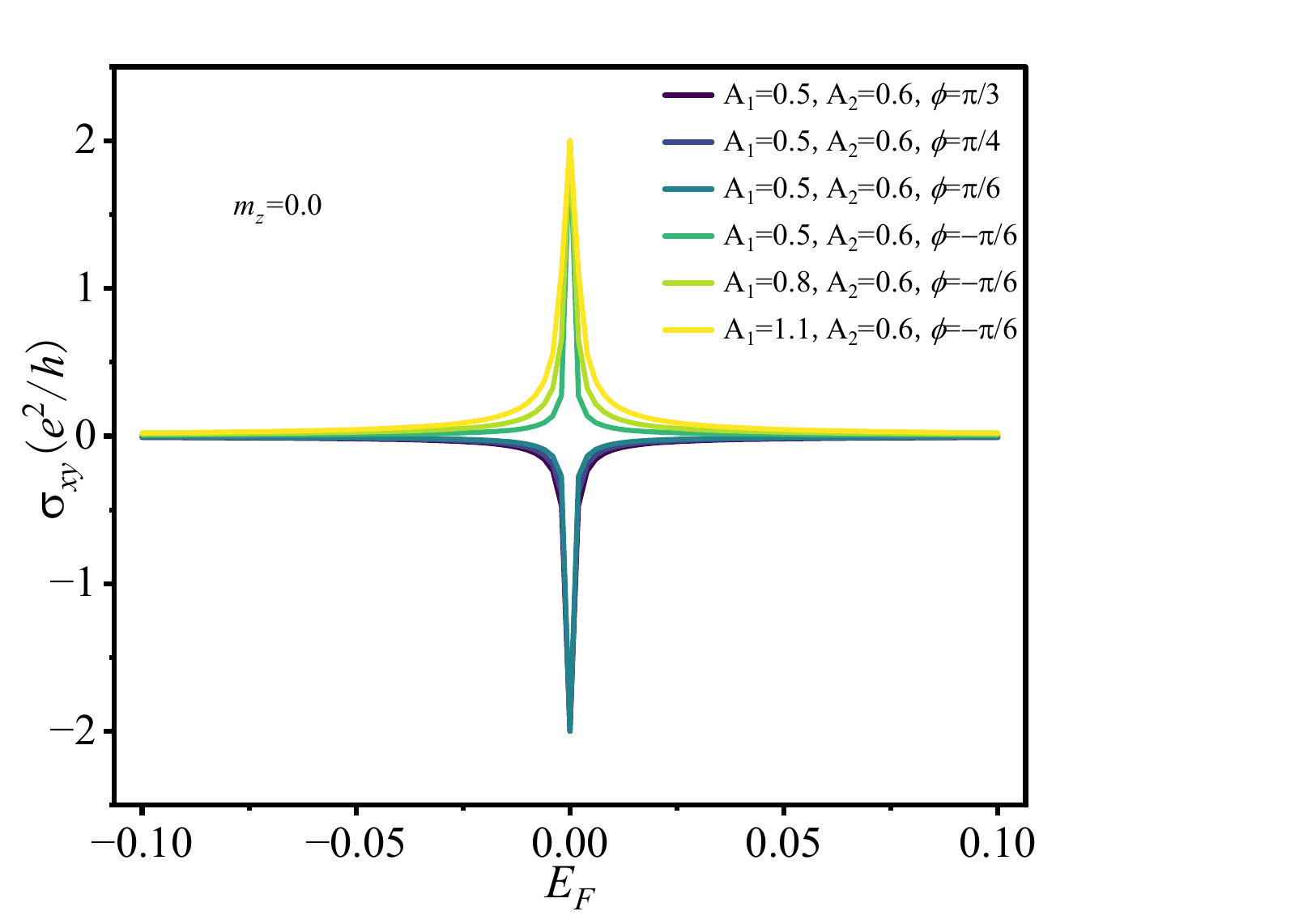}}\\
\vspace{-0.3cm}
\subfigure[]
{\includegraphics[width=2.8in]{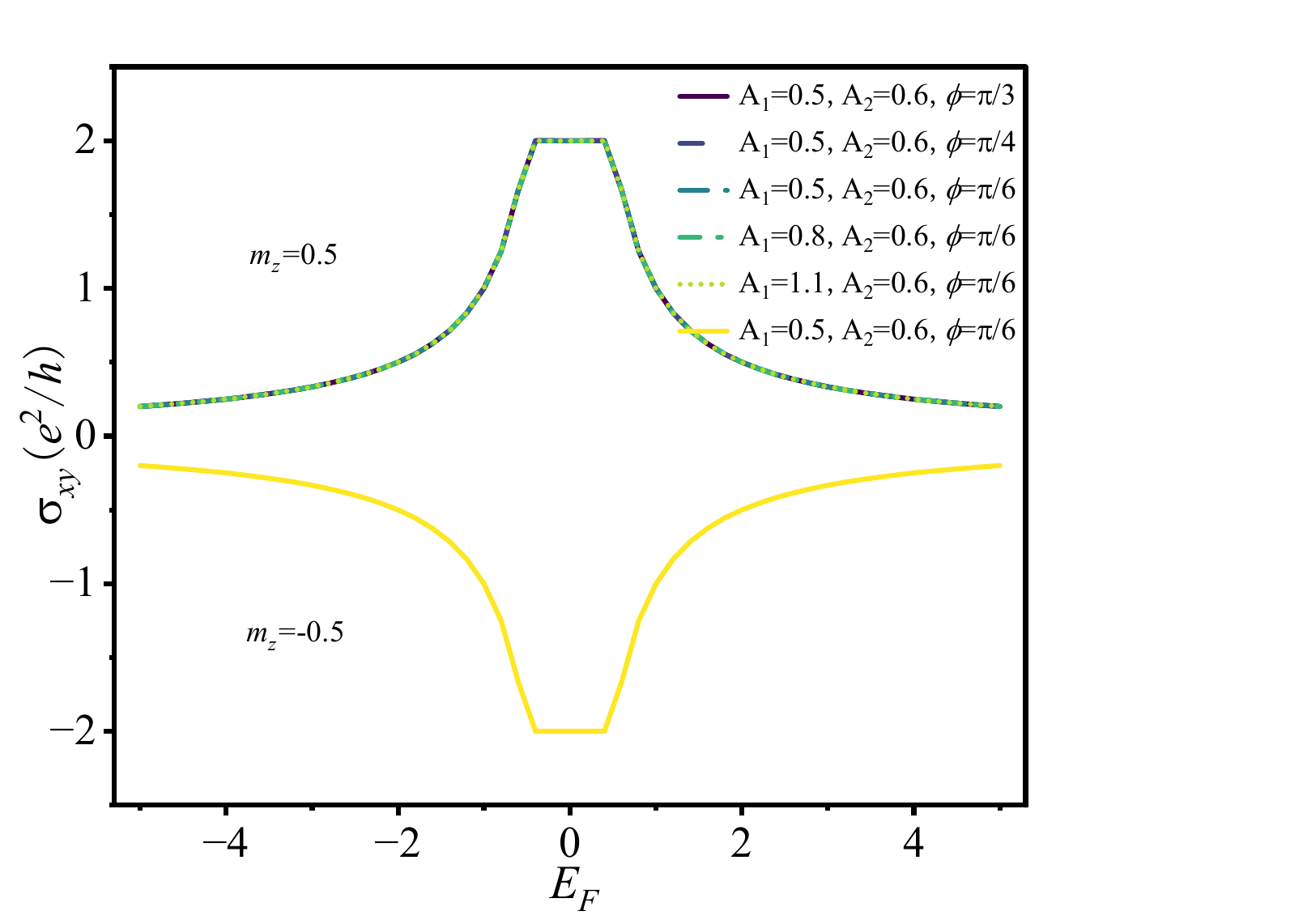}}
\hspace{-0.1cm}
\subfigure[]
{\includegraphics[width=2.8in]{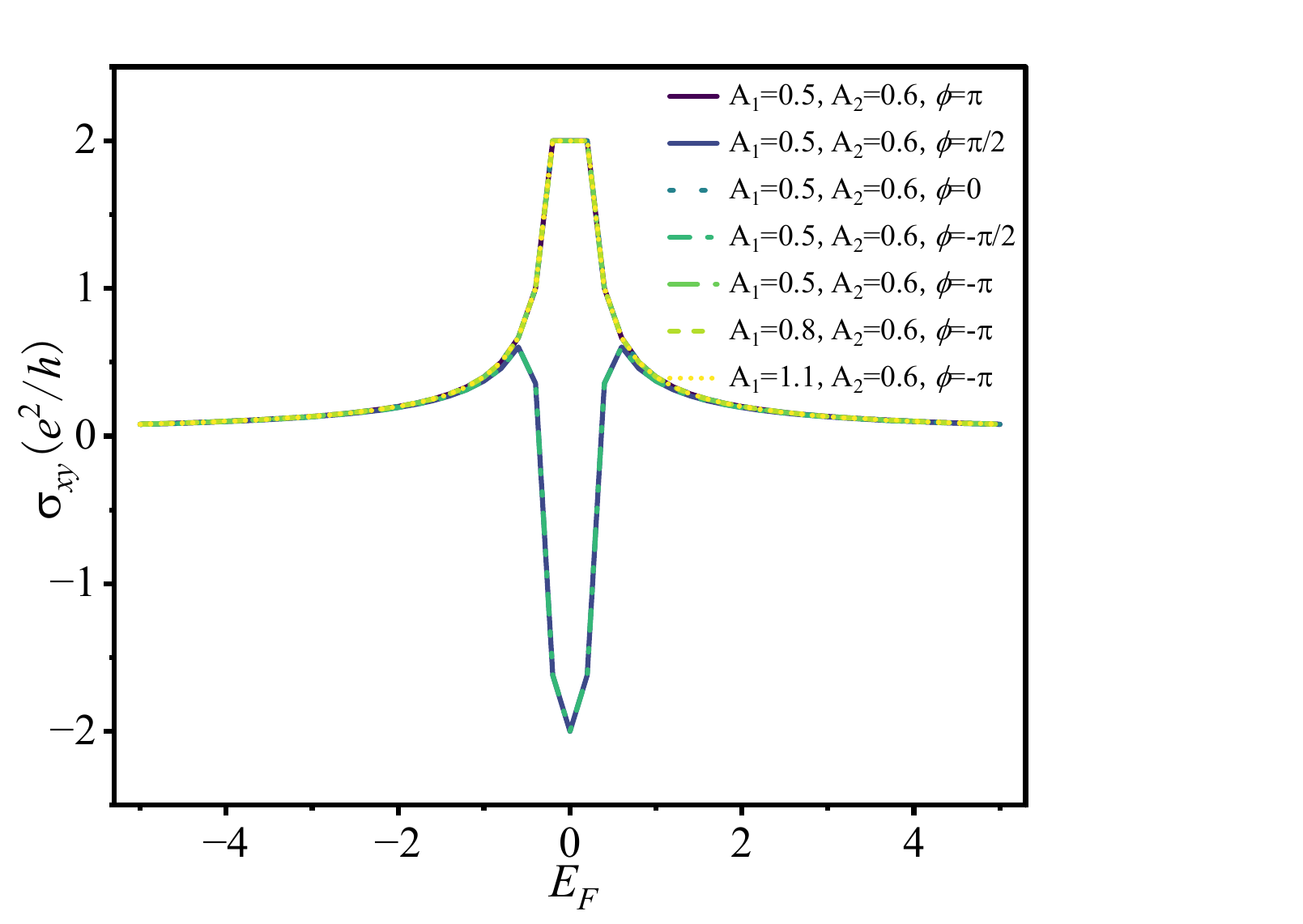}}
\vspace{-0.2cm}
\caption{(Color online) AHC $\sigma_{xy}$ as a function of the Fermi energy $E_F$.
(a) Perturbative scenario for representative values of $\Delta_x$, $\Delta_y$, and $m_z$.
Hybrid scenario with $g_1=0.5$, $g_2=1.0$, $A_2=0.6$, and $\omega=10^5$ for
(b) $m_z=0$ with varying TFOPL amplitude $A_1$ and phase $\phi$, (c) $\Delta_y=0$,
$\Delta_x=0.8$ with varying $\phi$ for $m_z=\pm0.5$, and (d) $\Delta_y=0$, $\Delta_x=-0.8$,
$m_z=0.2$ with varying $A_1$ and $\phi$.}
\label{fig:AHC}
\end{figure*}

In the perturbative scenario, the bulk gap is opened solely by the static mass $m_z$,
yielding the Chern number $C^{\mathrm{per}}$ given by Eq.~(\ref{eq:C_per}).
Combined with the TKNN relation, this gives rise to a quantized plateau at $\sigma_{xy}=\pm 2e^2/h$
as shown in Fig.~\ref{fig:AHC}(a), whose sign is determined entirely by $\mathrm{sgn}(m_z)$.
In comparison, the in-plane perturbations $\Delta_x$ and $\Delta_y$
only shift the band-node positions without altering the Chern number as discussed in Sec.~\ref{subsec_Chern_number},
so the plateau is topologically robust against these in-plane symmetry-breaking terms.
Away from the gap, Fig.~\ref{fig:AHC}(a) shows that $\sigma_{xy}$ evolves continuously
with \(E_F\) as the occupied manifold
samples regions of varying Berry curvature.

For the hybrid scenario, the transport properties are determined by the gap-closing
criterion in Eq.~(\ref{eq:condition}), and the resulting AHC depends on which mechanism opens the bulk gap.

When the gap is opened purely by the static mass ($\phi=0$ or $\pi$, so $\gamma=0$),
the system reduces to the perturbative limit with $\tilde{\Delta}_x=\Delta_x+\delta_x$ as
presented in Sec.~\ref{subsubsec_BC_hybrid}. As a consequence, the AHC is unchanged.

More interestingly, even without a static mass ($m_z=0$), a finite $\gamma$ combined with a nonzero $\Delta_y$
can open a Floquet-induced gap and stabilize a Chern-insulator phase shown in Fig.~\ref{fig:AHC}(b). The quantized
plateau $\sigma_{xy}=\mp 2e^2/h$ is then governed by $\mathrm{sgn}(\phi\Delta_y)$ rather than by $m_z$.
While the plateau height is topologically protected, stronger driving sharpens the Berry-curvature hot spots
and thereby accelerates the variation of the integrated Berry curvature as $E_F$ across the band edges.
This renders the off-gap evolution of $\sigma_{xy}$ highly sensitive to the TFOPL amplitudes $A_{1,2}$,
which is absent in the perturbative case.

Furthermore, when $\Delta_y=0$, the topology is dictated by the static parameters. For $\tilde{\Delta}_x>0$,
gap closing occurs only at $m_z=0$. Since the Chern number $C=\pm2\,\mathrm{sgn}(m_z)$ is fixed for $m_z\neq0$,
the AHC is insensitive to the Floquet parameters, as shown in Fig.~\ref{fig:AHC}(c). By contrast,
for $\tilde{\Delta}_x<0$, three distinct gap-closing points appear at $m_z=0,\pm\sqrt{-\tilde{\Delta}_x\gamma^2/g_1}$,
as derived in Sec.~\ref{subsub_coop_effects_three_roots}. As depicted in Fig.~\ref{fig:AHC}(d), tuning $\gamma$
shifts these critical points, and crossing a critical \(m_z\) drives topological transitions that flip the sign
of the AHC plateau. Within a gapped phase, however, the AHC versus $E_F$ remains insensitive to the TFOPL parameters,
which is required by the topological invariance of the Chern number.

Accordingly, in the hybrid scenario, Floquet driving can generate a quantized Hall plateau without
a static mass term or shift topological critical points while leaving the quantized transport values
inside gapped phases unchanged. By contrast, the perturbative parameters directly fix the quantized
values. Thus, the TFOPL controls where transitions occur, while static parameters determine
what quantized value is realized.

\subsection{Anomalous Nernst conductivity}\label{subsec_ANC}

\begin{figure*}
\centering
\subfigure[]{\includegraphics[width=2.8in]{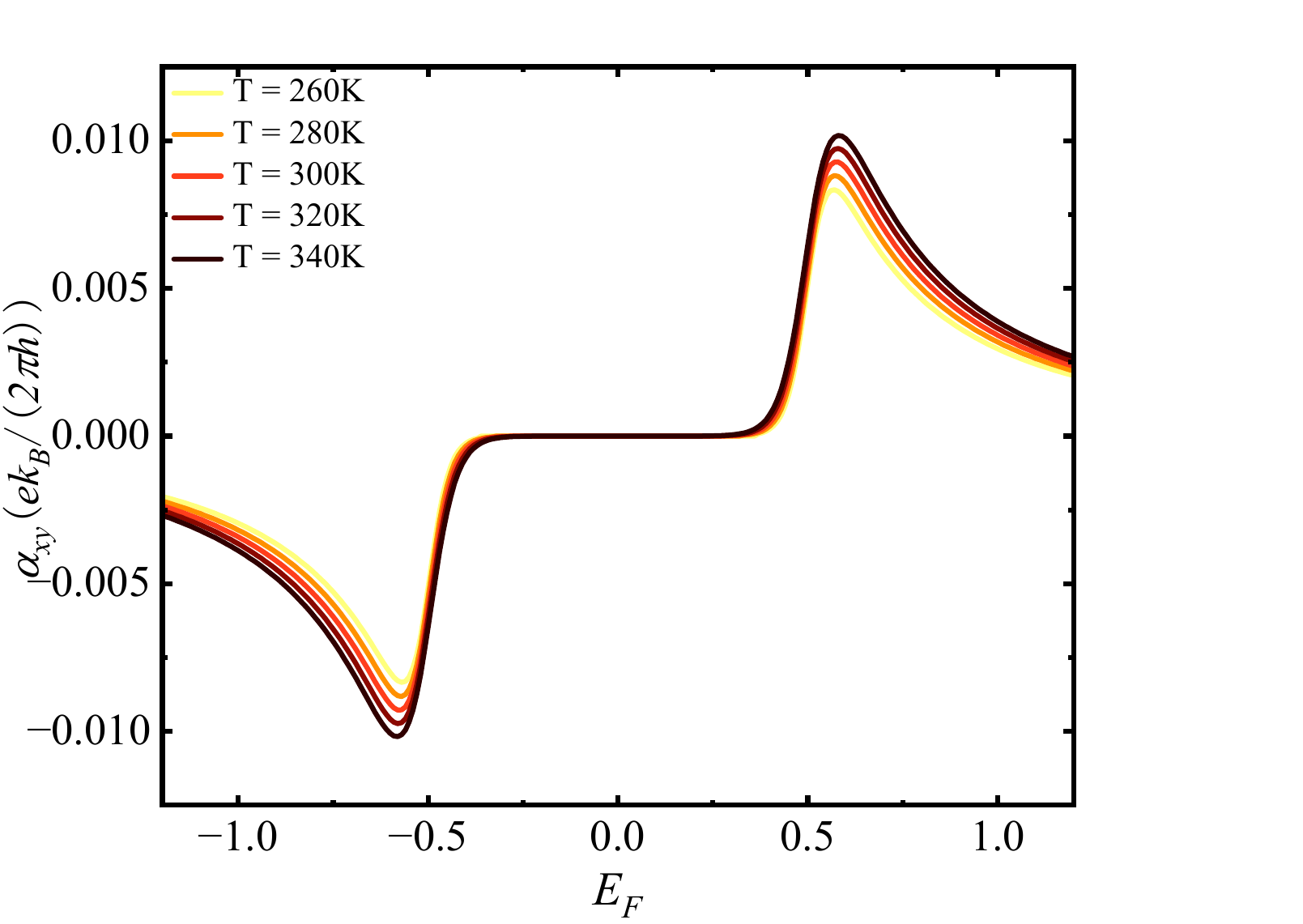}}
\hspace{-0.1cm}
\subfigure[]{\includegraphics[width=2.8in]{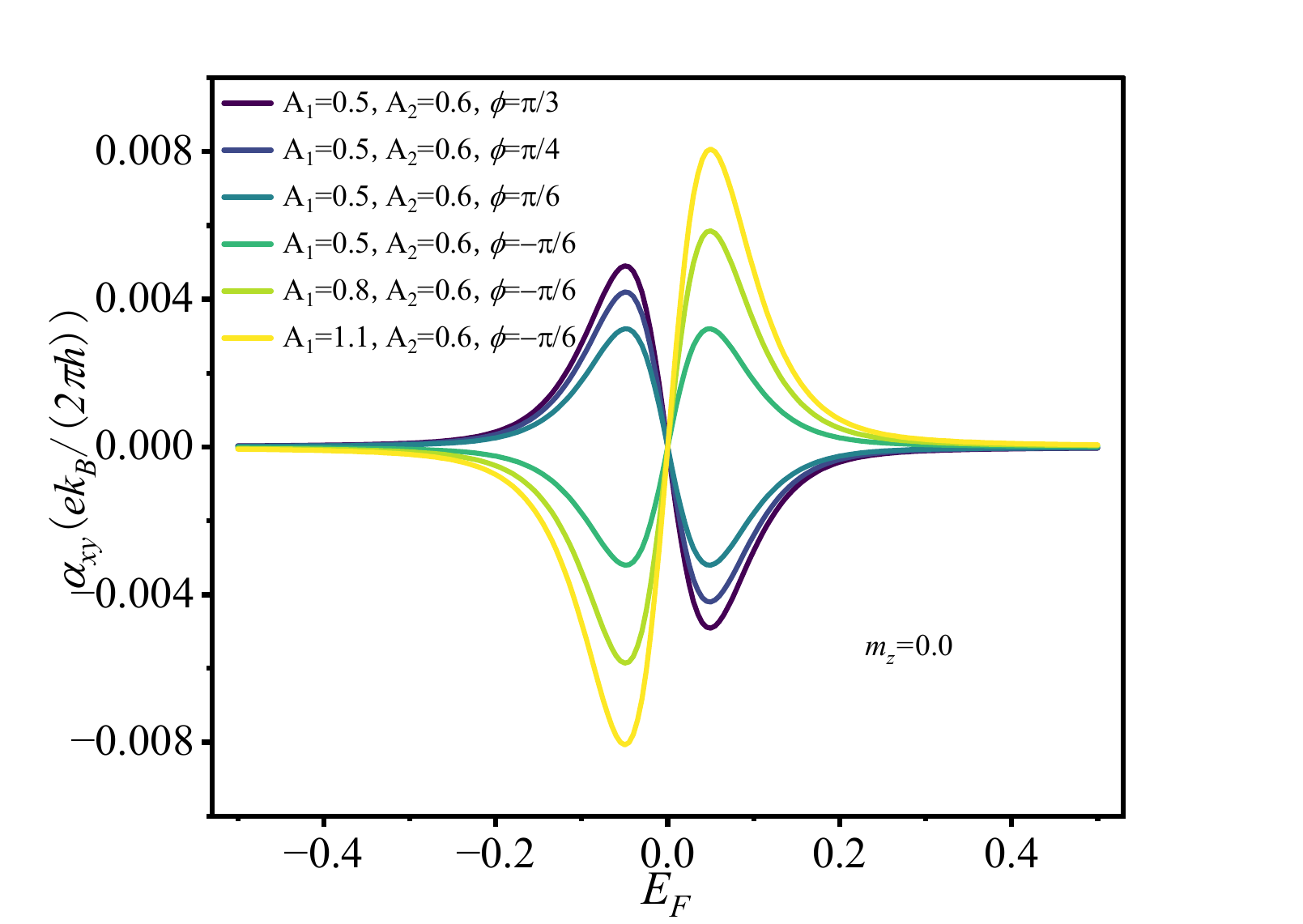}}\\
\vspace{-0.1cm}
\subfigure[]{\includegraphics[width=2.8in]{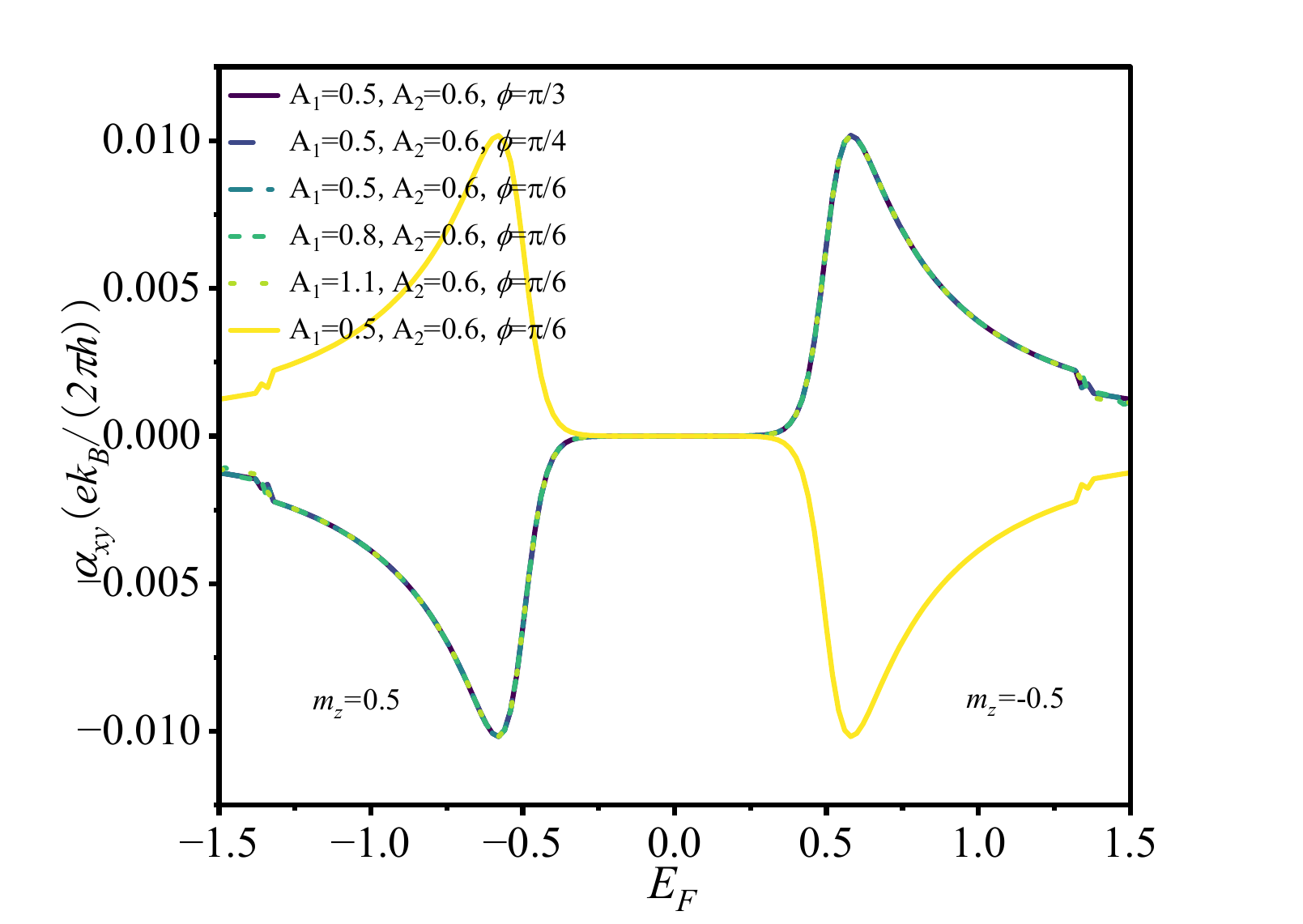}}
\hspace{-0.1cm}
\subfigure[]{\includegraphics[width=2.8in]{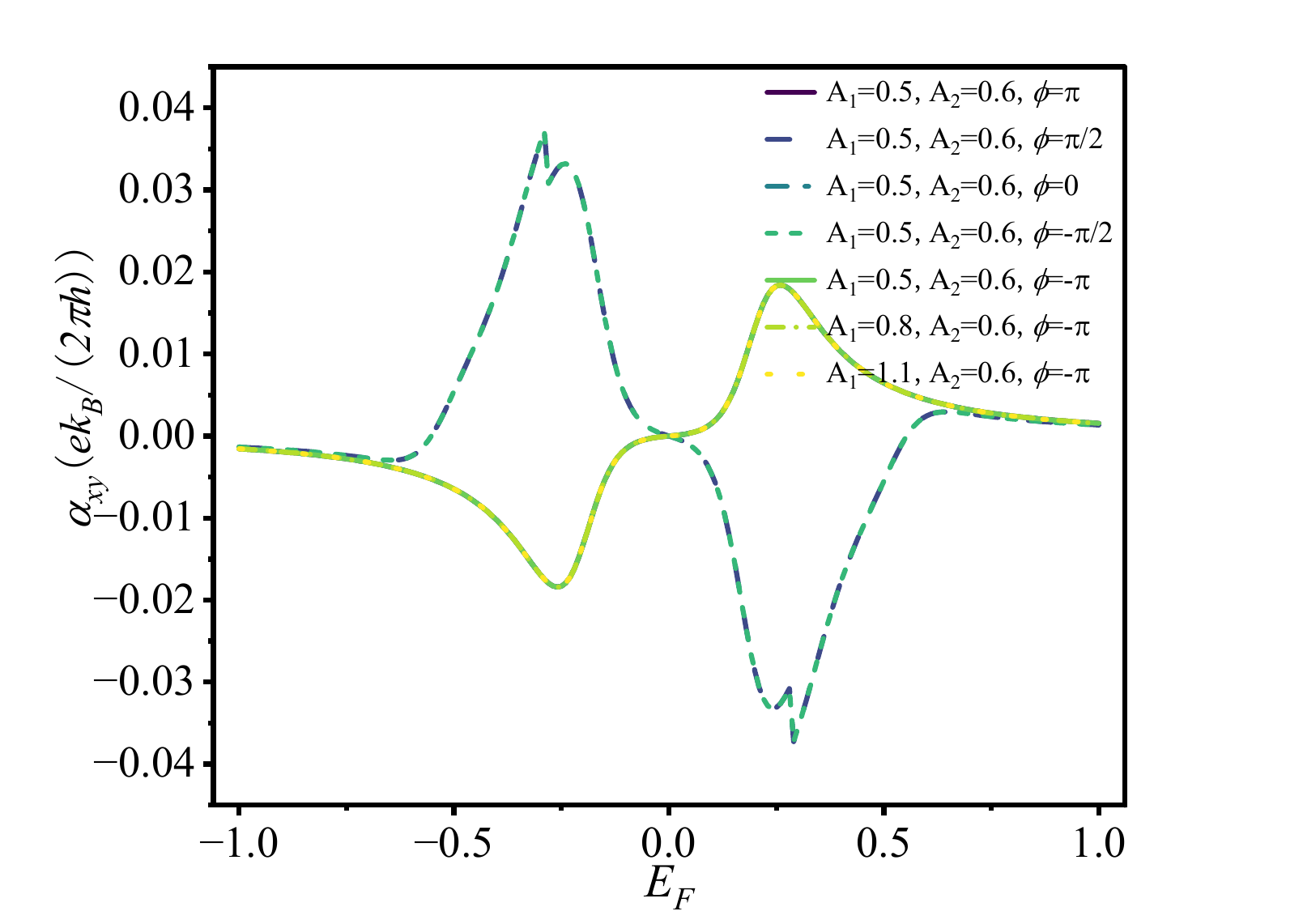}}
\vspace{-0.2cm}
\caption{(Color online) The ANC as a function of the Fermi energy $E_F$
in the (a) perturbative scenario for different temperatures $T$, with $\Delta_x = 0.4$,
$\Delta_y = 0.6$, and $m_z = 0.5$, and hybrid scenario at $T = 340$ K for (b) $m_z = 0$, $\Delta_x = 0.8$, $\Delta_y = 0.6$,
(c) $\Delta_y = 0$, $\Delta_x = 0.8$, $m_z=\pm 0.5$ and (d) $\Delta_y = 0$, $\Delta_x = -0.8$, $m_z = 0.2$.}
\label{fig:ANC}
\end{figure*}

The ANC is fundamentally a Fermi-surface property~\cite{Xiao2006PRL,Han2026arXiv}.
Physically, it describes the generation of a transverse charge current in response to a
longitudinal thermal gradient and takes the form of~\cite{Vargiamidis2025PRB,Xiao2006PRL,Han2026arXiv},
\begin{eqnarray}
\alpha_{xy}\equiv\frac{e}{\hbar}\int\frac{d^2\mathbf{k}}{(2\pi)^2}\sum_n S_n(\mathbf{k})\Omega_z^n(\mathbf{k}),\label{Eq_alpha_xy}
\end{eqnarray}
where the $n$ serves as the energy band index. Here, $\Omega_z^n$ specifies the BC
and $S_n(\mathbf{k})$ is the fermionic entropy defined as~\cite{Han2026arXiv}
\begin{eqnarray}
S_n(\mathbf{k})\equiv-k_B\left[f_n\log f_n+(1-f_n)\log (1-f_n)\right],
\end{eqnarray}
which describes the entropy density of the fermions.

To proceed, the ANC can be expressed in a form that directly reveals its dependence on the
BC and the Fermi-Dirac distribution. Using integration by parts, the definition~(\ref{Eq_alpha_xy})
can be rewritten in the exact form~\cite{Maryam2026arXiv}
\begin{eqnarray}
\alpha_{xy}=\frac{ek_B}{\hbar}\!\sum_n\!\int \!\frac{d^2\mathbf{k}}{(2\pi)^2}
\Omega_z^n(\mathbf{k})\ln\left(\!1+e^\frac{E_F-E}{k_B T}\!\right).\label{eq:ANC}
\end{eqnarray}
In the low-temperature limit $k_B T \ll E_F$, the ANC is well approximated by the Mott relation~\cite{Maryam2026arXiv},
\begin{eqnarray}
\alpha_{xy}\sim -\frac{\pi k_B^2 T}{3e}\frac{\partial \sigma_{xy}(E)}{\partial E}\bigg|_{E=E_F}.\label{eq:MR}
\end{eqnarray}
This relation connects the ANC to the energy derivative of the AHC. A key distinction from the AHC
is that the ANC is not strictly quantized inside the bulk gap. Instead, it is governed by the local
Berry-curvature distribution near the Fermi surface, making it a sensitive probe of low-energy
band geometry.

Combining Eq.~(\ref{eq:ANC}) with the TKNN relation~\cite{Thouless1982PRL} and the Chern number results from Sec.~\ref{subsec_Chern_number}, we can now examine the behavior of the ANC.

In the perturbative scenario, the Chern number is given by Eq.~(\ref{eq:C_per}),
which indicates that the sign of $\alpha_{xy}$ is governed by \(\mathrm{sgn}(m_z)\),
In addition, the in-plane perturbations $\Delta_x$ and $\Delta_y$ affect only the peak
and trough positions, leaving the overall amplitude unchanged, which reflects the topological
robustness of the integrated Berry curvature. At low temperature, $\alpha_{xy}$ varies
linearly with $T$ as shown in Fig.~\ref{fig:ANC}(a), which is expected from
the Mott relation, Eq.~(\ref{eq:MR}).

For the hybrid scenario, the ANC is a Fermi-surface property whose behavior depends on the
gap-opening mechanism. When the gap is opened purely by the TFOPL field with $m_z=0$, a finite ANC
emerges whose amplitude increases with $A_{1,2}$ or $|\phi|$, as clearly illustrated in Fig.~\ref{fig:ANC}(b)],
and whose sign follows $\mathrm{sgn}(\phi\Delta_y)$. Since the ANC is a Fermi-surface property, it is nonzero
only when $E_F$ lies within the bands and vanishes inside the light-induced gap~\cite{Xiao2006PRL,Saha2018EPJB,Ferreiros2017PRB,Alisultanov2025arXiv}. This
provides a clear experimental signature of Floquet-generated gaps even in the absence of a static mass term.
In comparsion, when the gap is instead solely governed by a nonzero static mass at $\Delta_y=0$ and $\tilde{\Delta}_x\geq0$,
Fig.~\ref{fig:ANC}(c) shows that $\alpha_{xy}$ reproduces the perturbative behavior and is nearly insensitive to
the Floquet parameters. For $\tilde{\Delta}_x<0$  as depicted in Fig.~\ref{fig:ANC}(d), however, multiple
light-tunable gap-closing points obtained from Sec.~\ref{subsub_coop_effects_three_roots} give rise to
a strongly nonmonotonic $\alpha_{xy}(E_F)$ with successive sign reversals, reflecting crossings of
multiple Floquet-shifted topological boundaries.

Compared with AHC in Sec.~\ref{subsec_AHC}, the ANC provides a local probe that
complements the global information obtained from the AHC. By resolving Fermi-surface-weighted
local Berry-curvature features, the ANC can detect light-induced gaps even when the global topological
invariants remain unchanged. This makes it a valuable tool for characterizing Floquet-induced phases.

\subsection{Anomalous thermal Hall conductivity}\label{subsec_ATHC}

The ATHC describes the generation of a transverse heat current in response to a longitudinal temperature gradient,
$J_y^Q = -\kappa_{xy}^{(0)} \partial_x T$~\cite{Vargiamidis2025PRB, Han2026arXiv}. To evaluate this quantity,
we can consider it as a purely intrinsic response arising from the BC of the occupied bands,
and then is given by the $c_2$-function formula derived via Luttinger's gravitational potential approach, ~\cite{Vargiamidis2025PRB,Maryam2026arXiv,Luttinger1964PR,Qin2011PRL},
\begin{eqnarray}
\kappa_{xy}^{(0)}\equiv-\frac{k_B^2T}{\hbar}\!\sum_n\!\!\int \!\!\frac{d^2\mathbf{k}}{(2\pi)^2}\!\Omega_z^n(\mathbf{k})c_2\left[f(E_n-E_F)\right],\label{Eq_kappa_xy}
\end{eqnarray}
where the function $c_2(f)$ is defined as
\begin{eqnarray}
c_2(f)\equiv\int_0^f\left(\ln\frac{1-t}{t}\right)^2dt,
\end{eqnarray}
which satisfies $c_2(0)=0$ and $c_2(1)=\pi^2/3$.

An interesting feature of the ATHC in a gapped system is its universal quantization at low temperature.
In this regime, when the chemical potential lies inside the bulk gap, all occupied states satisfy $f\to1$,
so that $c_2(f)\rightarrow\pi^2/3$. The ATHC therefore reduces to the simple quantized form
\begin{eqnarray}
\frac{\kappa_{xy}^{(0)}}{T}\bigg|_{T\to 0}\!=-\frac{\pi^2k_B^2}{3\hbar}
\times\frac{C}{2\pi}=-\frac{\pi^2k_B^2}{3h}C,\label{eq:ATHC}
\end{eqnarray}
where $C$ is the Chern number. This universal thermal quantization provides a temperature-dependent
fingerprint of the Chern number that is independent of electrical transport probes.

\begin{figure*}
\centering
\subfigure[]{\includegraphics[width=2.8in]{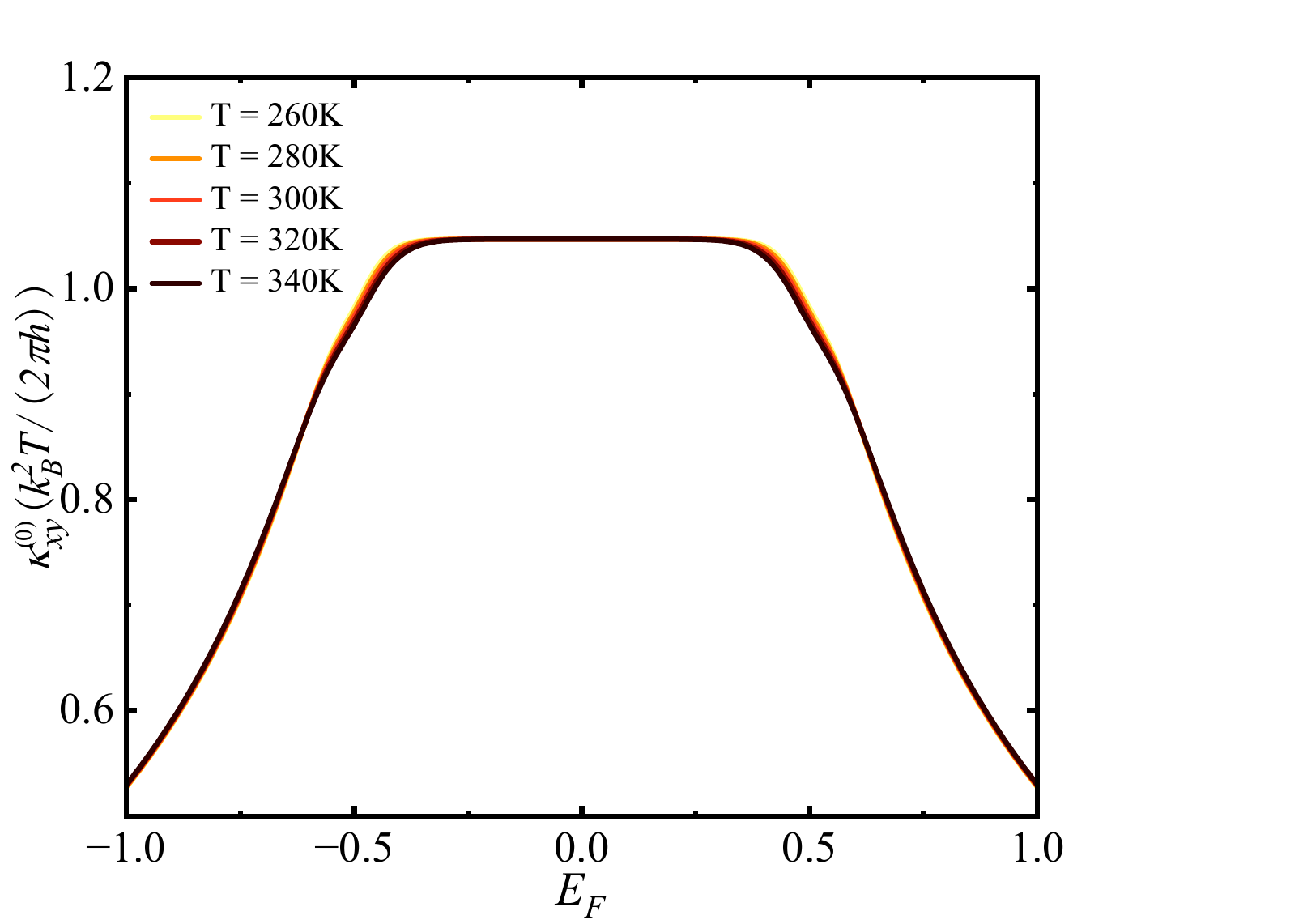}}
\hspace{-0.1cm}
\subfigure[]{\includegraphics[width=2.8in]{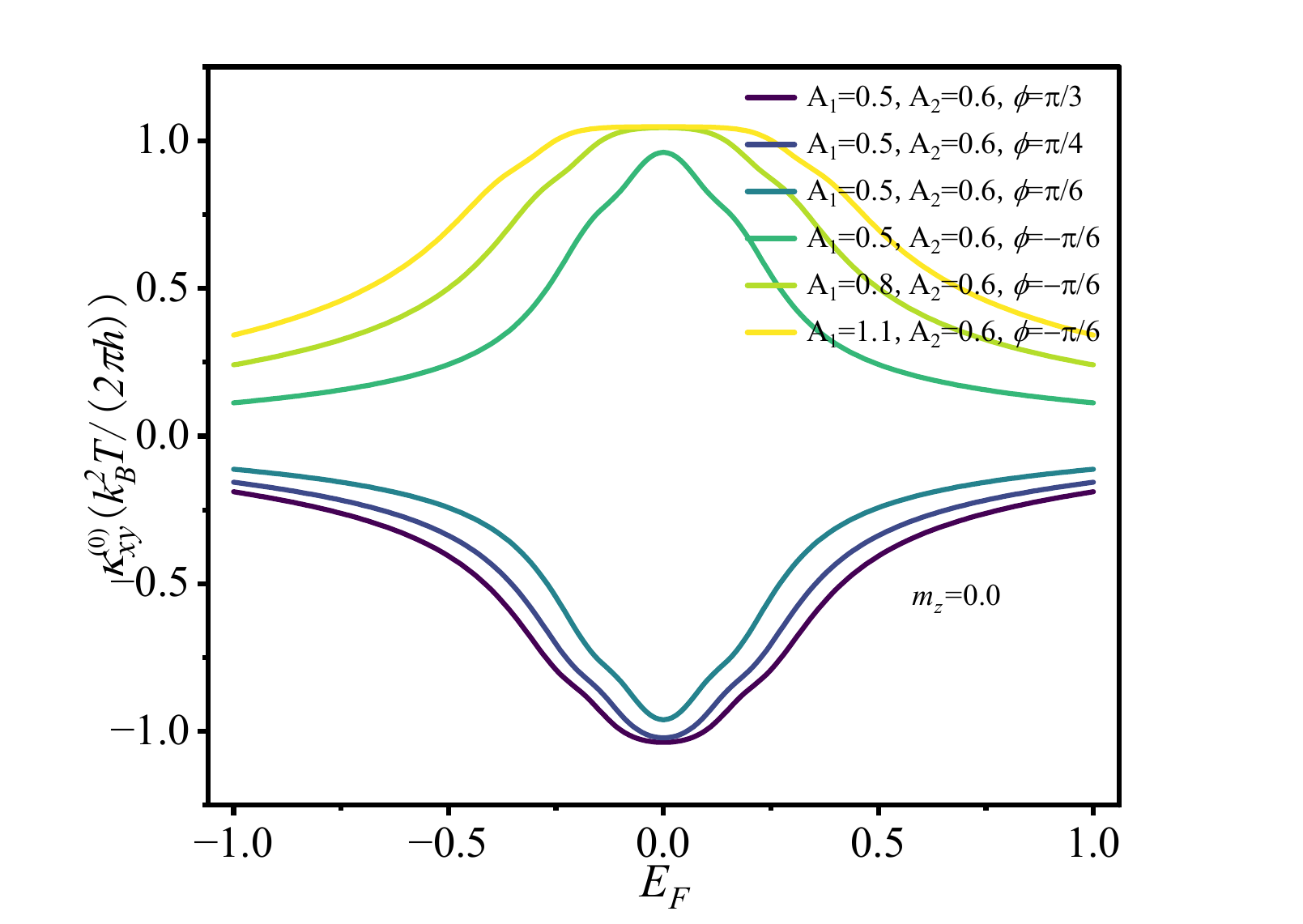}}\\
\vspace{-0.1cm}
\subfigure[]{\includegraphics[width=2.8in]{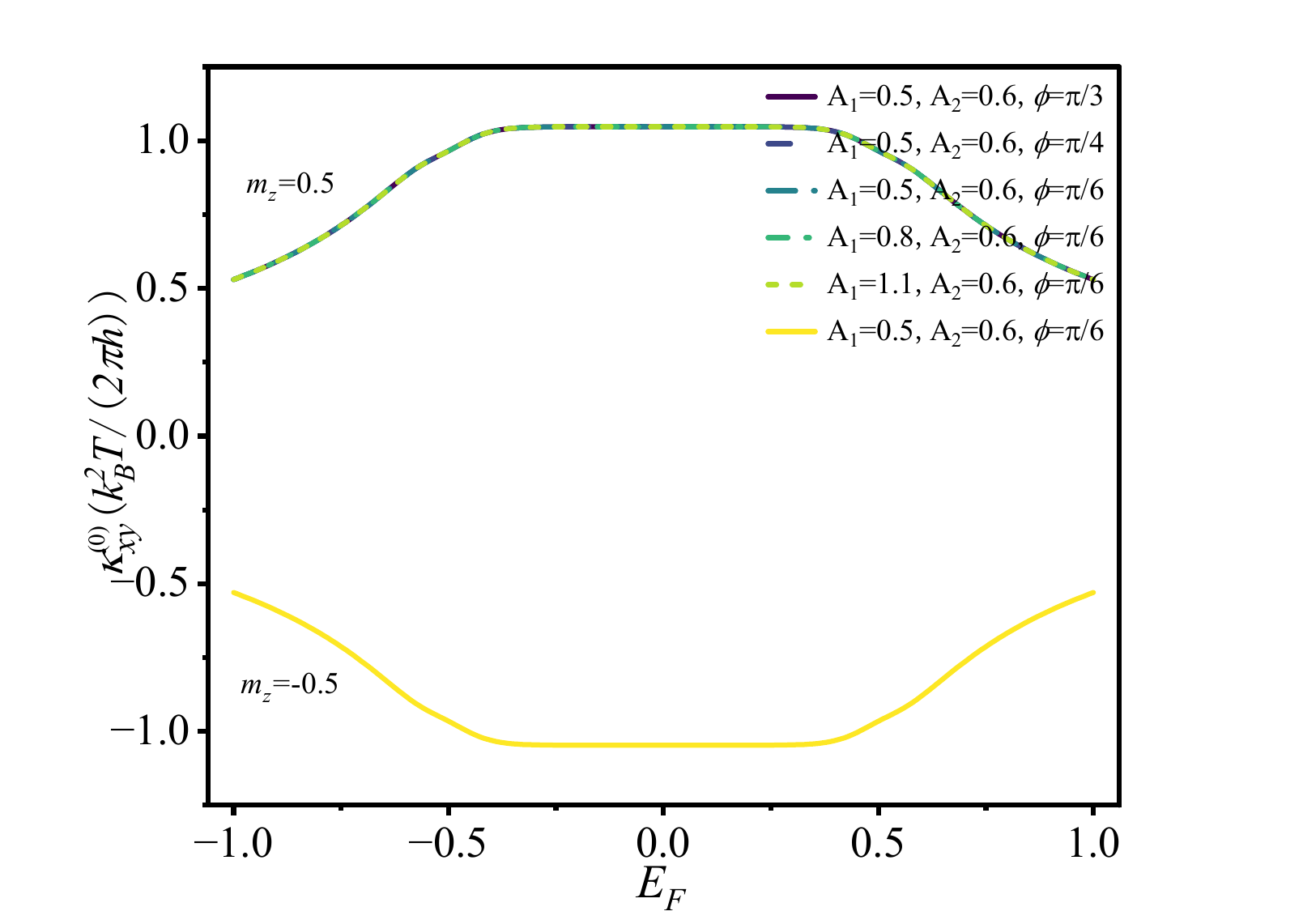}}
\hspace{-0.1cm}
\subfigure[]{\includegraphics[width=2.8in]{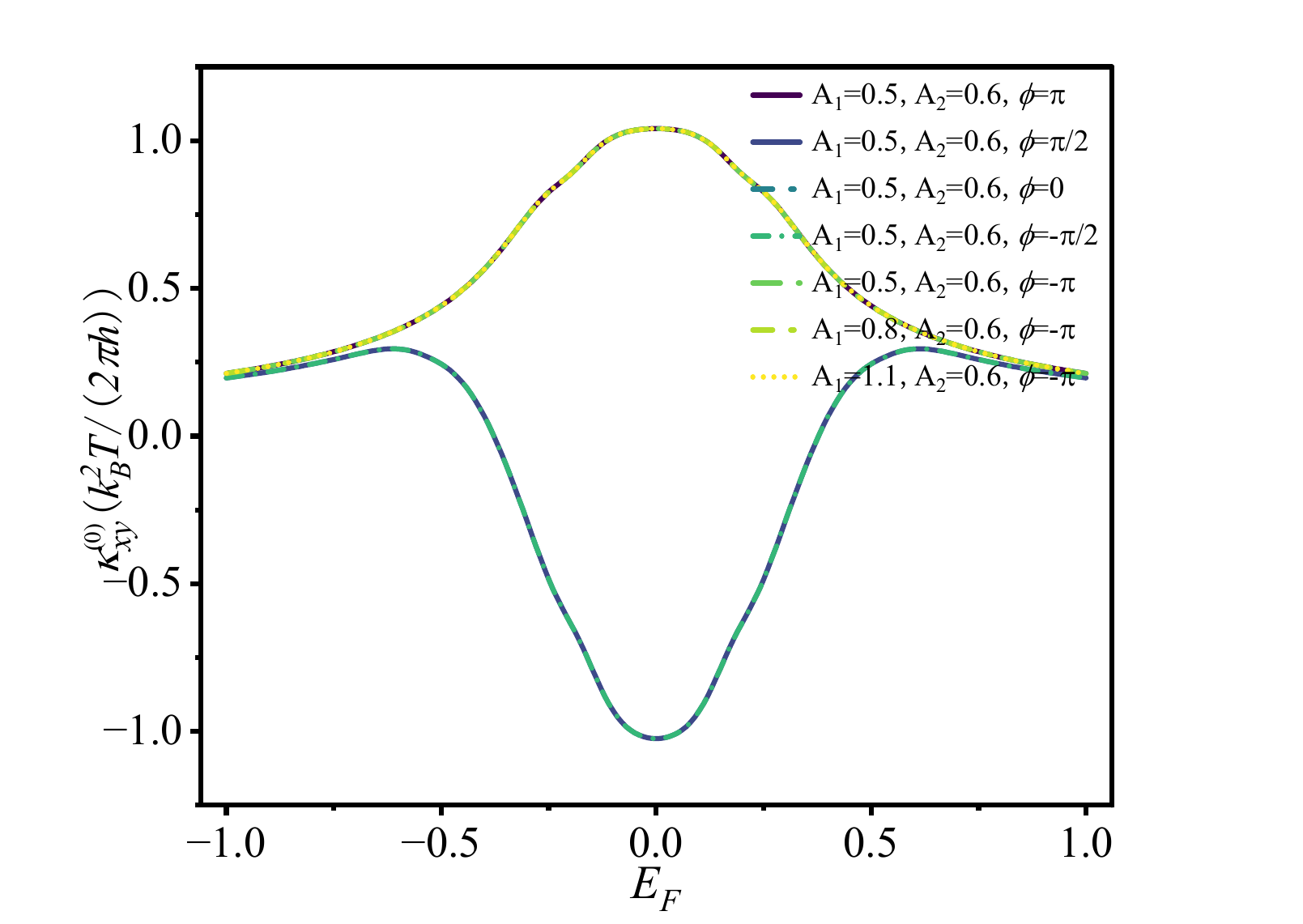}}
\vspace{-0.2cm}
\caption{(Color online) The ATHC as a function of the Fermi energy $E_F$ in the (a) perturbation scenario for different $T$ with $\Delta_x=0.4$, $\Delta_y=0.6$, $m_z=0.5$, and hybrid scenario at $T=340$K for (b) $m_z=0$, $\Delta_x=0.8$, $\Delta_y=0.6$, (c) $\Delta_y=0$, $\Delta_x=0.8$, $m_z=\pm0.5$ and (d) $\Delta_y=0$, $\Delta_x=-0.8$, $m_z=0.2$.}
\label{fig:ATHC}
\end{figure*}

Using the BC results from Sec.~\ref{subsec_BC} together with Eq.~(\ref{Eq_kappa_xy}),
we can now investigate the behavior of the ATHC.

In the perturbative case, the sign of the quantized plateau in $\kappa_{xy}^{(0)}/T$
is determined by $\mathrm{sgn}(m_z)$. This quantization survives at finite temperature provided
the thermal energy remains small compared to the gap. For the temperature range
considered here $(260\text{¨C}340\ \mathrm{K})$, the condition $k_BT \ll |m_z|$
is satisfied, and the chemical potential lies well inside the gap, so thermal
broadening is negligible. Consequently, the plateau remains essentially undistorted
as shown in Fig.~\ref{fig:ATHC}(a).

We then consider the hybrid scenario. For $m_z=0$, as discussed
in Sec.~\ref{subsub_coop_effects_three_roots}, the cooperative effects
enable the TFOPL to induce a gap, which produces a quantized plateau as
shown in Fig.~\ref{fig:ATHC}(b). Compared with the purely static-mass
case in Fig.~\ref{fig:ATHC}(a), the plateau height deviates from the
ideal universal value at finite temperature whenever the light-induced
gap becomes comparable to \(k_BT\). Increasing the light intensity
enlarges the effective gap, thereby restoring the ideal quantized magnitude.
The sign of the plateau remains determined by $\mathrm{sgn}(\phi\Delta_y)$,
which is consistent with the corresponding Chern number in Eq.~(\ref{Eq_hyb_C}).
This provides a distinct Floquet thermal topological response, in which the
quantization quality can be continuously tuned by the TFOPL driving strength.

For $\Delta_y=0$, the ATHC exhibits two distinct regimes depending on the sign
of $\tilde{\Delta}_x$. For $\tilde{\Delta}_x\ge0$, it remains insensitive to
the Floquet parameters deep inside the gapped phase, as shown in Fig.~\ref{fig:ATHC}(c).
For $\tilde{\Delta}_x<0$, by contrast, the behavior parallels the ANC case of
Sec.~\ref{subsec_ANC}, as shown in Fig.~\ref{fig:ATHC}(d), the multiple light-tunable
gap-closing points obtained in Sec.~\ref{subsub_coop_effects_three_roots} generate
a sequence of sign-reversed thermal Hall plateaus separated by gapless semimetallic windows.
However, for fixed parameters within a single gapped topological phase, the ATHC versus
$E_F$ curve remains unaffected by quantitative variations in the optical amplitudes or phase.

To recapitulate, the AHC and ATHC provide quantized signatures of Chern-insulator phases, while the ANC provides complementary Fermi-surface information. In the perturbative case, topological transitions are controlled solely by $m_z$.
In the hybrid scenario, quantized Hall plateaus emerge at $m_z=0$ with sign $\mathrm{sgn}(\phi\Delta_y)$
and light-tunable quantization quality, and for $\tilde{\Delta}_x<0$, multiple light-tunable
gap-closing points lead to sequential phase transitions as $m_z$ is varied. Owing to topological
protection, these responses remain robust against Floquet-parameter variations deep inside the gapped phases.
Combined AHC, ANC, and ATHC measurements can therefore confirm the cooperative effects between the
perturbative and Floquet scenarios.

\section{Summary}\label{Sec_summary}

In summary, we present a comprehensive study of the topological and anomalous
transport properties of 2D type-II SDSMs under static perturbations, Floquet
driving by TFOPL, and their cooperative interplay, corresponding to the perturbative, Floquet,
and hybrid scenarios, respectively.

We begin by introducing the models. The free type-II SDSM Hamiltonian in Eq.~(\ref{eq:Hamiltonian})
is characterized by an off-diagonal cross term $g_2 k_x k_y \sigma_y$ that couples the two momentum
components, leading to a strongly anisotropic band structure with a single nodal point at $\Gamma$.
In the perturbative scenario, the Hamiltonian in Eq.~(\ref{eq:H_eff}) incorporates in-plane pseudospin
shifts $(\Delta_x,\Delta_y)$ and a spin-orbit mass term $m_z\sigma_z$. Driving the Floquet scenario
with TFOPL produces a static shift $\delta_x\sigma_x$ and, for $\phi\neq0,\pi$,
a momentum-dependent Zeeman term $\gamma k_x\sigma_z$ with $\gamma\propto\sin\phi$.
The hybrid scenario combines static perturbations and Floquet driving, with the total
Hamiltonian in Eq.~(\ref{Eq_h_xyz}) governed by their cooperative interplay.

Next, we find that these models give rise to distinct topological properties.
The free model intrinsically breaks both spatial-inversion and time-reversal symmetries,
rendering the system nonreciprocal and topological trivial. In the perturbative scenario,
the mass term $m_z\sigma_z$ opens a full bulk gap of size $2|m_z|$ and drives the system
into a Chern insulator with $C=\pm2\,\mathrm{sgn}(m_z)$. By contrast, the
spectrum remains gapless at $\mathbf{k}=(0,\delta_x/v)$ in the Floquet scenario,
so no well-defined Chern number exists. However, the hybrid scenario clearly reveals
the cooperative effects. On the one hand, the TFOPL acts as a catalyst for mass generation, namely
for $m_z=0$, any $\gamma\neq0$ and $\Delta_y\neq0$ are able to open a gap
through an effective mass $m_{\mathrm{eff}}\propto\gamma\Delta_y$, yielding
a Chern insulator with $C=\mp2\,\mathrm{sgn}(\phi\Delta_y)$.
On the other hand, for $\Delta_y=0$ and $\tilde{\Delta}_x<0$, the gap-closure
criterion~(\ref{eq:condition}) admits up to three real roots for $m_z$, producing
multiple gap-closing points and a sequence of insulator¨Csemimetal
transitions with the Chern number alternating between $\pm2$ as shown in Fig.~\ref{fig:condition}
and Fig.~\ref{fig:Cnumber}. The cubic term $-v\gamma^3\Delta_y$,
of purely hybrid origin, breaks the $\Delta_y\to-\Delta_y$ symmetry and reflects
the Floquet-induced chirality in momentum space. Fig.~\ref{fig:Cnumber} and Table~\ref{tab:Cnumber}
collect the topological phase transitions and
Chern numbers for different situations.

Subsequently, these topological distinctions are imprinted on the band geometry
through the BC ($\Omega_z(\mathbf{k})$), and BCD ($D_{x,y}$) as illustrated in Fig.~\ref{fig:bc_bcd}
and Fig.~\ref{fig:BCD}. In the perturbative case, the BC is concentrated around the shifted
Dirac points, forming two lobes of opposite sign whose magnitudes are governed by $m_z$.
The $k_x\rightarrow-k_x$ mirror symmetry that is present at $\Delta_y=0$ forces the BC to
be even in $k_x$, yielding $D_x=0$ but $D_y\neq0$. Introducing a finite $\Delta_y$ breaks
this symmetry and skews the BC distribution, rendering both components nonzero.
In the Floquet scenario, the light-induced $\gamma k_x\sigma_z$ term renders the BC odd in $k_x$,
so that its positive and negative lobes are distributed antisymmetrically about $k_x=0$,
with amplitudes scaling as $\gamma\propto\sin\phi$. Consequently, $D_y$
vanishes identically while $D_x\propto\sin\phi$, vanishing at $\phi=0,\pi$
and tunable via the light intensity. In sharp contrast, the hybrid scenario with generic $\phi\neq0,\pi$,
as shown in Fig.~\ref{fig:BCD}, breaks all reflection symmetries of the BC, producing a fully asymmetric distribution
with sharp hot spots near gap-closing manifolds and a characteristic double-peak structure
in $D_x$ and $D_y$. These features enable topological switching across phase boundaries
and a synchronized sign reversal of the full BCD vector $\mathbf{D}=D_x\mathbf{e}_x+D_y\mathbf{e}_y$
(or its $y$-component $D_y$) under an external parameter.

Having established the topological phase diagram through the Chern number and the momentum-space structure
characterized by the BC and BCD, we furthermore demonstrate that these properties are reflected in the
anomalous transport coefficients shown in Figs.~\ref{fig:AHC}¨C\ref{fig:ATHC}. In the perturbative
scenario, since the ANC is a Fermi-surface probe, it vanishes when the Fermi level lies inside the gap,
whereas the AHC and ATHC exhibit quantized plateaus determined by $\mathrm{sgn}(m_z)$, with
$\sigma_{xy}=e^2C/h$ and $\kappa_{xy}^{(0)}/T=-(\pi^2k_B^2/3h)C$,
respectively. In the Floquet scenario, the BC possesses odd parity under $k_x\rightarrow-k_x$,
which forces its BZ integral to vanish and therefore produces no contribution to
these transport coefficients. In the hybrid scenario, the AHC and ATHC can develop quantized plateaus
even at $m_z=0$, with signs governed by $\mathrm{sgn}(\phi\Delta_y)$, and the TFOPL preserves the
quantized values inside a gapped phase but shifts the critical points. For $m_z\neq0$ and
$\tilde{\Delta}_x\geq0$, the ATHC plateau sign depends on $m_z$. By contrast, for $\tilde{\Delta}_x<0$,
a cascade of sign-reversed thermal Hall plateaus appears, separated by gapless windows. Meanwhile,
the ANC develops a strongly nonmonotonic, sign-reversing structure for $\tilde{\Delta}_x<0$ as
multiple light-tunable gap-closing points are crossed.

To recapitulate, the combined static and Floquet controls offer versatile experimental handles,
including laser amplitude, frequency, phase ($\phi$), and gate-tunable ($\Delta_x,\Delta_y,m_z$),
for tuning topological phase transitions and transport signatures. These results provide valuable
insights for further studies on topological properties in 2D type-II SDSMs and related systems.

\section*{ACKNOWLEDGEMENTS}

J.W. is supported by Tianjin Natural Science Foundation Project (25JCYBJC01640) and
was partially supported by the National Natural
Science Foundation of China under Grant No. 11504360.


\appendix

\section{Derivation of the Floquet Effective Hamiltonian}\label{AppendixA}

Here, we provide a detailed derivation of the Floquet effective Hamiltonian (\ref{eq:H_eff}).
To incorporate Floquet driving, we introduce the TFOPL field defined in Eq.~(\ref{eq:light_vector})
into the 2D SDSM Hamiltonian of Eq.~(\ref{eq:Hamiltonian}) via the minimal substitution
$\mathbf{k}\to\mathbf{k}+e\mathbf{A}(t)$, which yields the effective Hamiltonian.
\begin{eqnarray}
\mathcal{H}(\mathbf{k},t)
&=&
[g_1(k_x+eA_x)^2-v(k_y+eA_y)]\sigma_x\nonumber\\
&&+g_2(k_x+eA_x)(k_y+eA_y)\sigma_y.
\end{eqnarray}
After expanding and separating the static, the linear, and the quadratic terms, we are left with
\begin{eqnarray}
\mathcal{H}(\mathbf{k},t)
&=&
\mathcal{H}(\mathbf{k})+e(2g_1A_xk_x-vA_y)\sigma_x\nonumber\\
&&
+eg_2(A_yk_x+A_xk_y)\sigma_y\nonumber\\
&&
+e^2 g_1A_x^2\sigma_x+e^2g_2A_xA_y\sigma_y,\label{Eq_H_k_t}
\end{eqnarray}
where $\mathcal{H}(\mathbf{k})$ is defined in Eq.~(\ref{eq:Hamiltonian}).

For later convenience, we express the vector potential~(\ref{eq:light_vector}) in complex form,
\begin{eqnarray}
A_x(t)
&=&
\frac{A_1}{2}(e^{i\omega t}+e^{-i\omega t}),\\
A_y(t)
&=&
\frac{A_2}{2}(e^{i\phi}e^{i2\omega t}+e^{-i\phi}e^{-i2\omega t}),
\end{eqnarray}
To proceed, we expand $\mathcal{H}(\mathbf{k},t)$ in Eq.~(\ref{Eq_H_k_t}) as a Fourier series in $e^{-im\omega t}$
and define the Fourier components $H_m(\mathbf{k})$ by~\cite{Wang2023PRA,Valagiannopoulos2007PIER}
\begin{eqnarray}
\mathcal{H}(\mathbf{k},t)=\sum_{m=-\infty}^\infty H_m(\mathbf{k})e^{-im\omega t}.
\end{eqnarray}
The only nonzero Fourier components are those with $m=0,\pm1,\pm2,\pm3$. This is because the components $A_x$ and $A_y$
contribute frequency components at $\omega$ and $2\omega$, and their product generates a $3\omega$ component. Consequently,
the expansion reduces to a finite sum over $m=-3,\ldots,3$.

For the static part $\mathcal{H}(\mathbf{k})$ in $\mathcal{H}(\mathbf{k},t)$, it contributes only to
the $m=0$ term. As to the linear term in $\mathcal{H}(\mathbf{k},t)$, we can further write it as,
\begin{eqnarray}
\mathcal{H}^{\mathrm{lin}}(\mathbf{k},t)
&=&
e(2g_1A_xk_x-vA_y)\sigma_x\nonumber\\
&&+eg_2(A_yk_x+A_xk_y)\sigma_y\nonumber\\
&=&
H_1^{\mathrm{lin}}e^{-i\omega t}+H_{-1}^{\mathrm{lin}}e^{i\omega t}\nonumber\\
&&
+H_2^{\mathrm{lin}}e^{-i2\omega t}+H_{-2}^{\mathrm{lin}}e^{i2\omega t},
\end{eqnarray}
with
\begin{eqnarray}
H_1^{\mathrm{lin}}
&\equiv&
eg_1A_1k_x\sigma_x+e\frac{g_2A_1k_y}{2}\sigma_y,\\
H_2^{\mathrm{lin}}
&\equiv&
-e\frac{vA_2}{2}e^{-i\phi}\sigma_x+e\frac{g_2A_2k_x}{2}e^{-i\phi}\sigma_y,
\end{eqnarray}
and $(H_{-1,-2}^{\mathrm{lin}})^\dagger=H_{1,2}^{\mathrm{lin}}$.
Similarly, the quadratic term in $\mathcal{H}(\mathbf{k},t)$ can be rewritten as
\begin{eqnarray}
\mathcal{H}^{\mathrm{qua}}(\mathbf{k},t)
&=&
H_0^{\mathrm{qua}}+H_1^{\mathrm{qua}}e^{-i\omega t}+H_{-1}^{\mathrm{qua}}e^{i\omega t}\nonumber\\
&&
+H_2^{\mathrm{qua}}e^{-i2\omega t}+H_{-2}^{\mathrm{qua}}e^{i2\omega t}\nonumber\\
&&+H_3^{\mathrm{qua}}e^{-i3\omega t}+H_{-3}^{\mathrm{qua}}e^{i3\omega t},
\end{eqnarray}
with $(H_{-1,-2,-3}^{\mathrm{qua}})^\dagger=H_{1,2,3}^{\mathrm{qua}}$ and
\begin{eqnarray}
H_0^{\mathrm{qua}}&\equiv&e^2\frac{g_1A_1^2}{2}\sigma_x,\\
H_1^{\mathrm{qua}}&\equiv&e^2\frac{g_2A_1A_2}{4}e^{-i\phi}\sigma_y,\\
H_2^{\mathrm{qua}}&\equiv&e^2\frac{g_1A_1^2}{4}\sigma_x,\\
H_3^{\mathrm{qua}}&\equiv&e^2\frac{g_2A_1A_2}{4}e^{-i\phi}\sigma_y.
\end{eqnarray}

Therefore, by summing the above results over the indices $m$ and $-m$, we obtain,
\begin{eqnarray}
H_0\!&=&\!\left(\!g_1k_x^2-vk_y+e^2\frac{g_1A_1^2}{2}\!\right)\!\sigma_x+g_2k_xk_y\sigma_y,\\
H_1&=&eg_1A_1k_x\sigma_x+\bigg(e\frac{g_2A_1k_y}{2}\nonumber\\
&&
+e^2\frac{g_2A_1A_2}{4}e^{-i\phi}\bigg)\sigma_y,\label{Eq_app_H_1}\\
H_2&=&\left(-e\frac{vA_2}{2}e^{-i\phi}+e^2\frac{g_1A_1^2}{4}\right)\sigma_x\nonumber\\
&&
+e\frac{g_2A_2k_x}{2}e^{-i\phi}\sigma_y,\\
H_3&=&e^2\frac{g_2A_1A_2}{4}e^{-i\phi}\sigma_y,\label{Eq_app_H_2}
\end{eqnarray}
with $H_m^\dagger=H_{-m}$.

Under the high-frequency driving, where the photon energy $\hbar\omega$ is much larger than the original band-gap energy, the effective Hamiltonian to first order in $1/\omega$ is~\cite{Wang2023PRA}
\begin{eqnarray}
H_{\mathrm{eff}}=H_0+\sum_{m\neq0}\frac{[H_{-m},H_{m}]}{m\omega}+\mathcal{O}\left(\frac{1}{\omega^2}\right).\label{Eq_app_18}
\end{eqnarray}

Considering $m=1$, we rewrite Eq.~(\ref{Eq_app_H_1}) as
\begin{eqnarray}
H_1\equiv a_1\sigma_x+b_1\sigma_y,
\end{eqnarray}
with
\begin{eqnarray}
a_1&\equiv&eg_1A_1k_x,\\
b_1&\equiv&e\frac{g_2A_1k_y}{2}+e^2\frac{g_2A_1A_2}{4}e^{-i\phi}.
\end{eqnarray}
This yields $H_{-1}=a_1^*\sigma_x+b_1^*\sigma_y$ for $m=-1$ due to $H_m^\dagger=H_{-m}$.
Accordingly, we can obtain
\begin{eqnarray}
[H_{-1},H_1]&=&2i(a_1^*b_1-a_1b_1^*)\sigma_z\nonumber\\
&=&
e^3 g_1g_2A_1^2A_2 k_x\sin\phi~\sigma_z.\label{Eq_app_22}
\end{eqnarray}
Paralleling the similar analysis for $m=\pm2,\pm3$ leads to
\begin{eqnarray}
[H_{-2},H_2]&=&e^3\frac{g_1g_2A_1^2A_2k_x}{2}\sin\phi~\sigma_z,\label{Eq_app_23}\\
\left[H_{-3},H_3\right]&=&0.\label{Eq_app_24}
\end{eqnarray}
After combining Eq.~(\ref{Eq_app_18}) and Eqs.~(\ref{Eq_app_22})-Eqs.~(\ref{Eq_app_24}),
the corrections by the TFOPL field to the Hamiltonian can be read as
\begin{eqnarray}
H'=\frac{5e^3g_1g_2A_1^2A_2k_x\sin\phi}{4\omega}\sigma_z.\label{Eq_app_25}
\end{eqnarray}

On the basis of Eq.~(\ref{Eq_app_18}) and Eq.~(\ref{Eq_app_25}).  We finally get the Floquet effective Hamiltonian,
\begin{eqnarray}
\mathcal{H}_{\rm eff}(\mathbf{k})
&=&
(g_1k_x^2-v k_y+\delta_x)\sigma_x\nonumber\\
&&+g_2k_xk_y\sigma_y+\gamma k_x\sigma_z,
\end{eqnarray}
where the coefficients $\delta_x$ and $\gamma$
are defined in Eq.~(\ref{Eq_delta_gamma}) of maintext.


\end{document}